\documentclass{article}
\RequirePackage{etex}
\usepackage{graphicx} 
\usepackage[hmargin=1in,vmargin={1.3in,1.3in}]{geometry}
\usepackage{amsthm}
\usepackage{amsmath}
\usepackage{amsfonts}
\usepackage{amssymb}
\usepackage[affil-it]{authblk}

\usepackage{csquotes}
\usepackage{color}
\usepackage{amsmath, amssymb, enumitem, verbatim, color, tikzsymbols, tabularx, dcolumn,longtable, array, pdflscape, soul, bbm, subcaption, rotating}
\usepackage{url}

\usepackage[titletoc,toc,title]{appendix}
\usepackage{verbatimbox} 
\usepackage{etoolbox}
\usepackage{environ}
\NewEnviron{myresizeenv}{\resizebox{\linewidth}{!}{\BODY}}

\usepackage{tocloft}
\makeatletter
\renewcommand{\numberline}[1]{{\@cftbsnum #1\@cftasnum~}\@cftasnumb}
\makeatother

\usepackage{booktabs}
\usepackage{multirow, multicol}
\usepackage{float}

\usepackage{subcaption, caption} 
\usepackage[style=authoryear, citetracker=true, maxcitenames=2, hyperref=true, url=true, isbn=false, doi=true, natbib=true, backend=biber,bibstyle=authoryear, maxbibnames=30, sorting=nyt, uniquename=full, uniquelist=false]{biblatex}
\usepackage[colorlinks=true, allcolors=blue, hyperindex,breaklinks]{hyperref}
\usepackage{setspace}
\usepackage{placeins, adjustbox}
\usepackage{ragged2e}

\title{Revisiting \citet{hotte2025national}: A Companion Analysis with Extended Evidence from UK Inter-Industry Payment Data, 2017–2024}

\author[a]{Kerstin Hötte\footnote{Corresponding author: kerstin.hotte@kedgebs.com}}
\tiny 
\affil[a]{KEDGE Business School, Paris}
\date{\today}
\begin{document}

\maketitle
\begin{abstract}
	In 2025, the UK Office for National Statistics released a novel dataset of monthly inter-industry payment flows during January 2017 to November 2024 at the 5-digit SIC level \citep{ons2025interindustrymethods}, covering $>$3.1 million UK organizations. Annual aggregates amount to 490 million transactions with an aggregate value of over £3.1 trillion in 2023. 
	Such publicly available data are unprecedented by their granularity and timeliness, providing a rich basis for economic research and real-time policy advice. \citet{hotte2025national} provided an empirical validation supplemented with conceptual discussions for using such bottom-up collected data in macroeconomic, industry-level, and economic network studies based on an earlier non-public and smaller version of the data. The novel data features much greater coverage, along with several methodological improvements. 
	This paper gives an update on the earlier empirical results. It summarizes the major methodological changes of data construction, discusses key empirical observations, and their differences and consistencies relative to \citet{hotte2025national}. It concludes by discussing the implications for using payment data and outlining remaining challenges and expected future developments. 
	
\end{abstract}
\vspace{0.5cm} \noindent
\textbf{JEL codes:} C67, C8, D57, E01

\noindent
\vspace{0.2cm}
\textbf{Keywords:} National accounts, real-time data, payment data, economic networks, input-output table

\newpage
	\section*{Acknowledgments}
	The author would like to thank her colleagues from the ONS, especially Dragos Cozma, Keith Lai, and Joseph Colliass for technical advice and data development.  
	The author also acknowledges valuable contributions and feedback by Andreina Naddeo, Johannes Lumma, and Fran\c{c}ois Lafond to preceding research. Further gratitude is owed to the participants of the regular Payment Data Research Seminar for comments and insightful discussions. 
    \section*{Suggested citation}
    Please cite the original article when referring to this paper: H\"otte, Kerstin. ``Mapping the disaggregated economy in real-time: using granular payment network data to complement national accounts.'' Economic Systems Research (2025): 1-28. 
	
	\newpage
	%

\section{Introduction}
In 2025, the UK Office for National Statistics released a novel dataset of monthly inter-industry payment flows covering the period January 2017 to November 2024 at the 5-digit SIC level \citep{ons2025interindustrymethods, ons2025interindustry}, spanning more than 3.1 million UK organizations. Annual aggregates amount to 490 million transactions with a total value exceeding £3.1 trillion in 2023.
This kind of publicly available data is unprecedented in its granularity and timeliness, offering a rich foundation for economic research and real-time policy advice. \citet{hotte2025national} provided an empirical validation, supplemented with conceptual discussions, of using such bottom-up data in macroeconomic, industry-level, and economic-network studies, based on an earlier non-public, smaller version of the dataset.\footnote{At the time of writing \citet{hotte2025national}, a public version, released in December 2023, was available. This version had been much more aggregated, covering only about 40 industries and containing many missing entries due to Statistical Disclosure Control (SDC).}
Since then, ONS has achieved major improvements, expanding coverage and addressing several conceptual challenges related to classification.

This paper updates the earlier empirical results. It summarizes the main methodological changes in data construction, discusses key empirical observations, and highlights their differences and consistencies relative to the earlier release.

The paper follows the structure of \citet{hotte2025national}, beginning with benchmarking against UK payment statistics (Sec. \ref{sec:data}) and key macroeconomic variables including GDP, monetary aggregates, and inflation (Sec. \ref{sec:macro_benchmarking}).
Sec. \ref{sec:national_accounts_benchmarking} continues with a systematic comparison to national accounts, focusing on input–output tables (IOTs) at both the network and industry levels.
To assess the validity of empirical patterns at the 5-digit network level, Sec. \ref{sec:stylised_facts_5digit} evaluates the consistency of the new data with established stylized facts from the network-economics literature.
Most empirical figures and tables are reproduced from \citet{hotte2025national}, supplemented with additional descriptive statistics. For technical details, conceptual discussions, and a deeper understanding of the indicators presented, readers are referred to the original article.

Overall, the empirical evidence confirms significant improvements in data coverage and shows high levels of consistency between the new and earlier datasets. The new release often displays greater conceptual alignment with established national-accounts indicators and stylized facts.
The paper concludes by outlining implications for the use of the dataset in economic measurement and applied research, as well as expected future developments.

This paper is not intended to be self-contained. Rather, it is designed as a companion to \citet{hotte2025national}, focusing on key updates and novel empirical and conceptual insights. For further detail and the broader conceptual background, readers are referred to the original article.

\FloatBarrier
\section{Data}
\label{sec:data}
The dataset is derived from the infrastructure of two major UK payment systems, enabling real-time monitoring of transactions. An introduction to payment system data and business payments (with a focus on the UK) is provided in \citet{hotte2025national}. 

The key differences of the 2025 data release compared to the previous version are the following: 
\begin{itemize}
    \item \textbf{Greater coverage for two major reasons:} 
    (1) The algorithm for business classification has improved. In the previous version, about 118,000 business accounts could be matched to Companies House (CH). The 2025 dataset includes 3.142 million organizations, more than half of all organizations registered in the UK in 2023 (5.656 million, of which about 5.1 million were coded as `active').\footnote{\url{https://www.gov.uk/government/statistics/companies-register-activities-statistical-release-2022-to-2023/companies-register-activities-2022-to-2023\#other-statistics-in-this-release} [accessed in August 2025]} 
    (2) Transactions made through the Faster Payment System (FPS) have been added. The earlier version only covered the Bankers' Automated Clearing System (Bacs). The dataset now includes both major payment systems used by UK businesses for domestic payments. Unlike Bacs, which includes Direct Debit (DD) collections, FPS can only be used for payments initiated by the payer. It is similar to Bacs Direct Credits but also supports standing payments for regular transactions.\footnote{In addition to the CH-matched data with a valid SIC code, the raw dataset also contains transactions identified as B2B payments that either could not be matched to CH or are missing SIC information. These residual payments are assigned a dummy code `0'.}
    
    \item \textbf{Qualitative changes in the classifiers:} Previously, businesses with multiple SIC codes were classified by their first code. Now, transactions are classified by all 5-digit SIC codes attached to a business in its CH entry, with equal weight given to each code. 
    
    \item \textbf{Different time period covered:} The new dataset covers monthly transactions from January 2017 to November 2024. The previous non-public dataset covered August 2015 to December 2023, while the earlier public release covered January 2016 to October 2023.\footnote{Publishing agreements with the data owner do not allow ONS to extract longer time periods from the data for publication. However, future research can construct longer time series by combining subsequent releases on a rolling basis.} 
    
    \item \textbf{Disaggregation by industry:} The 2025 data are available at the 5-digit level, distinguishing 712 different sectors. The earlier non-public data had the same level of disaggregation, but the publicly available version distinguished only about 40 sectors. The broader coverage of businesses in the new release has enabled publication of highly disaggregated data without breaching Statistical Disclosure Control (SDC) requirements.\footnote{The non-public data used in \citet{hotte2025national} had been subject to less restrictive SDC compared to the publicly available versions, yet the impact of different SDC procedures on the qualitative composition of firms and transactions covered is expected to be negligible.} 
\end{itemize}
Supplementary discussions of the data changes and the underlying methodology are provided in \citet{ons2025interindustrymethods, ons2025interindustry, hotte2025national}. 

\begin{figure}[h!]
    \caption{Monthly time series of payment data and major UK schemes}
    \label{fig:ts_UKpayments_benchmarking} 
    \centering
    \begin{subfigure}[b]{\textwidth}
        \includegraphics[width=\textwidth]{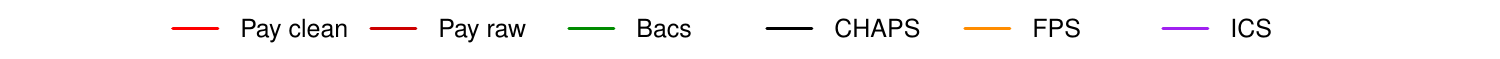}
    \end{subfigure}
    
    \begin{subfigure}[t]{0.32\textwidth}
        \centering     
        \includegraphics[width=0.8\textwidth]{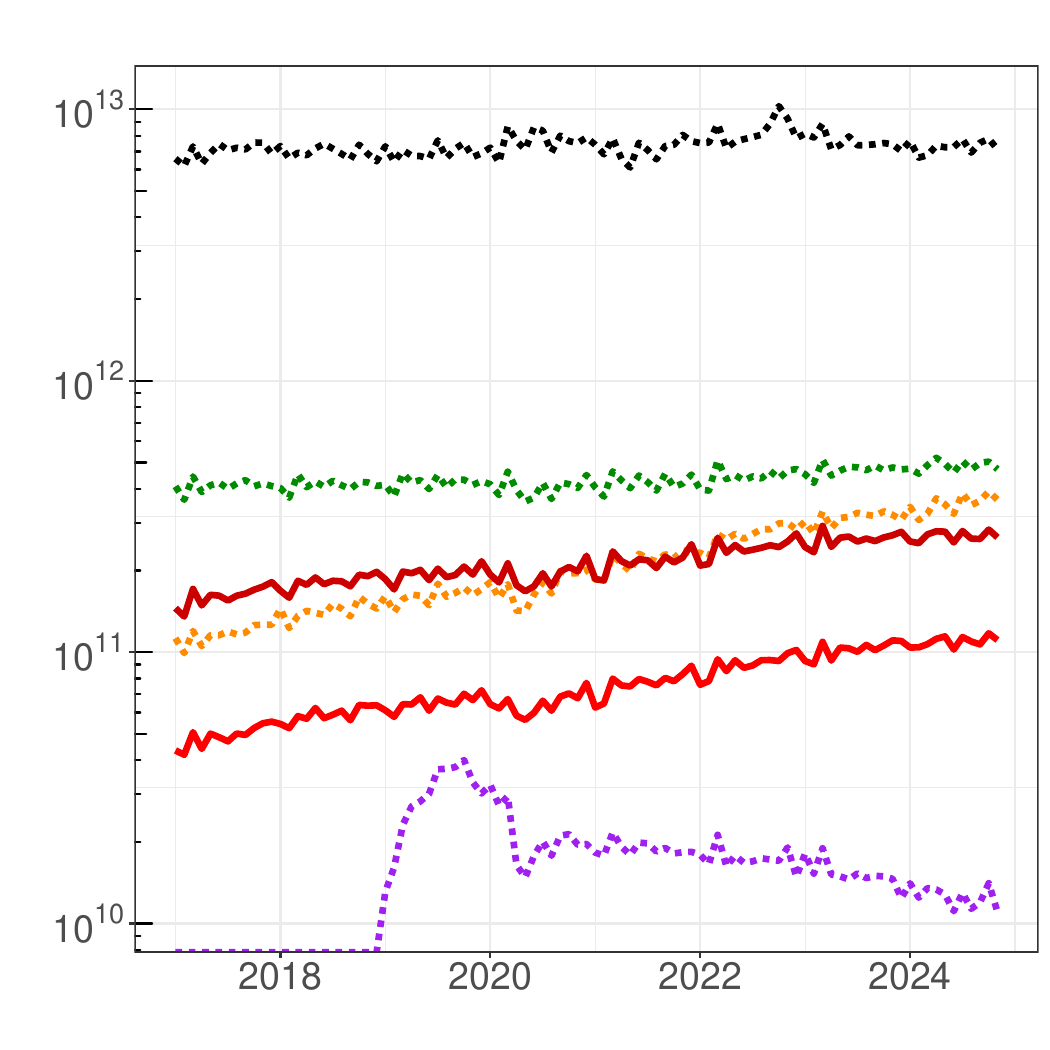}
        \caption{Values}
        \label{fig:ts_UKschemes_values}
    \end{subfigure}  
    \begin{subfigure}[t]{0.32\textwidth}
        \centering        
        \includegraphics[width=0.8\textwidth]{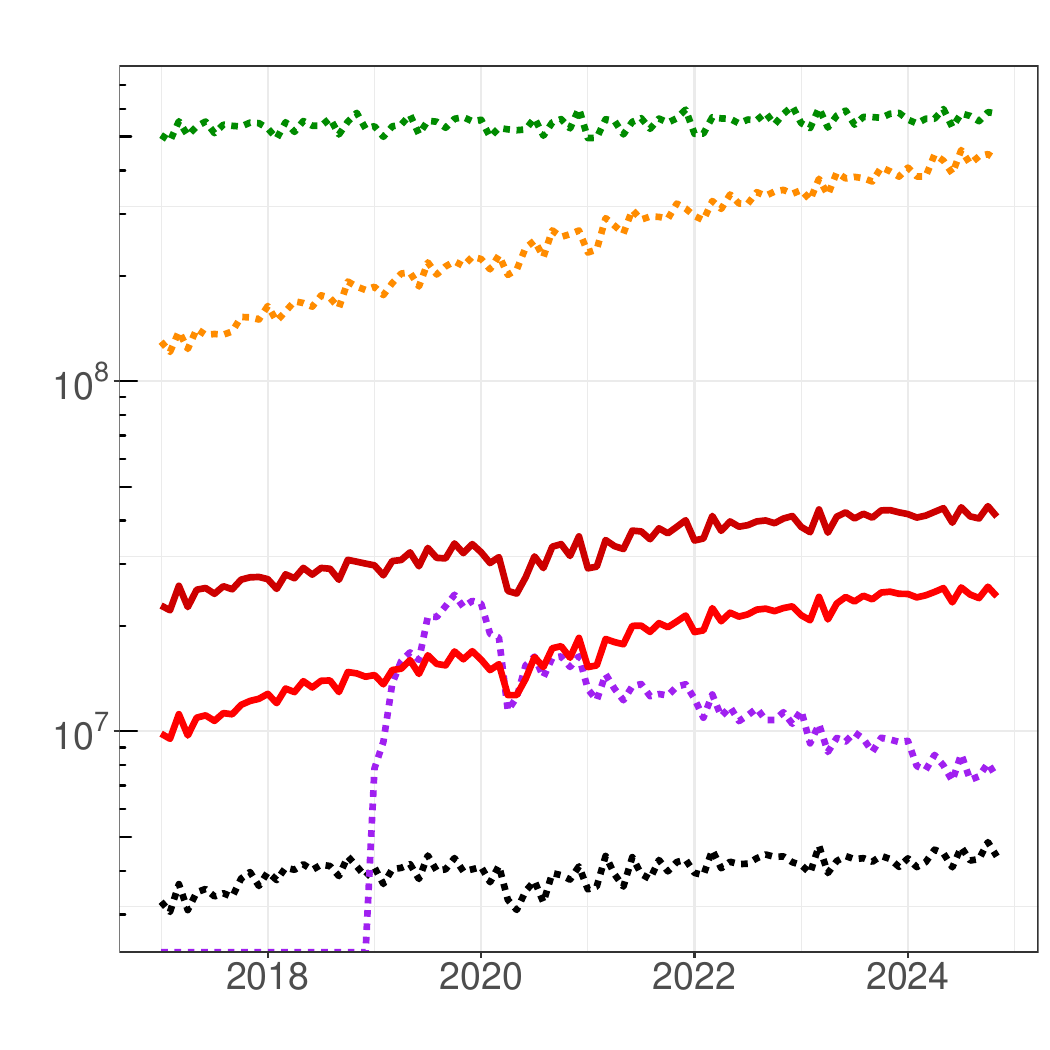}
        \caption{Counts}
        \label{fig:ts_UKschemes_counts}
    \end{subfigure}
    \begin{subfigure}[t]{0.32\textwidth}
        \centering        
        \includegraphics[width=0.8\textwidth]{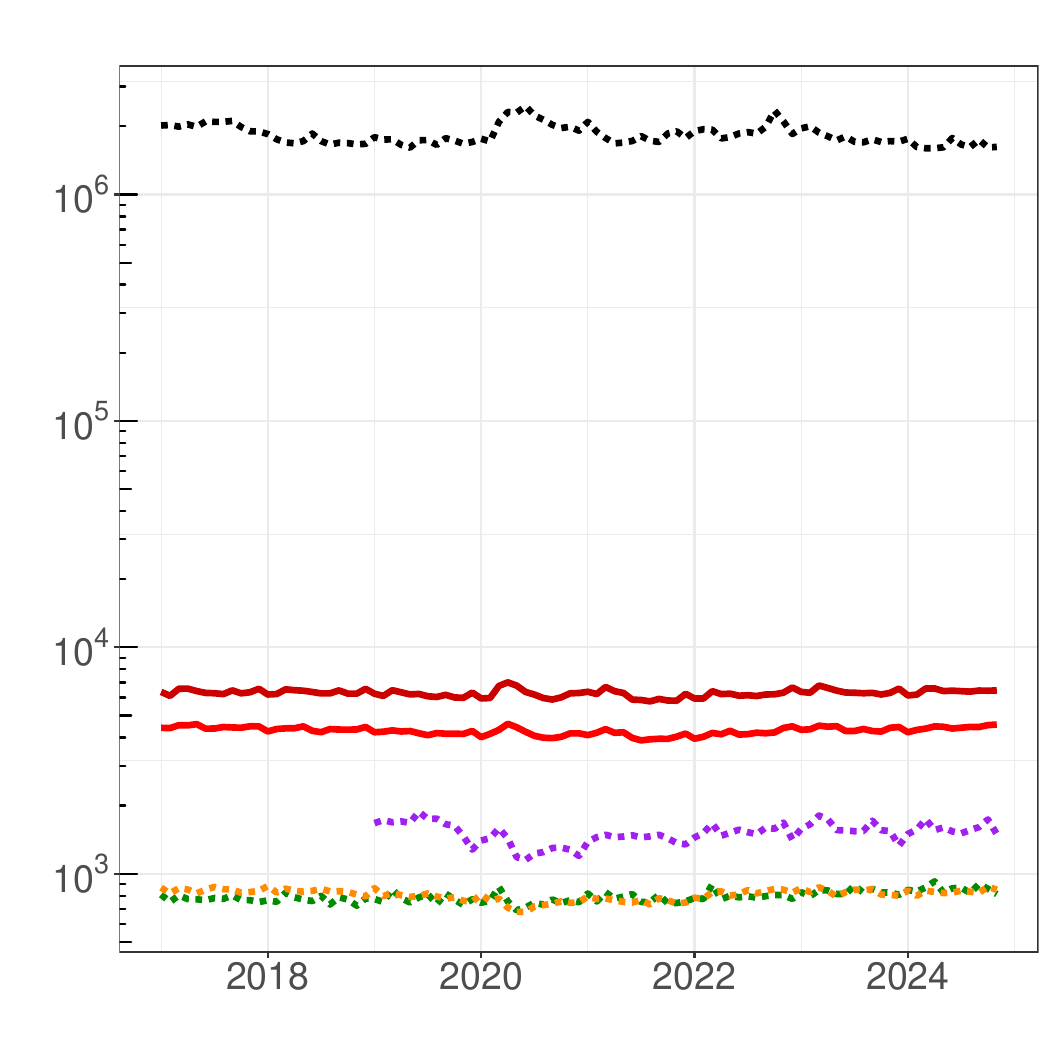}
        \caption{Average value}
        \label{fig:ts_UKschemes_avg}
    \end{subfigure}
    
    \justifying \scriptsize \noindent
    Notes: The vertical axis is on a log-10 scale. Payments (red) are monthly aggregates of our data. The Bacs, CHAPS, FPS, and Image Clearing System data are downloaded from \citet{payuk2023historicaldata}. 
\end{figure}

Figure \ref{fig:ts_UKpayments_benchmarking} benchmarks aggregate payment values, volumes, and average transaction values of the dataset against other major UK payment schemes. Analogous figures for the previous release are provided in \ref{app:subsec:payment_benchmarking}. Unlike the results in \citet{hotte2025national}, the new figures show two distinct time series: `cleaned' and `raw' data. The cleaned data exclude all payment flows where the payer or payee could not be matched to CH or where the SIC code could not be mapped to any of the 104 CPA codes used in the official national accounts IOTs \citep{ons2024blue}. The raw data include all transactions.\footnote{Non-identified payers/payees are assigned to `0' in the data. Non-matchable codes include, for example, activities of extra-territorial organizations or non-trading and dormant companies, which are tagged as `74990' and `99999' in CH, respectively.} Results in \citet{hotte2025national} reported only aggregates of the raw data. 

The key empirical observations are the following: 
\begin{itemize}
    \item \textbf{Unmatchable payment flows:} A large share (about 60\%) of the transaction network remains unclassified or assigned to codes that cannot be matched to CPA codes, with the payer and/or payee assigned to a non-CPA matchable code. 
    
    \item \textbf{Coverage in numbers:} Coverage by aggregate value and transaction counts is significantly higher than in the previous dataset. In 2023, the raw data recorded an annual aggregate value of £3.131 trillion and 490 million transactions. The cleaned data show £1.227 trillion and 281 million transactions. Average transaction values were £6,394 (raw) and £4,366 (cleaned). By comparison, the previous dataset recorded £1.25 trillion, 77 million transactions, and an average transaction value of £16,200. Monthly averages in 2023 were £261 billion (raw) and £102 billion (cleaned) for values, and 41 million (raw) and 23 million (cleaned) for counts. 
    
    \item \textbf{Lower average transaction values and types of payments:} The lower average transaction value (about £4,400) compared to the earlier dataset (£16,200) may reflect the inclusion of more smaller businesses in the new release. It may also be related to the integration of FPS payments, which are capped at £1 million since 2022 (and £250,000 before that), whereas Bacs allows up to £20 million. The two systems also differ in accessibility, and businesses may use FPS and Bacs for different purposes \citep[see][]{hotte2025national}. Average transaction values are relatively stable across schemes. Notably, the average value is higher in the raw data than in the cleaned data, suggesting that many high-value transactions are associated with non-matchable accounts. This is unexpected, as large organizations with high transaction values would typically be easier to identify. A possible explanation is that international trade and foreign organizations with UK Bacs accounts execute many high-value transfers. These potential explanations warrant further investigation in future research. 
    
    \item \textbf{Qualitative consistency:} The new and earlier data show qualitative similarity in several respects: (1) Both capture transactions with relatively high values. This is unsurprising, as most payments by count in Bacs and FPS are from or to consumers, while the dataset only captures B2B transfers. (2) As before, the drop during Covid-19 is more pronounced for counts. (3) An upward overall trend is observed, which appears stronger in the new dataset. 
\end{itemize}
\FloatBarrier

\section{Macroeconomic benchmarking}
\label{sec:macro_benchmarking}
This section repeats the macroeconomic benchmarking exercise presented in \citet{hotte2025national}, but extends it with additional results. These include the distinction between raw and cleaned data, correlations with producer price inflation (PPI), and correlations with the OECD Composite Leading Indicator (CLI). The CLI captures business and consumer confidence and serves as a real-time indicator often used in economic nowcasting.

\begin{table}[!htbp] \centering 
	\caption{Correlations with other payments and macro aggregates (including Covid-19 period)} 
	\label{tab:macro_benchmarking_5_digit} 
	\scriptsize 
	\begin{tabular}{lccccccccccc} 
		\\[-1.8ex]\hline 
		\hline \\[-1.8ex] 
		& Pay & Bacs & FPS & CHAPS & GDP nsa & GDP sa & M1 & M3 & CPI & PPI & CLI \\ 
		\hline \\[-1.8ex] 
		
		\multicolumn{12}{l}{\emph{Raw payment data including non-classified payment flows}}\\
		
		\hline \\[-1.8ex] 
		Share in 2023 & $2.550$ & $0.556$ & $0.836$ & $0.034$ & $1.234$ & & & & & & \\ 
		Yearly (value) & $0.999$ & $0.916$ & $0.994$ & $0.695$ & $-0.074$ & $0.605$ & $0.894$ & $0.941$ & $0.969$ & $0.962$ & $-0.325$ \\ 
		Monthly (value) & $0.986$ & $0.854$ & $0.959$ & $0.432$ & $0.390$ & $0.570$ & $0.793$ & $0.880$ & $0.924$ & $0.916$ & $0.011$ \\ 
		Yearly (count) & $0.998$ & $0.968$ & $0.984$ & $0.829$ & $-0.044$ & $0.591$ & $0.919$ & $0.953$ & $0.938$ & $0.937$ & $-0.323$ \\ 
		Monthly (count) & $0.990$ & $0.686$ & $0.956$ & $0.813$ & $0.467$ & $0.618$ & $0.828$ & $0.897$ & $0.899$ & $0.901$ & $0.059$ \\ 
		Yearly (avg) & $0.917$ & $0.314$ & $0.657$ & $0.134$ & $-0.113$ & $0.061$ & $-0.336$ & $-0.231$ & $0.142$ & $0.098$ & $0.052$ \\ 
		Monthly (avg) & $0.817$ & $0.361$ & $0.256$ & $0.108$ & $-0.453$ & $-0.273$ & $-0.113$ & $-0.031$ & $0.153$ & $0.107$ & $-0.283$ \\ 

		\hline \\[-1.8ex] 
		\multicolumn{12}{l}{\emph{Cleaned payment data excluding payments that could not be matched to CPA codes}}\\
		
		\hline \\[-1.8ex] 
		
		Share in 2023 & $0.392$ & $0.218$ & $0.328$ & $0.013$ & $0.484$ & & & & & & \\ 
		Yearly (value) & $0.999$ & $0.924$ & $0.997$ & $0.691$ & $-0.127$ & $0.600$ & $0.889$ & $0.939$ & $0.977$ & $0.965$ & $-0.312$ \\ 
		Monthly (value) & $0.986$ & $0.810$ & $0.986$ & $0.379$ & $0.311$ & $0.570$ & $0.796$ & $0.897$ & $0.966$ & $0.947$ & $0.037$ \\ 
		Yearly (count) & $0.998$ & $0.967$ & $0.991$ & $0.811$ & $-0.174$ & $0.561$ & $0.924$ & $0.960$ & $0.947$ & $0.938$ & $-0.326$ \\ 
		Monthly (count) & $0.990$ & $0.624$ & $0.982$ & $0.761$ & $0.318$ & $0.574$ & $0.844$ & $0.924$ & $0.938$ & $0.929$ & $0.044$ \\ 
		Yearly (avg) & $0.917$ & $0.296$ & $0.827$ & $0.052$ & $0.321$ & $0.208$ & $-0.494$ & $-0.406$ & $-0.026$ & $-0.032$ & $0.286$ \\ 
		Monthly (avg) & $0.817$ & $0.290$ & $0.480$ & $-0.020$ & $-0.266$ & $-0.055$ & $-0.279$ & $-0.149$ & $0.190$ & $0.137$ & $-0.046$ \\ 
		
		\hline \\[-1.8ex]  
		\hline \\[-1.8ex] 
		\multicolumn{8}{l}{\textbf{\emph{Growth rates}}}\\
		\hline \\[-1.8ex] 
		
		\multicolumn{12}{l}{\emph{Raw payment data including non-classified payment flows}}\\
		
		\hline \\[-1.8ex] 
		
		Yearly (value) & $0.993$ & $0.737$ & $0.943$ & $-0.286$ & $0.917$ & $0.902$ & $-0.198$ & $-0.185$ & $0.435$ & $0.630$ & $0.108$ \\ 
		Monthly (value) & $0.977$ & $0.767$ & $0.899$ & $0.038$ & $0.860$ & $0.746$ & $-0.033$ & $-0.058$ & $0.334$ & $0.472$ & $0.209$ \\ 
		Yearly (count) & $0.965$ & $0.667$ & $0.565$ & $0.840$ & $0.984$ & $0.921$ & $-0.016$ & $-0.082$ & $0.094$ & $0.410$ & $0.231$ \\ 
		Monthly (count) & $0.976$ & $0.560$ & $0.713$ & $0.820$ & $0.934$ & $0.856$ & $0.010$ & $-0.061$ & $0.073$ & $0.281$ & $0.416$ \\ 
		Yearly (avg) & $0.874$ & $0.054$ & $0.224$ & $0.365$ & $-0.811$ & $-0.433$ & $-0.357$ & $-0.168$ & $0.646$ & $0.259$ & $-0.324$ \\ 
		Monthly (avg) & $0.892$ & $0.043$ & $0.141$ & $0.514$ & $-0.816$ & $-0.625$ & $-0.028$ & $0.085$ & $0.384$ & $0.167$ & $-0.557$ \\ 
		
		\hline \\[-1.8ex] 
		\multicolumn{12}{l}{\emph{Cleaned payment data excluding payments that could not be matched to CPA codes}}\\
		
		\hline \\[-1.8ex] 
		
		Yearly (value) & $0.993$ & $0.708$ & $0.933$ & $-0.330$ & $0.868$ & $0.863$ & $-0.224$ & $-0.202$ & $0.420$ & $0.581$ & $0.140$ \\ 
		Monthly (value) & $0.977$ & $0.716$ & $0.879$ & $-0.032$ & $0.833$ & $0.758$ & $-0.083$ & $-0.095$ & $0.342$ & $0.467$ & $0.244$ \\ 
		Yearly (count) & $0.965$ & $0.613$ & $0.703$ & $0.937$ & $0.910$ & $0.812$ & $-0.020$ & $-0.082$ & $0.009$ & $0.316$ & $0.193$ \\ 
		Monthly (count) & $0.976$ & $0.513$ & $0.721$ & $0.861$ & $0.889$ & $0.804$ & $0.024$ & $-0.041$ & $0.024$ & $0.246$ & $0.386$ \\ 
		Yearly (avg) & $0.874$ & $0.511$ & $0.555$ & $0.171$ & $-0.926$ & $-0.017$ & $-0.456$ & $-0.264$ & $0.895$ & $0.516$ & $-0.130$ \\ 
		Monthly (avg) & $0.892$ & $0.162$ & $0.350$ & $0.363$ & $-0.750$ & $-0.382$ & $-0.208$ & $-0.076$ & $0.624$ & $0.345$ & $-0.423$ \\ 

		\hline \\[-1.8ex] 
		
	\end{tabular}

	\justifying \scriptsize \noindent
	Notes: 
	This table shows Pearson correlations between annual (monthly) payments and other UK payment schemes and macroeconomic aggregates (GDP, M1, M3, Prices) during 2017 and 2024 (01/2017 and 11/2024), excluding the Covid-19 period, proxied by 2020 to 2022 (03/2020 to 12/2022). 
    `sa' (`nsa') is short for (non-) seasonally adjusted in all other rows. Our payment data and other payment aggregates are compared by aggregate values, counts, and average values (short `avg') given by value divided by count. 
	Growth rates are calculated as percentage growth compared to the (same month of the) previous year (for monthly data). Bacs, FPS, and CHAPS data are obtained from \citet{payuk2023historicaldata}. 
	Monthly GDP is proxied by indicative (non-)seasonally adjusted monthly `Total Gross Value Added' index data published by the ONS \citep{ons2023indicativeGDPdata, ons2023indicativeGDPadjusted}. GDP nsa data end in 2021, rendering the calculation of annual correlations when removing the Covid-19 period meaningless. Consumer price inflation (CPI) and producer price inflation (PPI) data are obtained from ONS (PDID: D7BT and GB7S). 
	M1, M3 and composite leading indicator (CLI) data are obtained from the OECD Key Economic Indicators (KEI) and Main Economic Indicators (MEI) dataset \citep{oecd2023MEIdata, oecd2023KEIdata}.
	
\end{table} 

\begin{table}[!htbp] \centering 
	\caption{Correlations with other payments and macro aggregates (excluding Covid-19 period)} 
	\label{tab:macro_benchmarking_5_digit_noCovid} 
	\scriptsize
	\begin{tabular}{lccccccccccc} 
		\\[-1.8ex]\hline 
		\hline \\[-1.8ex] 
		& Pay & Bacs & FPS & CHAPS & GDP nsa & GDP sa & M1 & M3 & CPI & PPI & CLI \\ 
		\hline \\[-1.8ex] 
		
		\multicolumn{12}{l}{\emph{Raw payment data including non-classified payment flows}}\\
		
		\hline \\[-1.8ex] 
		Yearly (value) & $0.999$ & $0.984$ & $0.993$ & $0.921$ & $0.985$ & $0.976$ & $0.986$ & $0.985$ & $0.984$ & $0.984$ & $-0.571$ \\ 
		Monthly (value) & $0.991$ & $0.901$ & $0.971$ & $0.533$ & $0.781$ & $0.940$ & $0.939$ & $0.955$ & $0.961$ & $0.957$ & $-0.018$ \\ 
		Yearly (count) & $0.999$ & $0.999$ & $0.991$ & $0.824$ & $0.997$ & $0.989$ & $0.973$ & $0.971$ & $0.970$ & $0.969$ & $-0.625$ \\ 
		Monthly (count) & $0.995$ & $0.714$ & $0.976$ & $0.821$ & $0.743$ & $0.959$ & $0.934$ & $0.949$ & $0.955$ & $0.950$ & $-0.055$ \\ 
		Yearly (avg) & $0.884$ & $0.430$ & $0.931$ & $0.459$ & $-0.925$ & $0.001$ & $0.360$ & $0.372$ & $0.368$ & $0.378$ & $0.628$ \\ 
		Monthly (avg) & $0.777$ & $0.471$ & $0.438$ & $0.064$ & $-0.094$ & $0.069$ & $0.262$ & $0.276$ & $0.275$ & $0.279$ & $0.274$ \\ 

		\hline \\[-1.8ex] 
		\multicolumn{12}{l}{\emph{Cleaned payment data excluding payments that could not be matched to CPA codes}}\\
		
		\hline \\[-1.8ex] 
		Yearly (value) & $0.999$ & $0.989$ & $0.997$ & $0.934$ & $0.982$ & $0.968$ & $0.992$ & $0.991$ & $0.989$ & $0.989$ & $-0.541$ \\ 
		Monthly (value) & $0.991$ & $0.876$ & $0.988$ & $0.486$ & $0.716$ & $0.949$ & $0.949$ & $0.971$ & $0.983$ & $0.976$ & $0.043$ \\ 
		Yearly (count) & $0.999$ & $1.000$ & $0.995$ & $0.812$ & $0.991$ & $0.982$ & $0.982$ & $0.980$ & $0.979$ & $0.978$ & $-0.593$ \\ 
		Monthly (count) & $0.995$ & $0.666$ & $0.990$ & $0.786$ & $0.691$ & $0.963$ & $0.947$ & $0.967$ & $0.976$ & $0.971$ & $-0.004$ \\ 
		Yearly (avg) & $0.884$ & $0.127$ & $0.965$ & $0.800$ & $-0.999$ & $-0.373$ & $-0.008$ & $0.002$ & $0.006$ & $0.010$ & $0.905$ \\ 
		Monthly (avg) & $0.777$ & $0.255$ & $0.415$ & $0.234$ & $-0.378$ & $-0.136$ & $0.152$ & $0.186$ & $0.211$ & $0.194$ & $0.577$ \\ 
		
		\hline \\[-1.8ex]

		\hline \\[-1.8ex] 
		\multicolumn{8}{l}{\textbf{\emph{Growth rates}}}\\
		\hline \\[-1.8ex] 
		
		\multicolumn{12}{l}{\emph{Raw payment data including non-classified payment flows}}\\
		
		\hline \\[-1.8ex]

		Yearly (value) & $1.000$ & $0.956$ & $0.996$ & $0.972$ & & $0.869$ & $0.992$ & $0.995$ & $0.987$ & $0.994$ & $0.988$ \\ 
		Monthly (value) & $0.976$ & $0.799$ & $0.909$ & $0.679$ & $0.473$ & $0.551$ & $0.820$ & $0.847$ & $0.851$ & $0.863$ & $0.036$ \\ 
		Yearly (count) & $0.994$ & $1.000$ & $0.997$ & $-0.100$ & & $0.899$ & $0.998$ & $0.999$ & $0.995$ & $0.999$ & $0.996$ \\ 
		Monthly (count) & $0.977$ & $0.603$ & $0.902$ & $0.361$ & $0.574$ & $0.603$ & $0.802$ & $0.812$ & $0.801$ & $0.827$ & $-0.052$ \\ 
		Yearly (avg) & $0.961$ & $0.816$ & $1.000$ & $0.353$ & & $0.745$ & $0.943$ & $0.951$ & $0.932$ & $0.948$ & $0.932$ \\ 
		Monthly (avg) & $0.881$ & $0.561$ & $0.667$ & $0.222$ & $-0.029$ & $0.251$ & $0.562$ & $0.621$ & $0.655$ & $0.630$ & $0.235$ \\ 
		
		\hline \\[-1.8ex] 
		\multicolumn{12}{l}{\emph{Cleaned payment data excluding payments that could not be matched to CPA codes}}\\
		
		\hline \\[-1.8ex] 
		
		Yearly (value) & $1.000$ & $0.952$ & $0.994$ & $0.968$ & & $0.862$ & $0.990$ & $0.993$ & $0.985$ & $0.992$ & $0.985$ \\ 
		Monthly (value) & $0.976$ & $0.738$ & $0.947$ & $0.608$ & $0.335$ & $0.544$ & $0.868$ & $0.908$ & $0.923$ & $0.931$ & $0.106$ \\ 
		Yearly (count) & $0.994$ & $0.996$ & $0.983$ & $0.012$ & & $0.844$ & $0.985$ & $0.989$ & $0.979$ & $0.987$ & $0.979$ \\ 
		Monthly (count) & $0.977$ & $0.504$ & $0.934$ & $0.338$ & $0.334$ & $0.586$ & $0.851$ & $0.877$ & $0.872$ & $0.898$ & $0.002$ \\ 
		Yearly (avg) & $0.961$ & $0.944$ & $0.956$ & $0.598$ & & $0.901$ & $0.998$ & $0.999$ & $0.996$ & $0.999$ & $0.996$ \\ 
		Monthly (avg) & $0.881$ & $0.530$ & $0.586$ & $0.338$ & $0.101$ & $0.208$ & $0.552$ & $0.623$ & $0.691$ & $0.636$ & $0.430$ \\ 
		
		\hline \\[-1.8ex] 
	\end{tabular}

	\justifying \scriptsize \noindent
	Notes: 
	This table shows Pearson correlations between annual (monthly) payments and other UK payment schemes and macroeconomic aggregates (GDP, M1, M3, Prices) during 2017 and 2024 (01/2017 and 11/2024), excluding the Covid-19 period, proxied by 2020 to 2022 (03/2020 to 12/2022). 
    `sa' (`nsa') is short for (non-)seasonally adjusted in all other rows. Our payment data and other payment aggregates are compared by aggregate values, counts, and average values (short `avg') given by value divided by count. 
	Growth rates are calculated as percentage growth compared to the (same month of the) previous year (for monthly data). Bacs, FPS, and CHAPS data are obtained from \citet{payuk2023historicaldata}. 
	Monthly GDP is proxied by indicative (non-)seasonally adjusted monthly `Total Gross Value Added' index data published by the ONS \citep{ons2023indicativeGDPdata, ons2023indicativeGDPadjusted}. GDP nsa data end in 2021, rendering the calculation of correlations among annual growth rates meaningless. Consumer price inflation (CPI) and producer price inflation (PPI) data are obtained from ONS (PDID: D7BT and GB7S). 
	M1, M3 and composite leading indicator (CLI) data are obtained from the OECD Key Economic Indicators (KEI) and Main Economic Indicators (MEI) dataset \citep{oecd2023MEIdata, oecd2023KEIdata}. 
\end{table}

Table \ref{tab:macro_benchmarking_5_digit} shows an analysis of raw correlations between monthly and annual payments (raw and cleaned data), other UK payment schemes (Bacs, FPS, CHAPS) and various macroeconomic indicators (GDP, inflation and others). 
The upper (bottom) two panels in the table show correlations of data in levels (transformed into year-on-year growth rates) using raw and clean payment aggregates. The first rows in the upper two panels relate the aggregate annual value of the raw and cleaned payment data to annual aggregates from other payment schemes and GDP. The first column in the table correlates the raw and clean payment data. 

Analogous results excluding the Covid-19 period are shown in Table \ref{tab:macro_benchmarking_5_digit_noCovid} and results from \citet{hotte2025national} are provided in \ref{app:subsec:macro_benchmarking} to ease the comparison. 

\begin{figure}[!h]
	\centering
	\begin{subfigure}[b]{0.9\textwidth}
		\includegraphics[width=\textwidth]{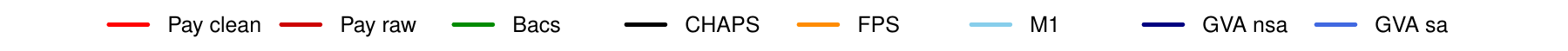}
	\end{subfigure}
	
	\begin{subfigure}[b]{0.32\textwidth}
		\centering
		\includegraphics[width=\textwidth]{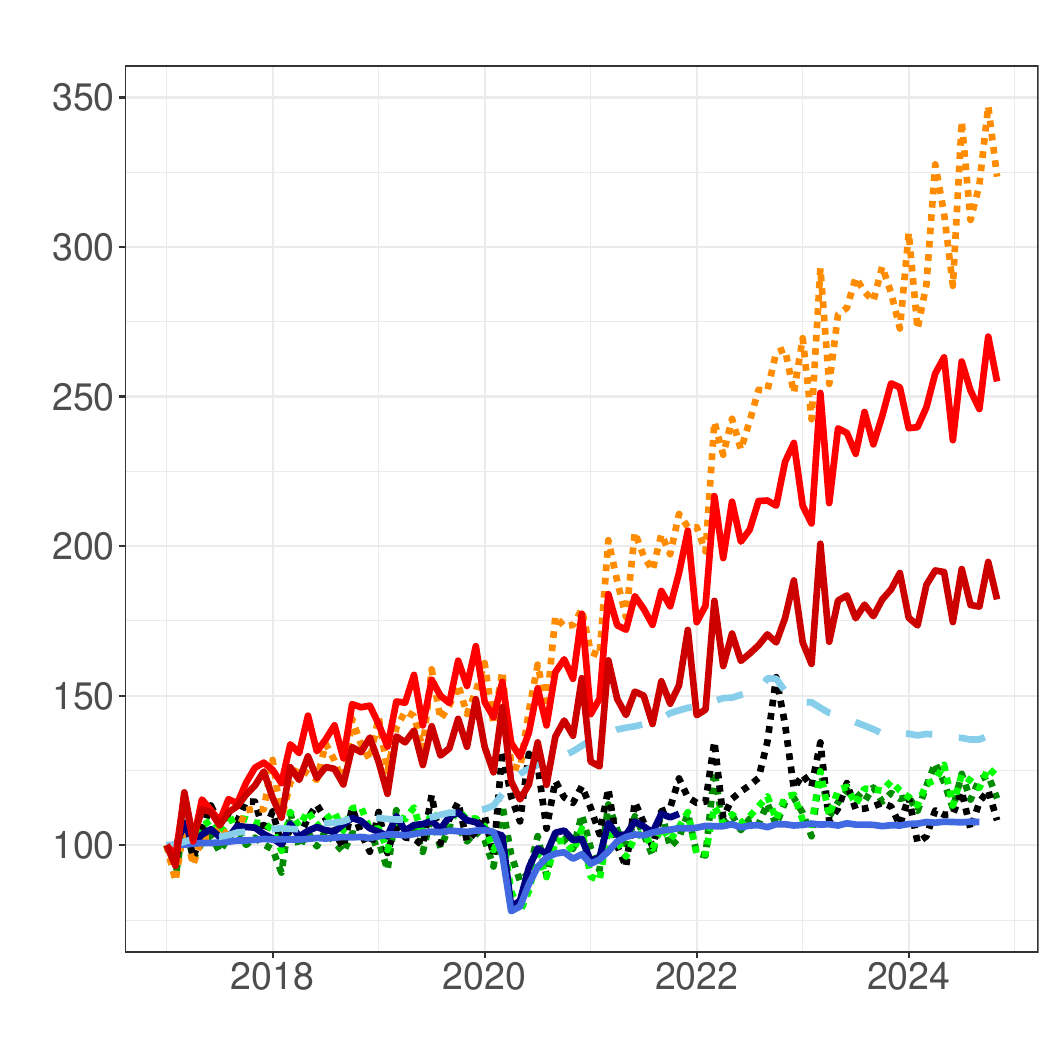}
		\caption{Value}
	\end{subfigure}
	\begin{subfigure}[b]{0.32\textwidth}
		\centering
		\includegraphics[width=\textwidth]{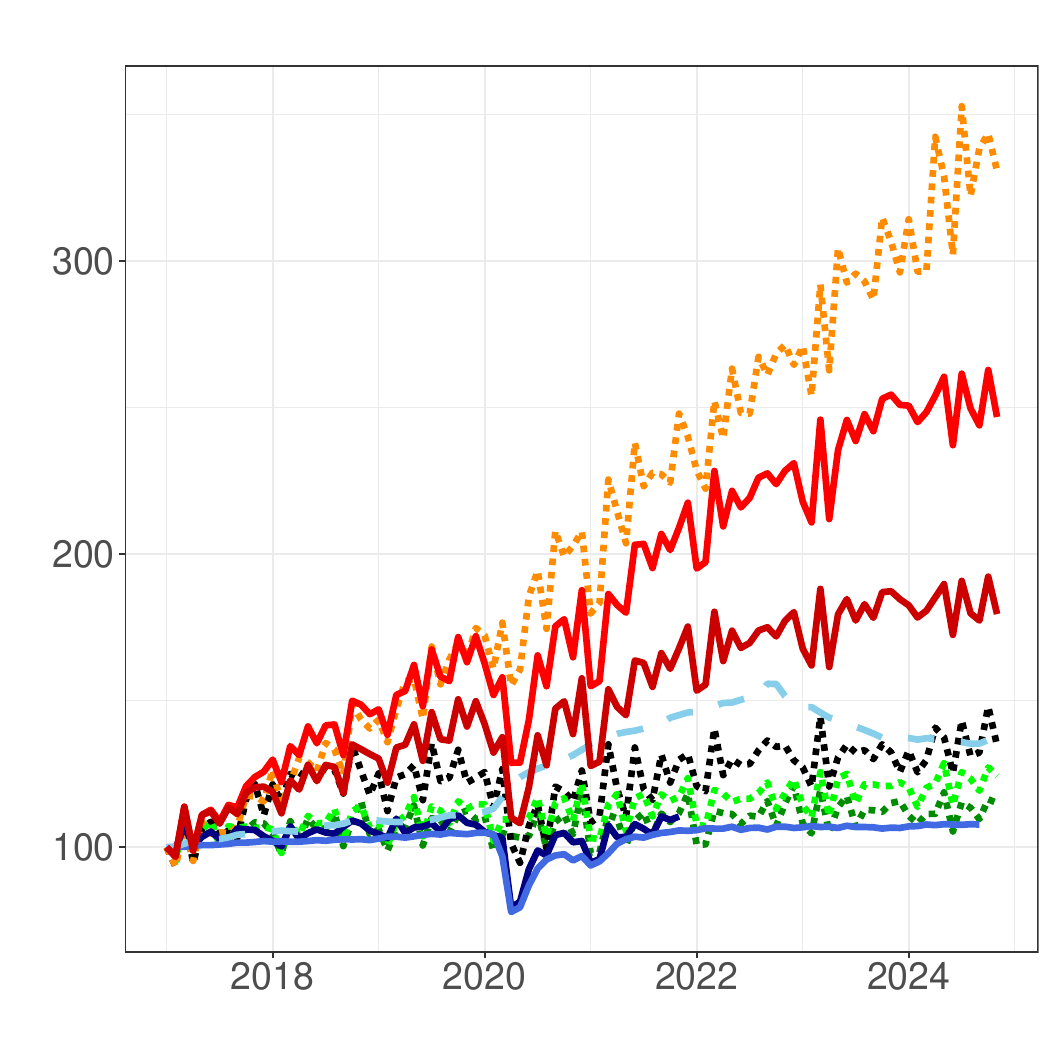}
		\caption{Counts}
	\end{subfigure}
	\begin{subfigure}[b]{0.32\textwidth}
		\centering
		\includegraphics[width=\textwidth]{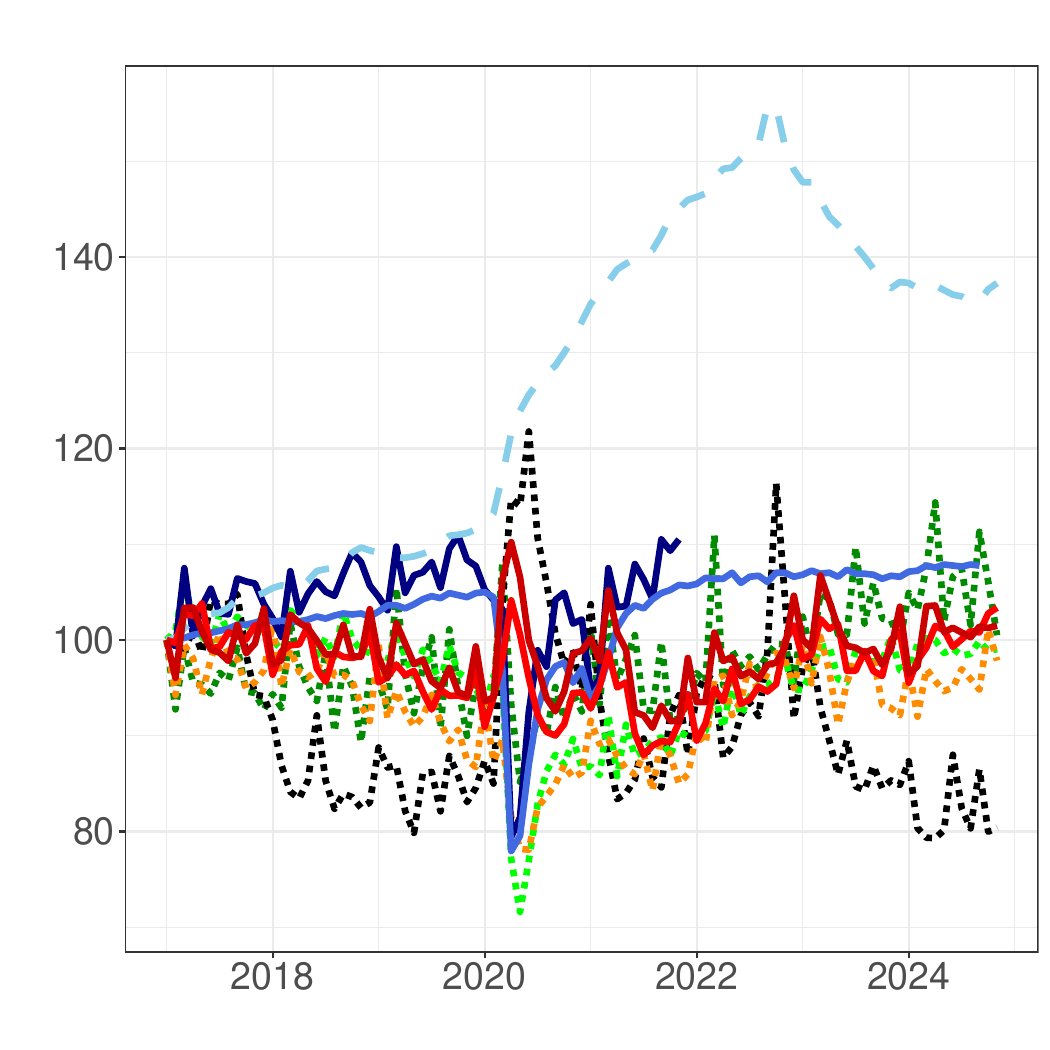}
		\caption{Average value}
	\end{subfigure}
	\caption{Monthly UK payments, GDP and M1}
	\label{fig:timeseries_aggr_index_payments_vs_gdp}
	
	\justifying \scriptsize
	\noindent
	Notes: These figures show monthly time series (indexed to 01/2017 = 100) for payments, the major UK payment schemes, and indicative (non-)seasonally adjusted monthly 'Total Gross Value Added' (GVA) data published by the ONS \citep{ons2023indicativeGDPadjusted, ons2023indicativeGDPdata}. Average values are obtained by dividing total values by counts. `Pay raw' refers to payment data that include non-classified payment flows. `Pay clean' excludes all payments that could not be matched to CPA codes.
	
\end{figure}

Table \ref{tab:macro_benchmarking_5_digit} and \ref{tab:macro_benchmarking_5_digit_noCovid} reveal the following: 
\begin{itemize}
    \item \textbf{Relative values:} In 2023, the aggregate value of the raw data was about 2.55 times that of the cleaned data, and about 1.23 times nominal GDP. The cleaned data amounted to roughly 50\% of GDP.\footnote{Payments are conceptually not comparable to GDP, which only capture value added, whereas payments are gross transaction values. The comparison is made purely to gauge the scale of the dataset.} In the previous raw dataset, annual aggregates corresponded to roughly 40\% of GDP. In general, raw and cleaned data correlate almost perfectly by values and counts ($>$97–99\%), but slightly less by average transaction value (82–89\%), as seen in column 1. 
    
    \item \textbf{GDP correlations:} For monthly growth rates (bottom panels), correlations range between 75–86\% for deseasonalized GDP and 83–93\% for non-deseasonalized GDP, with typically higher values for (1) payment count data and (2) the raw data. These correlations are stronger than in the previous dataset, where they ranged between 65–86\% and 71–89\%. Consistent with the earlier data, correlations with non-deseasonalized GDP are higher when including the Covid-19 years. Excluding the Covid-19 period (Table \ref{tab:macro_benchmarking_5_digit_noCovid}), correlations are lower, ranging between 54–60\% for deseasonalized GDP and 33–57\% for non-deseasonalized GDP. This decline may partly reflect the shorter sample, especially for the non-deseasonalized series, which is only available until the end of 2021. Average transaction values show inconsistent correlation patterns and appear less meaningful for drawing conclusions about GDP. 
    
    \item \textbf{Inflation and monetary aggregates:} The inclusion or exclusion of the Covid-19 period has a large effect on observed correlations between monetary indicators and payment data. Excluding Covid-19, correlations between monthly growth rates of payments (values and counts) and consumer and producer price inflation (CPI and PPI) are high, at 85–93\%. Including the Covid-19 period, correlations fall sharply, to around 33–47\% for payment values. Correlations between payment counts and PPI are 25–28\%, while correlations with CPI are almost absent. In all settings, correlations are (1) higher for PPI than for CPI and (2) weaker for counts than for values. 
    
    The relationship with monetary aggregates (M1, M3) helps explain these differences. Excluding Covid-19, there is a clear statistical relationship, with correlations above 80\% (see also Fig. \ref{fig:timeseries_aggr_index_payments_vs_gdp}). Including Covid-19, correlations become slightly negative or vanish. The Covid-19 period was characterized by massive counter-cyclical monetary and fiscal interventions, leading to sharp increases in monetary supply captured by M1 and M3. 
    
    Interestingly, the relationship between average payment values in the cleaned data and CPI appears robust, with correlations of about 62–69\% both with and without Covid-19. PPI shows similar correlations when Covid-19 is excluded, but the relationship is much weaker otherwise.  
    
    \item \textbf{Payments and business/consumer confidence:} Payments also show positive correlations (stronger for counts) with the OECD Composite Leading Indicator (CLI), which measures business and consumer confidence. This relationship, however, is absent when excluding the Covid-19 period. Since the CLI is widely used in economic nowcasting and forecasting as a measure of expectations about future activity, these results suggest that realized transactions can be an indicator of expectations during times of crisis. 
\end{itemize}

Fig. \ref{fig:timeseries_aggr_index_payments_vs_gdp} illustrates fluctuations and trends in the monthly aggregates of payments, other UK payment schemes, M1, and GDP. The data are indexed to January 2017 = 100. Analogous figures for the previous data release are shown in \ref{app:subsec:macro_benchmarking}. 
The key observations are as follows: 
\begin{itemize}
    \item \textbf{Rising payments:} Payments in our sample show a stronger increase than aggregates of Bacs, CHAPS, M1, and GDP. Only FPS exhibits a stronger rise.\footnote{FPS is a relatively young scheme, introduced in 2008 and subject to changes. In 2022, its transaction limit was raised from £250,000 to £1 million, which may have contributed to its wider diffusion.} The rise also appears stronger compared to the previous version of the data, partly due to the inclusion of FPS transactions. This pattern holds for both values and counts, while average transaction values of most payment series appear relatively stable in the long run, with strong fluctuations during Covid-19. In the previous version, average transaction values rose noticeably. Several factors may explain this stability: financial innovations in payment interfaces have lowered access barriers to FPS and Bacs for small firms, and may also encourage firms to execute more frequent, small-value real-time transactions rather than bundling them. 
    
    \item \textbf{Steeper rise of cleaned than raw data:} The observed rise is stronger for the cleaned than for the raw data, though both series behave similarly in long-term trends of average value and in monthly fluctuations of counts and values. This may reflect a selection bias from relying on a current snapshot of the UK business register (CH). The register includes CPA-matchable codes for active businesses, while inactive firms are tagged as `non-trading' or `dormant' with SIC codes `74990' and `99999'. Firms that went bankrupt between 2017 and 2024 would be classified as dormant and excluded from the cleaned dataset, even if they were active with valid SIC codes earlier. Newly founded firms, by contrast, would be added. This issue could be addressed in future versions by using annual snapshots of CH.\footnote{Backdating is not possible due to the lack of historical snapshots.} The CH file also lists `foreign entities' without SIC codes. A deeper investigation of the discrepancy between cleaned and raw data is beyond the scope of this paper. 
    
    \item \textbf{Seasonality:} While no strong seasonal patterns are evident, payment series (including FPS, CHAPS, Bacs) show greater monthly fluctuations than deseasonalized GDP, similar to non-deseasonalized GDP. This is consistent with the earlier data and highlights the potential need for deseasonalization when working with payment data. 
\end{itemize}

\FloatBarrier
\section{Comparison to national accounts}
\label{sec:national_accounts_benchmarking}
This section benchmarks input-output tables (IOTs) constructed from the payment data against three types of official IOTs from the UK national accounts \citep{ons2024blue}: the intermediate consumption table from the Supply and Use Tables (SUT), and the analytical product-by-product (PxP) and industry-by-industry (IxI) IOTs. Technical and conceptual details of this exercise are provided in \citet{hotte2025national}.  
By design, the results presented here focus on the cleaned data, as the analysis only includes payment flows where both the payer and payee can be mapped to CPA codes.  

Several minor changes compared to the analyses in \citet{hotte2025national} should be noted:  
(1) The time coverage differs, and the analysis is extended through 2022. In addition, the availability of analytical PxP and IxI data has improved, as these tables are now published annually after 2018.\footnote{Note that the analytical IOTs are not backward revised, in contrast to the SUT, and their use as a time series comes with conceptual limitations \citep{ons2024blue}.}  
(2) The aggregation now includes 104 sectors instead of 105, because sector C254 was merged with the aggregate category C25OTHER in the 2024 release of \citet{ons2024blue}.\footnote{This may lead to minor deviations from \citet{hotte2025national} when presenting results on ONS tables, for example in Fig. \ref{tab:network_stats_2019_no_truncation}.}  
(3) This section also presents several additional descriptive statistics, which are not available in \citet{hotte2025national} and therefore cannot be benchmarked against the earlier data.

\subsection{Aggregate network statistics}
\label{subsec:aggregate_network}

\begin{table}[!htbp] \centering 
	\caption{Properties of the payment and ONS-based input-output networks in 2022} 
	\label{tab:network_stats_2022_no_truncation} 
	\footnotesize 
	\begin{tabular}{@{\extracolsep{8pt}} lccccc} 
		\\[-1.8ex]\hline 
		\hline \\[-1.8ex] 
		Variable & Value & Count & SUT & PxP & IxI \\ 
		\hline \\[-1.8ex]
		\underline{\emph{Raw transactions} }\\
		Density & $0.552$ & $0.404$ & $0.474$ & $0.718$ & $0.978$ \\ 
		Average degree & $57.423$ & $41.650$ & $48.343$ & $73.275$ & $99.735$ \\ 
		Average strength & $8,872.127$ & $2,314,359.000$ & $15,643.100$ & $12,790.050$ & $12,809.820$ \\ 
		Average weight & $154.505$ & $55,566.200$ & $323.585$ & $174.550$ & $128.438$ \\ 
		Reciprocity & $0.783$ & $0.753$ & $0.535$ & $0.788$ & $0.987$ \\ 
		Transitivity & $0.757$ & $0.682$ & $0.750$ & $0.860$ & $0.990$ \\ 
		Assortativity by degree & $-0.324$ & $-0.416$ & $-0.192$ & $-0.183$ & $-0.006$ \\ 
		\hline \\[-1.8ex] \underline{\emph{Input shares}} \\   
		Average strength & $0.867$ & $0.943$ & $0.733$ & $0.829$ & $0.834$ \\ 
		Average weight & $0.015$ & $0.023$ & $0.015$ & $0.011$ & $0.008$ \\ 
		\hline \\[-1.8ex]  \underline{\emph{Output shares}} \\ 
		Average strength & $0.877$ & $0.922$ & $0.724$ & $0.791$ & $0.817$ \\ 
		Average weight & $0.015$ & $0.022$ & $0.015$ & $0.011$ & $0.008$ \\ 
		\hline \\[-1.8ex] 
	\end{tabular} 
	
	\vspace{0.25cm}
	
	\justifying \noindent \scriptsize
	Notes: The first (second) column uses payment values (counts) as weights. The other columns represent official IOTs published by the ONS, where PxP is short for Product-by-Product, IxI for Industry-by-Industry, and SUT for Supply-and-Use Table. The data are aggregated into 104 distinct CPA codes \citep[see][Sec. 4.1]{hotte2025national}.
	Raw transaction data are shown in £ million. 
\end{table}

\begin{figure}[h]
	\centering
	
	\caption{Network density at different truncation thresholds}
	\label{fig:network_density_plot_2022}
	
	\begin{subfigure}[b]{\textwidth}
		\centering
		\includegraphics[width=\textwidth]{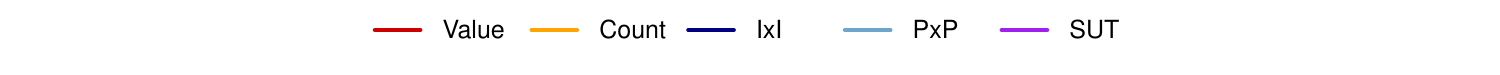}
	\end{subfigure}
	
	\begin{subfigure}[b]{0.44\textwidth}
		\centering
		\includegraphics[width=\textwidth]{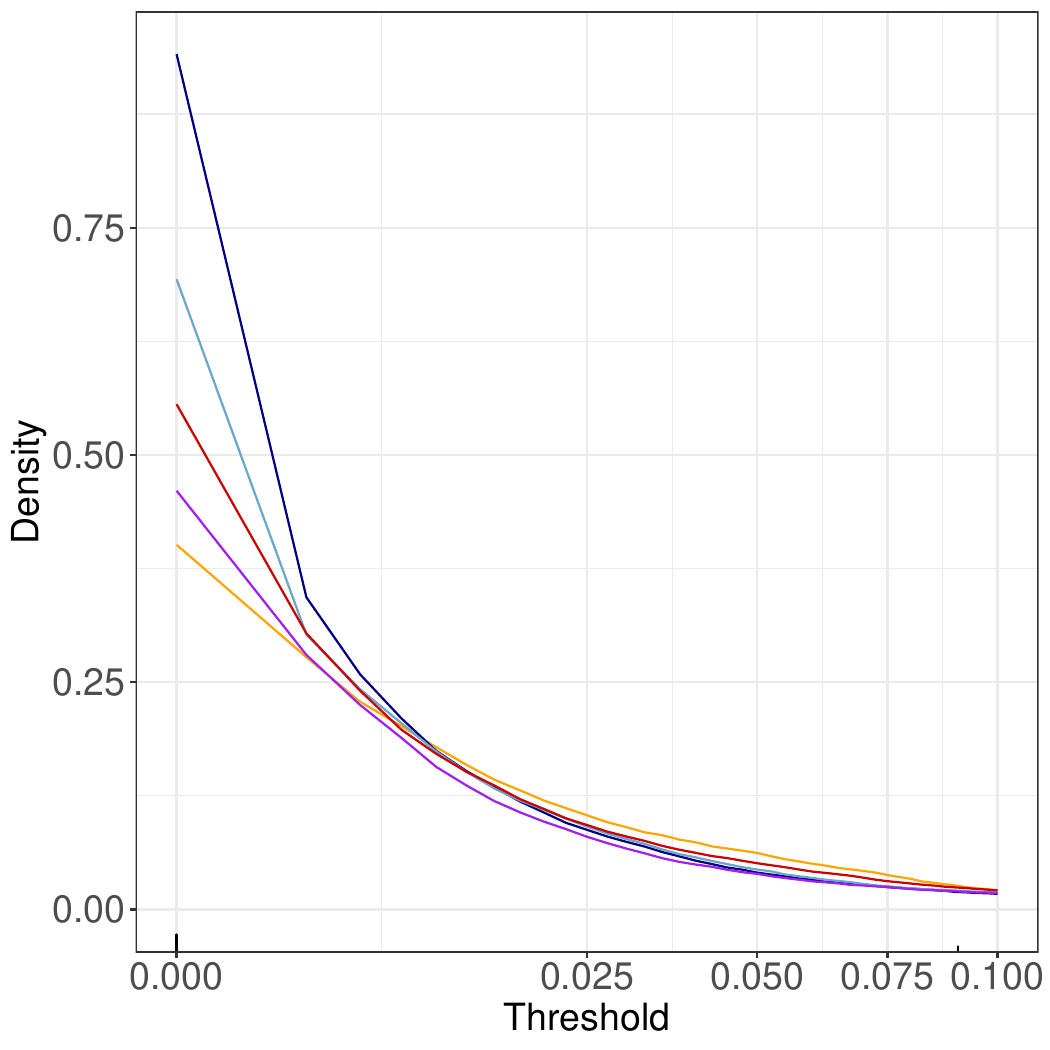}
		\caption{Input network}
	\end{subfigure}
	\begin{subfigure}[b]{0.44\textwidth}
		\centering
		\includegraphics[width=\textwidth]{inputs_Jan2025_data/CPA/Input_share_network_2022.pdf}
		\caption{Output network}
	\end{subfigure}

	\justifying \scriptsize \noindent
	Notes: This figure shows the effect of network truncation thresholds (x-axis) on the network density (y-axis). In the left (right) figure, a link is removed if the input (output) share is smaller than the threshold value. Figure uses 2022 data.
	
\end{figure}

Table \ref{tab:network_stats_2022_no_truncation} summarizes the aggregate network properties of the payment-based (for counts and values) and the three ONS IOTs (SUT, IxI, PxP), using an annual snapshot from 2022, which is the last date for which ONS IOTs are available in the 2024 Blue Book \citep{ons2024blue}. 
Supplementary descriptives using 2019 data and a network density plot are provided in Table \ref{tab:network_stats_2019_no_truncation} and Fig. \ref{fig:network_density_plot_2022}. 
Analogous results from the previous data release are shown in \ref{app:subsec:comparison_NA}. 

The key observations from this exercise are the following: 
\begin{itemize}
	\item \textbf{Network density:} The improved coverage of the novel data is associated with a higher density compared to the previous release. The density informs about the existence of transaction links. With 55\% it scores still lower than the observed density of IxI and PxP tables (71.18\% and 97.8\%), but higher than the density of SUTs (47.4\%). The count-based network has a lower density, which arises from the SDC procedure.\footnote{A bilateral transaction link can be non-zero by value but zero by counts if the number of counts is $<$50 \citep[see][]{ons2025interindustrymethods}}.
	
	Similarly to the previous data, the network density across datasets converges with network truncation (Fig. \ref{fig:network_density_plot_2022}), i.e. removing IO links if their weight given by the input or output share falls below a percentage threshold. Thresholds of $<$2.5\% are sufficient to achieve almost identical levels of density. The previous data has been less sensitive to truncation. This suggests that the improved coverage led to the addition of many small-weight transaction linkages, which is in line with the earlier observation of lower average transaction values and the interpretation of many more small firms having been integrated into the sample. Notably, SDC can operate in a similar way as network truncation. 
	
	\item \textbf{Coverage by transaction values:} By value, the payment data capture lower volumes compared to the official IOTs, as indicated by a relatively lower average strength. The average value of transactions traded by an industry is about £8.87 million, as measured by the strength. This is about two thirds of the value observed for the PxP and the IxI tables but 3 times higher than the value observed in the previous data release. 
	Normalizing the links in the IOTs into shares, the strength tends to be slightly higher compared to IxI and PxP. 
	
	The average weight indicates the value of an average transaction link in the network. This indicator is sensitive to the number of links being included in the network and varies across the IxI and PxP IOTs.
	
	\item \textbf{Structure of mutual connections:} The payment-based IOTs have slightly lower transitivity, reciprocity and degree assortativity. This may indicate that there are fewer or different  hub and clustering structures compared to the official tables. However, this has not been tested here. 
	
	\item \textbf{Stability:} The topological properties of the payment network and their qualitative relation to the official IOTs are relatively stable over time, as a comparison of the 2022 to 2019 tables reveals (see \ref{tab:network_stats_2019_no_truncation}). However, all IOTs show a rise of nominally weighted indicators (strength and weight), which is not surprising as all data types (excluding the counts) are nominal measures and subject to inflation. 
\end{itemize}

\FloatBarrier
\subsection{Auto- and cross-correlations}
\label{subsec:auto_cross_correlations}
The preceding subsection informs about the topological properties of the aggregate network, while being agnostic about the identity of sectors.\footnote{For example, two economic networks can have an identical topology from an aggregate view but for policy it makes a difference whether a critical industry as identified by the topology is, for example, financial services or machinery manufacturing.}
This subsection explores auto- and cross-correlation patterns of indicators constructed from the payment-based and national accounts IOTs. The correlation patterns are indicative of transaction- and industry-level similarities across the data sets. 

\subsubsection{Transaction-level correlations}
\begin{figure}[!h]
	
	\caption{Auto- \& cross-correlations at the edge level (2018-2022)}        
	\label{fig:correlations_edge_level_ONS_payments_raw_transactions_18_22}
	
	\centering
	
	\includegraphics[width=\textwidth]{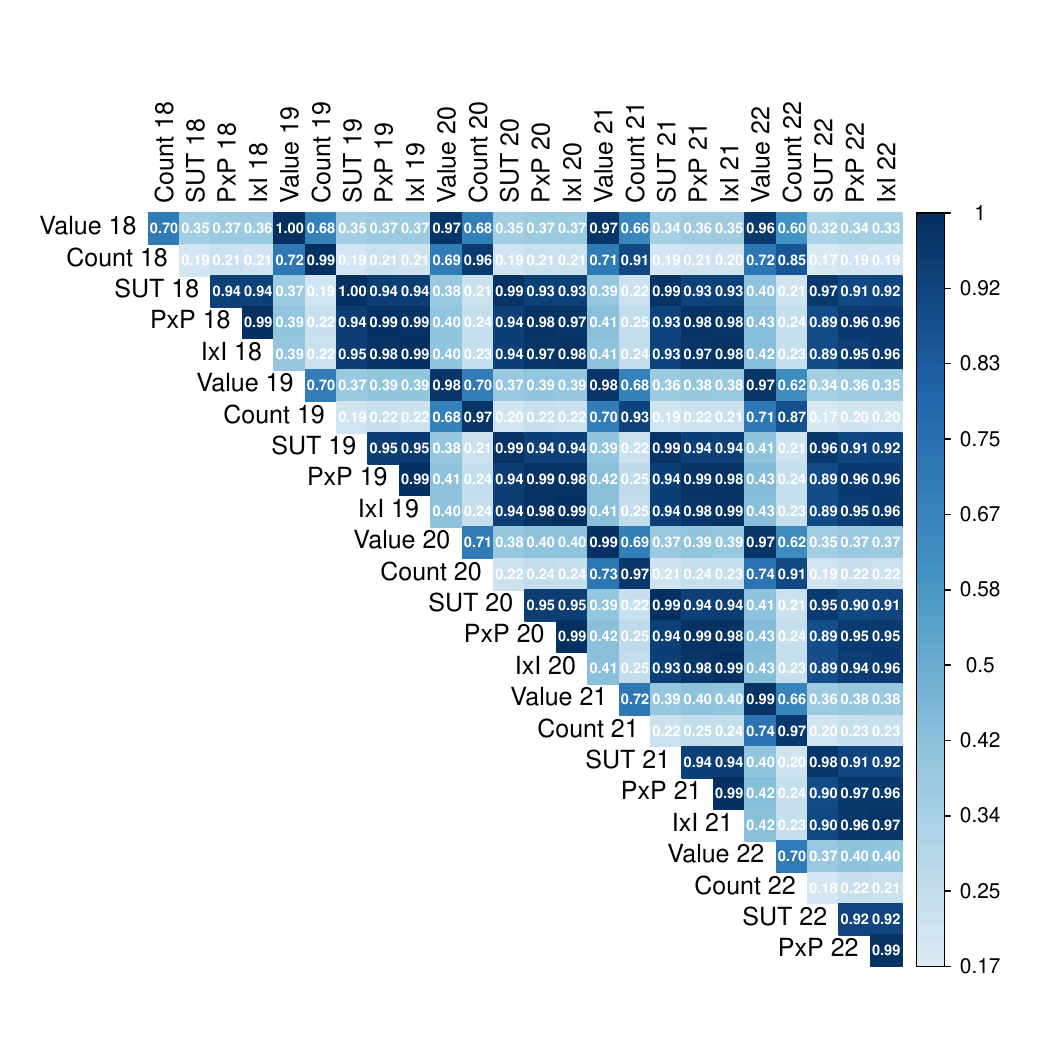}
	
	\justifying \scriptsize \noindent
	Notes: The correlations are measured by the Pearson correlation coefficient between raw transactions in the payment-based IOTs (values and counts) and the IxI, PxP, and SUTs. 
	
\end{figure}

\begin{figure}[!h]
	
	\caption{Auto- \& cross-correlations at the edge level (2021-2022)}        
	\label{fig:correlations_edge_level_ONS_payments_IO_shares_21_22}
	
	\centering
	
	\begin{subfigure}[b]{0.49\textwidth}
		\includegraphics[width=\textwidth]{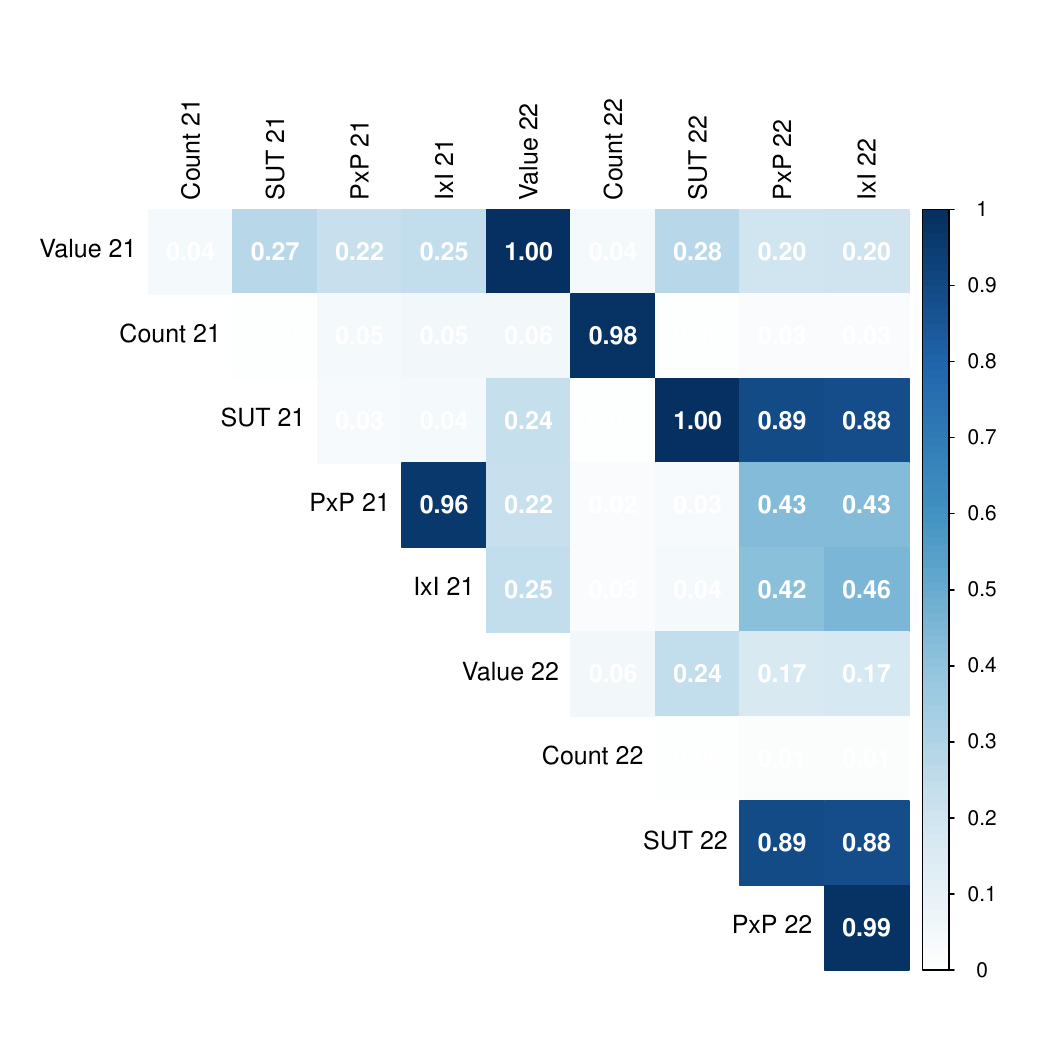}
		\caption{Input shares}
		\label{subfig:correlations_edge_level_ONS_payments_input_shares}
	\end{subfigure}    
	\begin{subfigure}[b]{0.49\textwidth}
		\includegraphics[width=\textwidth]{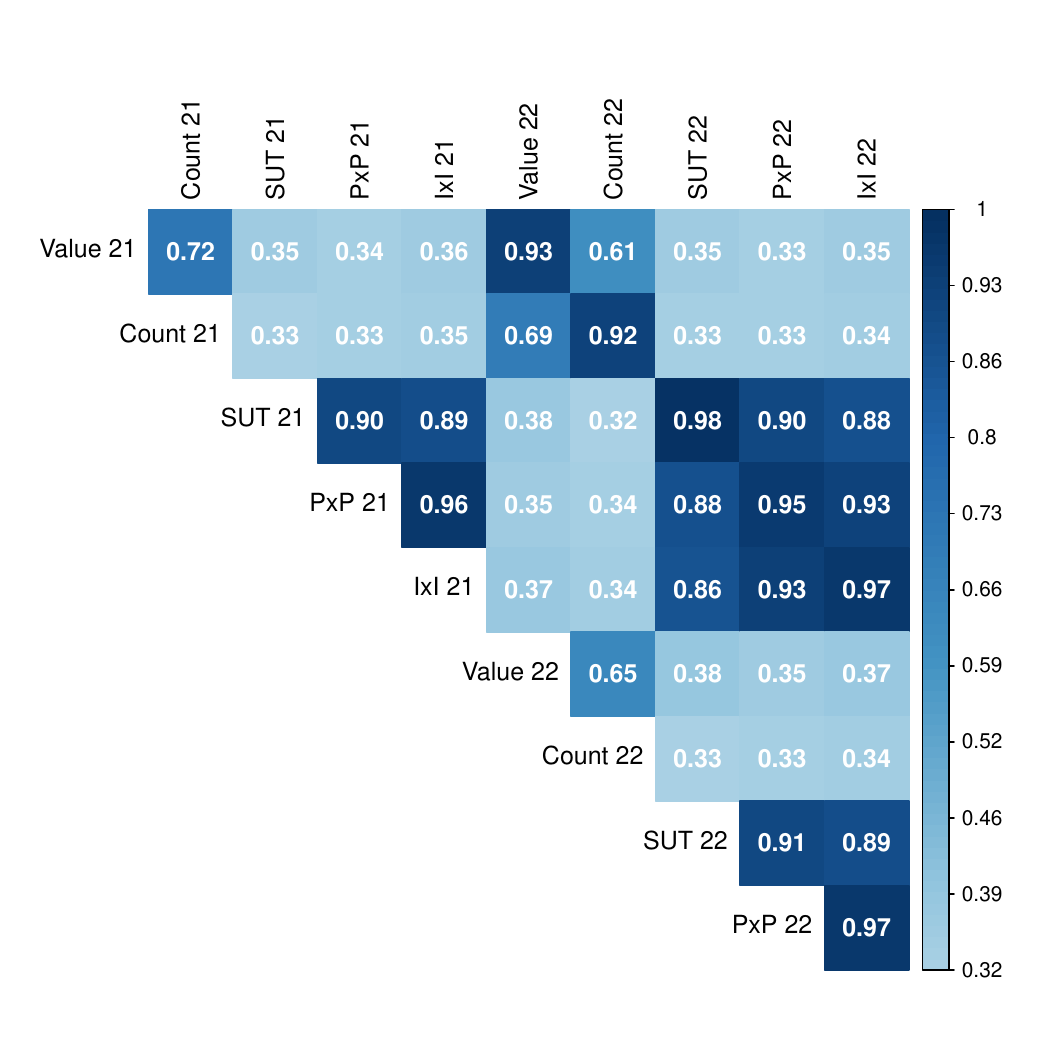}
		\caption{Output shares}
		\label{subfig:correlations_edge_level_ONS_payments_output_shares}
	\end{subfigure}
	
	\justifying \scriptsize \noindent
	Notes: The correlations are measured by the Pearson correlation coefficient between raw transactions, input and output shares in the payment-based IOTs (values and counts) and the IxI, PxP, and SUTs. 
	
\end{figure}

Fig. \ref{fig:correlations_edge_level_ONS_payments_raw_transactions_18_22}  and \ref{fig:correlations_edge_level_ONS_payments_IO_shares_21_22} illustrate correlations of IO links across the different network and across time. The former shows the correlations of links normalized as input- and output-shares, the latter uses raw transaction values without any data transformation such as taking logs, normalization or truncation. Results from the previous data release are provided in \ref{app:subsec:auto_cross_correlations}. 

The key observations are as follows: 
\begin{itemize}
	\item \textbf{Promising link-level similarity:} Cross-correlations between transaction links in the payment-value-based IOTs and the ONS IOTs score between 35-42\%, with slightly higher values for the analytical PxP and IxI tables than for SUTs (Fig. \ref{fig:correlations_edge_level_ONS_payments_raw_transactions_18_22}). These values are even higher when log-transforming the data, indicating skewness. 
	
	The cross-correlations are relatively stable when using lagged data, indicating the potential of using the real-time payment network to nowcast IOTs. 
	
	\item \textbf{Higher similarities on the output side:} Normalizing the data into input and output shares (Fig. \ref{fig:correlations_edge_level_ONS_payments_IO_shares_21_22}) reveals stronger cross-correlations for output shares than for input shares (34-35\% versus 20-28\%). This contrasts with the observations from the previous version of the data (\ref{OLD:fig:correlations_edge_level_ONS_payments_IO_shares}), although overall correlation levels are higher than before. Cross-correlations for input shares are weaker when using pre-Covid-19 data, while correlation patterns for output shares appear stable across time periods (\ref{fig:correlations_edge_level_ONS_payments_IO_shares}).

	Input-output shares provide an indication of the relative importance of industries as input suppliers or downstream customers. In the case of the payment data, they may be more informative about downstream linkages, possibly due to the underrepresentation of primary inputs from agriculture, mining, and energy in the data (see Fig. \ref{fig:time_series_A16_inputs} below).  
	
	\item \textbf{Payment counts and IOTs:} Payment counts poorly correlate with input shares, but show persistent correlations with output shares, at slightly lower but similar levels compared to payment value-based tables. They also reveal weaker correlations when comparing IOTs by raw transactions. In the previous data version, the correlations between counts and values were very similar (\ref{OLD:fig:correlations_edge_level_ONS_payments_IO_shares}). 
\end{itemize}

\FloatBarrier
\subsubsection{Industry-level similarities}
\label{subsec:industry_similarity}
\begin{figure}[hb]
	\centering
	
	\caption{Auto- and cross-correlations of inputs \& outputs at the node level}
	\label{fig:correlations_industry_level_input_output_raw}
	
	\includegraphics[width=0.8\textwidth]{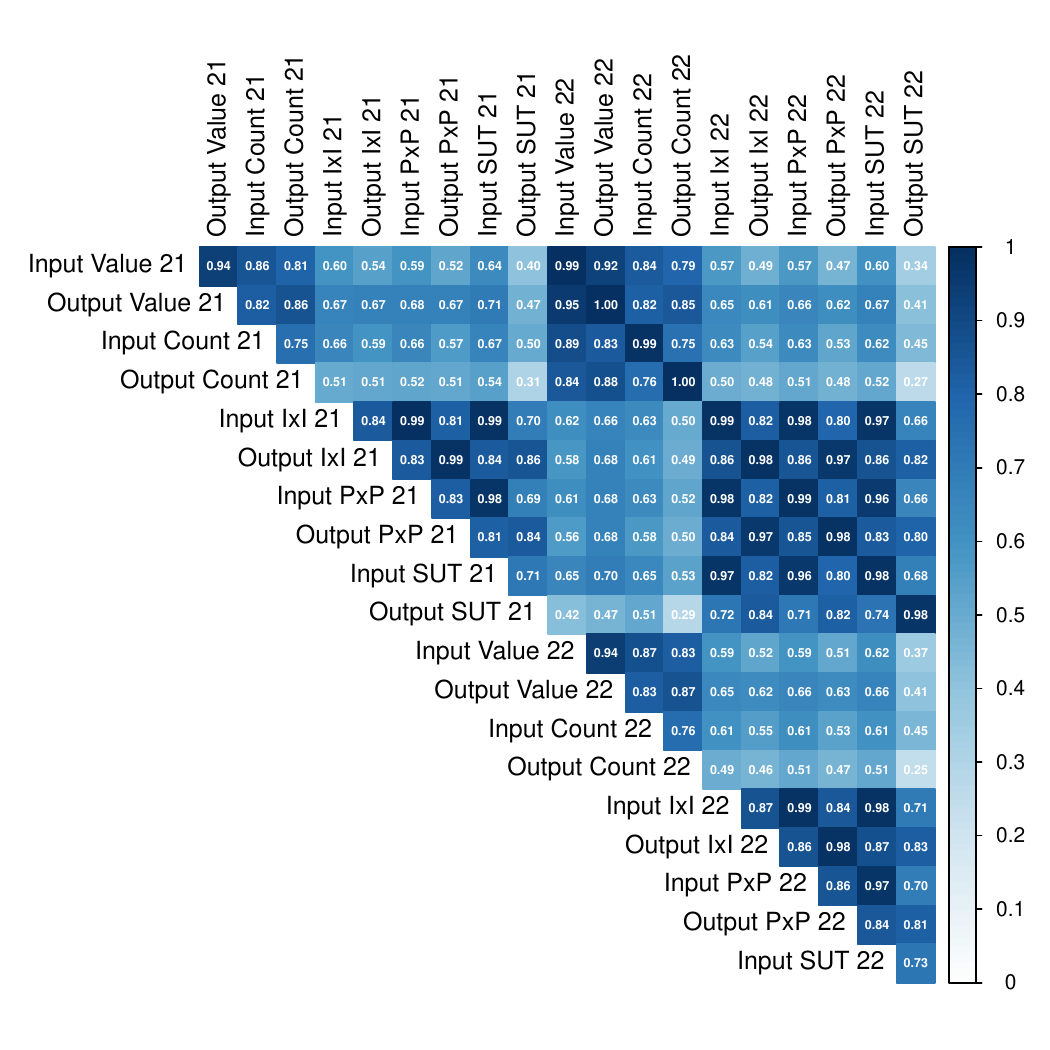}
	
	\justifying \scriptsize \noindent
	
	Notes: The correlations are measured by the Pearson correlation coefficient between industry-level annual outputs and inputs in 2021-2022 calculated by using raw transaction values and counts of the payment data and the row- and column sums of ONS IxI, PxP, and SUTs. 
\end{figure}

\begin{figure}[h]
	\centering
	\includegraphics[width=\textwidth]{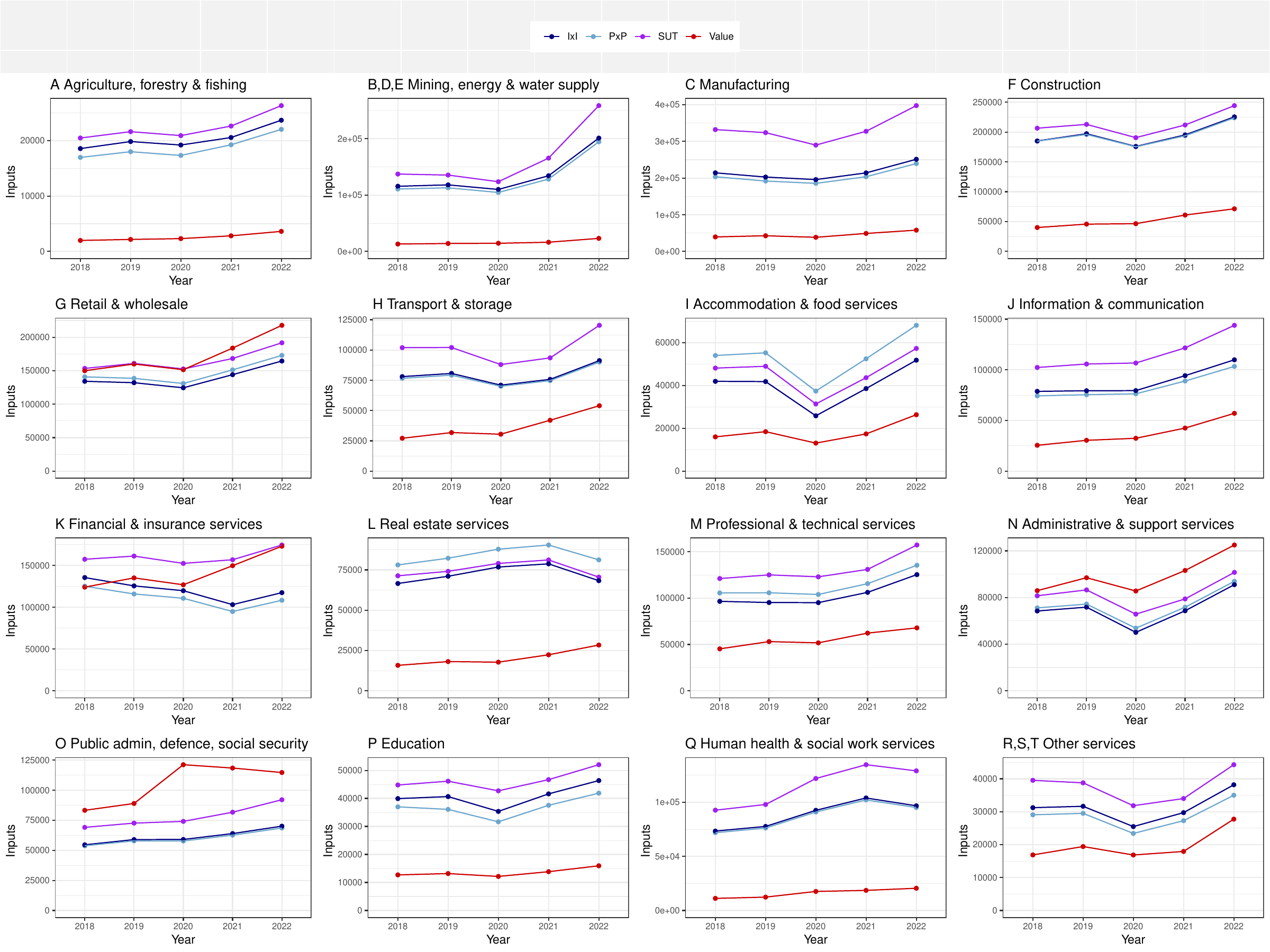}
	\caption{Comparison of aggregate industry sizes by sum of inputs over time}
	\label{fig:time_series_A16_inputs}

	\justifying \scriptsize
	\noindent
	Notes: The figure shows how industry level aggregates evolved over time for different data sets. Industries are grouped into sectors. Aggregates are calculated as sum over all input links for an industry group. Scales of the y-axes differ across plots.
\end{figure}

\begin{figure}[h]
	\centering
	\includegraphics[width=\textwidth]{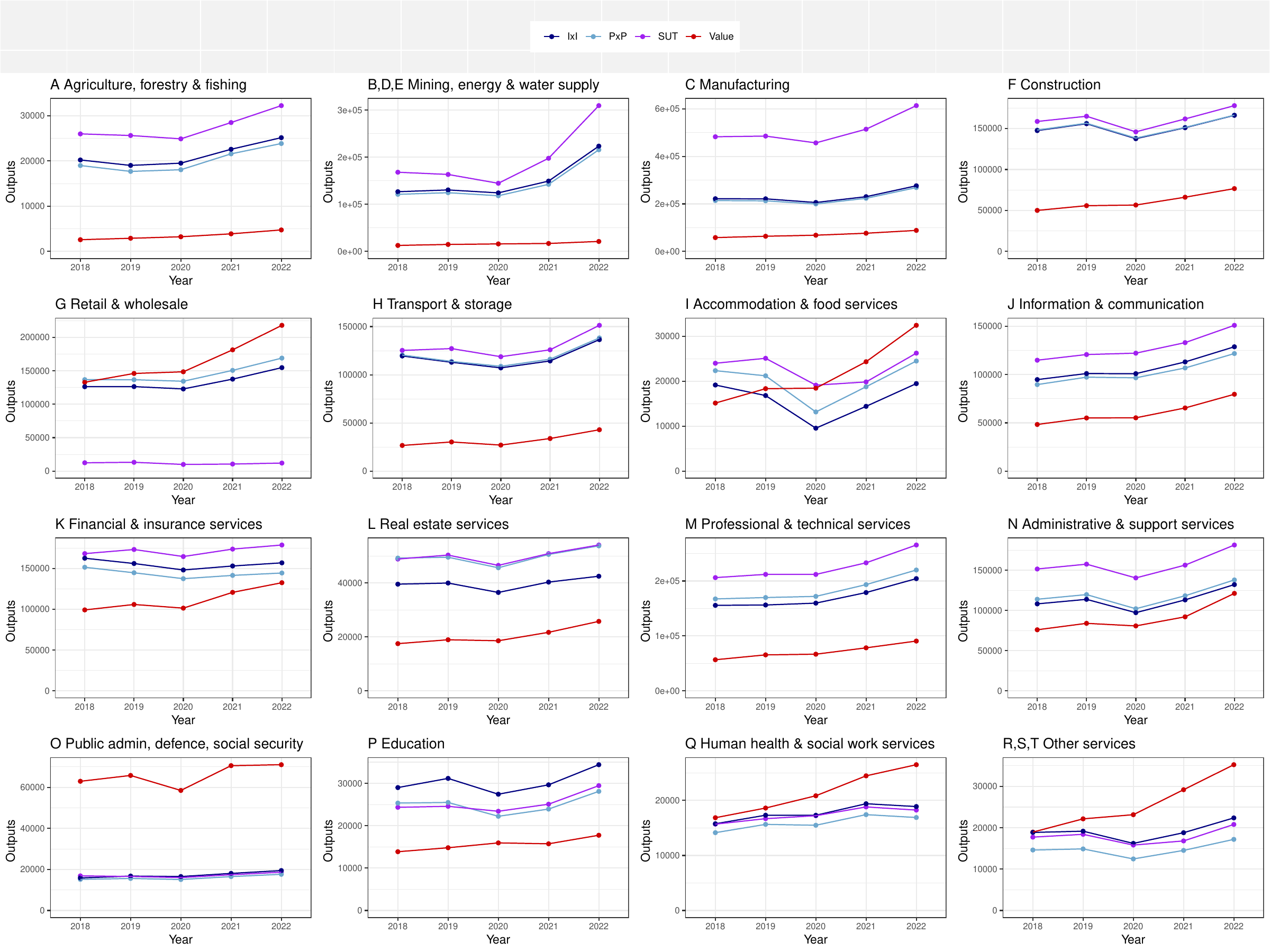}
	\caption{Comparison of aggregate industry sizes by sum of outputs over time}
	\label{fig:time_series_A16_outputs}

	\justifying \scriptsize
	\noindent
	Notes: The figure shows how industry level aggregates evolved over time for different data sets. Industries are grouped into sectors. Aggregates are calculated as sum over all output links for an industry group. Scales of the y-axes differ across plots.
\end{figure}

This subsection explores industry-level similarities showing correlations between payment-based and official IOTs using industry-level aggregates of input and output flows (Fig. \ref{fig:correlations_industry_level_input_output_raw}) and the relation of payments to other industry-level indicators, such as value added, salaries, final demand, and exports (Fig. \ref{fig:correlations_industry_level_SUT_raw}). Additional results for data in growth rates are available in \ref{app:subsec:comparison_NA}. 
Fig. \ref{fig:time_series_A16_inputs}, \ref{fig:time_series_A16_outputs} and \ref{fig:input_output_scatterplot_2022}, \ref{fig:time_series_CPA_inputs}, \ref{fig:time_series_CPA_outputs} illustrate the relative extent to which aggregate payment values cover the transaction values in the official tables and their evolution over time.

The key observations are as follows: 
\begin{itemize}
	\item \textbf{Inputs versus outputs:} Comparing how payment-based input and output aggregates correlate with those in the official IOTs, we observe stronger correlations for payments with inputs rather than outputs obtained from the IxI, PxP and SUTs (Fig. \ref{fig:correlations_industry_level_input_output_raw}) with correlations ranging between 51-71\% and 42-50\% for growth rates from 2021-22 (Fig. \ref{fig:correlations_industry_level_input_output_growth}). Growth rates during Covid-19 from 2020-2021 show no meaningful relationship.
	
	The opposite tends to hold for cross-correlations of payment value-based input and output aggregates: Payment-based output values correlate more strongly with most of the IxI, PxP and SUT aggregates than payment-based inputs. 
	In some cases, payment outputs more strongly correlate with PxP and IxI inputs rather than payment inputs, which indicates that input purchases tend to be incompletely captured by the payment data. 
	Payment counts tend to behave oppositely, with input counts being more strongly related to IOT aggregates than output counts. 
	
	\item \textbf{Relative coverage by sector:} 
	Compared to the previous version of the data (Fig. \ref{fig:input_output_scatterplot_2022} and  \ref{OLD:app:fig:input_output_scatterplot_2019}), coverage improved across all datasets: various industries that have been almost absent by value in the previous data are relatively well captured in the novel data, with few exceptions. Overall, the relative coverage appears more equally distributed than before. We also observe that a larger number of sector entries now overreports values in comparison to national accounts.  
	
	Supplementing the coverage analysis, Fig. \ref{fig:time_series_A16_inputs} and \ref{fig:time_series_A16_outputs} use a 16-sector aggregation and illustrate how the relative coverage of input and output transactions differs across sectors and time. Underreporting (on the input and output side) of the payment data appears to be strongest for agriculture, mining and energy, manufacturing and construction, which are typically more upstream in the supply chain. Despite their low coverage and as discussed below, some of these sectors show high levels of network centrality (Table \ref{app:tab:top10_influence_vector_no_services}) despite their relatively weak coverage. 
	
	We also observe significant underreporting in various service sectors, including transport and storage, real estate, communication and professional services. 
	Some sectors show significant underreporting on the input side but have a good or even excessive coverage on the output side, such as accommodation and food services and health and social services.
	Notably, the concern of an excessive overrepresentation of intermediary industries like retail and wholesale as well as financial services appears to be weakly justified at the 16-sector aggregation, as these show rather similar aggregate values as observed in national accounts.\footnote{Fig. \ref{fig:time_series_CPA_inputs} and \ref{fig:time_series_CPA_outputs} show additional time series figures at the granular level of 104 distinct CPA codes, illustrating heterogeneity at the granular level in coverage.} 
	Only public administration remains as an extreme outlier with significant overreporting at the output side. This likely arises from a conceptual issue as public administration entities receive payments by collecting fees and tax payments which would be excluded from the official IOTs.
	
	
\end{itemize}

\FloatBarrier
\section{Stylized facts of the granular data}
\label{sec:stylised_facts_5digit}
This section investigates the consistency of payment data at the most disaggregated 5-digit SIC level with stylized facts on highly granular production networks that have been documented in the literature. More details on the background of these analyses are provided in \citet{hotte2025national}.

\FloatBarrier
\subsection{Correlation of growth rates}
\label{sec:corr_growth_rates_main}

\begin{figure}[!h]
	\centering
	
	\caption{Correlations of growth rates}
	\label{fig:growth_correl_by_distance}
	
	\includegraphics[width=\textwidth]{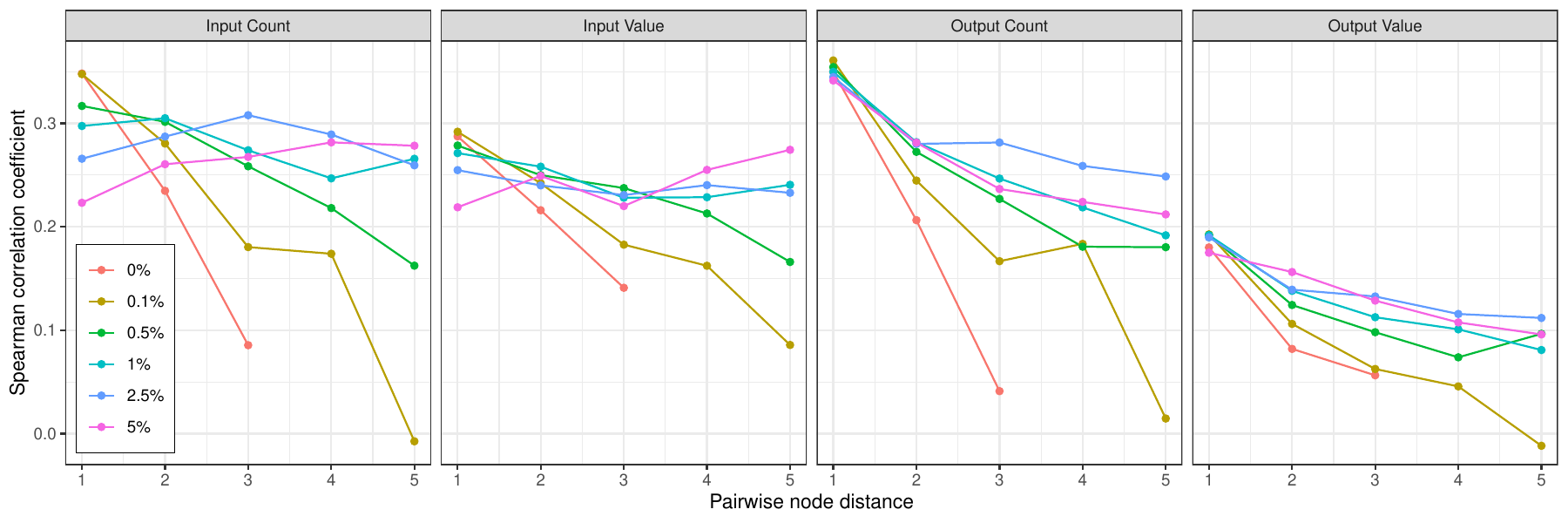}
	
	\justifying \scriptsize
	\noindent
	Notes: These figures illustrate the Spearman correlation coefficients between monthly (year on year) growth rates of directly and indirectly connected pairs of industries, using data from 2016 to 2018 and 2024. 
	The x-axis shows the distance of the industry pairs in annual network aggregates.\footnote{The distances are the shortest path (lowest number of steps) that connects the pair of industries in the network. A value of one indicates a direct link (one step) between the pair.} 
	The colors indicate truncation thresholds imposed on the network before calculating the distances. Links with a weight (input share) below the threshold are removed (see also Section \ref{subsec:aggregate_network}).
	
	\label{fig:corr_network}
\end{figure}

Fig. \ref{fig:growth_correl_by_distance} shows correlations of growth rates of a pair of industries at different stages of distance in the network and for different levels of network truncation. Additional results for the old data and figures excluding intermediary industries or using Pearson instead of Spearman correlations are provided in the Appendix \ref{app:subsec:stylized}. The key observations are the following:

\begin{itemize}
	\item \textbf{Correlations decrease with distance:} Growth rates correlate less when industries are more distant in the network, supporting a stylized fact from the literature. This pattern is robust when truncating the network at moderate levels and holds for growth rates of inputs and outputs measured as counts or values. 
	The levels of correlation are higher compared to the previous version of the data (Fig. \ref{OLD:fig:growth_correl_by_distance}) and robust when removing intermediary industries (retail and wholesale (G45, G46, G4), financial and insurance services (K64, K65, K66), public administration (O84)) (Fig. \ref{fig:growth_correl_by_distance_no_interm}).     
	\item \textbf{Effects of network truncation:} When strongly truncating the network by removing small links, correlations are not meaningful and we observe in some cases and a u-shaped pattern of correlations. This effect is stronger when removing intermediaries (Fig. \ref{fig:growth_correl_by_distance_no_interm}) or using the Pearson correlation coefficient (Fig. \ref{fig:growth_correl_by_distance_pearson}), which is more sensitive to outliers. Compared to the earlier version of the data, the pattern of a negative relationship between growth and distance is more robust against different approaches to data pre-processing. Yet it remains true that when too many links are removed, the network may no longer be a meaningful representation of the supply chain and thus any truncation based on threshold values must be carefully be thought through. 
\end{itemize}

\FloatBarrier
\subsection{Centrality distribution}
\label{sec:influence_vector}

\begin{figure}[!h]
	\centering
	\caption{CCDF of the Katz-Bonacich centrality}
	\label{fig:influence_vector}
	\includegraphics[width=0.8\textwidth]{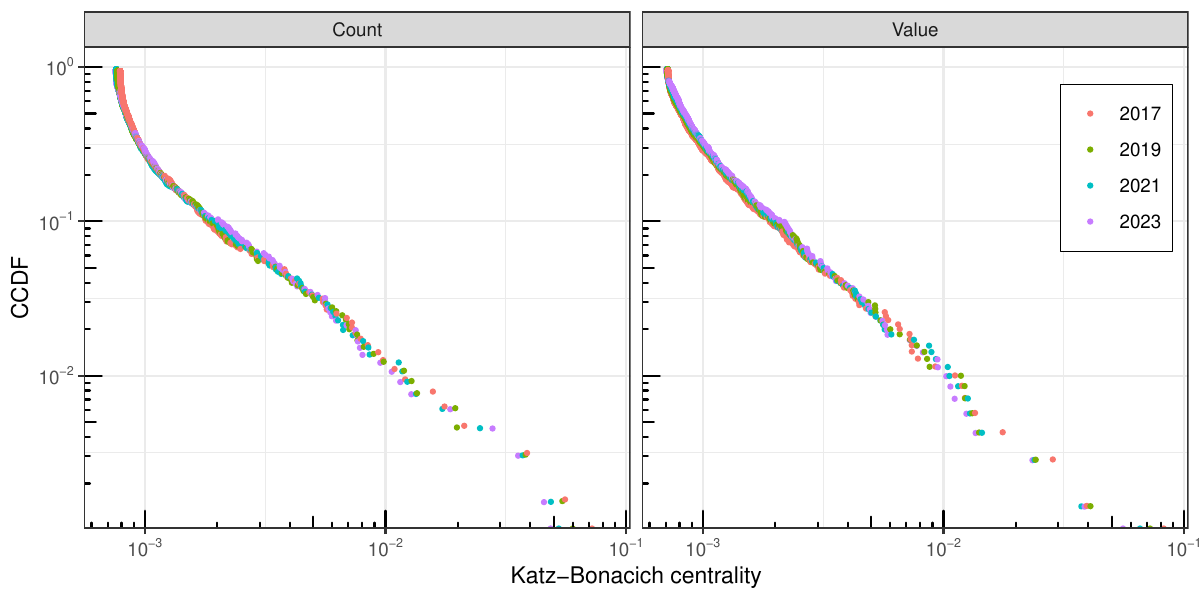}
	
	\justifying \scriptsize
	\noindent
	Notes: These figures illustrate the CCDF of the Katz-Bonacich centrality \citep[see][]{hotte2025national} for different years, using a labor share parameter of $\alpha_L = 0.5$ \citep{magerman2016heterogeneous} and payment-based input share matrices based on counts and values. 
	
\end{figure}

Fig. \ref{fig:influence_vector} and \ref{OLD:fig:influence_vector} show the complementary cumulative distribution function (CCDF) of the Katz-Bonacich centrality of the 5-digit industries. Robustness checks evaluate the impact of network truncation, and potential role of outliers by removing intermediary sectors (Fig. \ref{fig:influence_vector_no_intermediaries}) and services (Fig. \ref{fig:influence_vector_no_services}). Table \ref{tab:top10_influence_vector} shows a ranking of the top-10 sectors that rank highest by their influence vector and would be expected --according to theory-- to be the drivers of aggregate fluctuations. The upper (bottom) panel shows the ranking in the raw network data (for data where service sectors have been removed before centrality scores are calculated). 
Additional ranking tables for other years and samples excluding intermediaries and service sectors are shown in Table \ref{app:tab:top10_influence_vector}, \ref{app:tab:top10_influence_vector_no_intermed}, \ref{app:tab:top10_influence_vector_no_services}).

\begin{table}[!htbp] \centering 
	\footnotesize
	\caption{Top 10 industries by influence vector in 2023} 
	\label{tab:top10_influence_vector} 
	\begin{tabularx}{\textwidth}{@{\extracolsep{5pt}} ccX|ccX} 
		\\[-1.8ex]\hline 
		\hline \\[-1.8ex] 
		SIC & & Industry description & SIC & & Industry description\\ 
		\hline \\[-1.8ex] 
		\multicolumn{6}{l|}{\textbf{\underline{Raw network data}}}\\ 
		\multicolumn{3}{c|}{Value}&\multicolumn{3}{c}{Count}\\ 
		\hline \\[-1.8ex] 
		84110 & 0.0823 & General public administration & 61900 & 0.0722 & Other telecommunications activities \\ 
		82990 & 0.0394 & Other business support services n.e.c. & 82990 & 0.0557 & Other business support services n.e.c. \\ 
		64999 & 0.0285 & Financial intermediation n.e.c. & 84110 & 0.0388 & General public administration \\ 
		61900 & 0.0176 & Other telecommunications activities & 64999 & 0.0212 & Financial intermediation n.e.c. \\ 
		65110 & 0.0135 & Life insurance & 65110 & 0.0176 & Life insurance \\ 
		62090 & 0.0123 & Other information technology services & 62090 & 0.0157 & Other information technology services \\ 
		49410 & 0.0119 & Freight transport by road & 96090 & 0.0121 & Other service activities n.e.c. \\ 
		70100 & 0.0112 & Activities of head offices & 64910 & 0.0109 & Financial leasing \\ 
		62020 & 0.0092 & Information technology consultancy & 65120 & 0.0098 & Non-life insurance \\ 
		96090 & 0.0078 & Other service activities n.e.c. & 49410 & 0.0093 & Freight transport by road \\ 
		\hline \\[-1.8ex] 
		\hline \\[-1.8ex] 
		\multicolumn{6}{l|}{\textbf{\underline{Excluding services}}}\\ 
		\multicolumn{3}{c|}{Value}&\multicolumn{3}{c}{Count}\\ 
		\hline \\[-1.8ex] 
		32990 & 0.033 & Other manufacturing n.e.c. & 32990 & 0.0535 & Other manufacturing n.e.c. \\ 
		43999 & 0.0304 & Other specialised construction n.e.c. & 43999 & 0.0252 & Other specialised construction n.e.c. \\ 
		35140 & 0.023 & Trade of electricity & 36000 & 0.0219 & Water collection, treatment and supply \\ 
		35130 & 0.0198 & Distribution of electricity & 33200 & 0.0207 & Industrial machinery installation \\ 
		25990 & 0.0174 &Metal products manufacture n.e.c. & 35140 & 0.0204 & Trade of electricity \\ 
		35110 & 0.0154 & Production of electricity & 35220 & 0.0168 & Gas fuels distribution through mains \\ 
		35220 & 0.0146 & Gas fuels distribution through mains & 35130 & 0.0168 & Distribution of electricity \\ 
		43210 & 0.0142 & Electrical installation & 43210 & 0.0156 & Electrical installation \\ 
		22290 & 0.0117 & Manufacture of other plastic products & 25990 & 0.0139 &Metal products manufacture n.e.c. \\ 
		42990 & 0.0109 & Civil engineering construction n.e.c. & 43220 & 0.0122 & Plumbing/heat/air-condition installation \\  
		\hline \\[-1.8ex] 
	\end{tabularx}

	\justifying \noindent \scriptsize
	Notes: Transaction links to service-related sectors (G45-Q88, S94-U99) have been removed from the data before the centrality scores are calculated.
\end{table}

The key observations are as follows: 

\begin{itemize}
	\item \textbf{Power law-like cumulative distribution:} The CCDF exhibits a nearly log-linear shape beyond a tiny threshold value ($<$0.001). This indicates consistency with stylized facts from the literature. The near-linearity is visually stronger for data measured in values and stronger in comparison to the previous data (Fig. \ref{fig:influence_vector} and \ref{OLD:fig:influence_vector}). 
	This pattern is consistent across years and robust against network truncation (Fig. \ref{fig:influence_vector_truncated}), the exclusion of intermediaries (Fig. \ref{fig:influence_vector_no_intermediaries}), and service sectors (Fig. \ref{fig:influence_vector_no_services}). 
	However, for most data configurations, significance tests on whether the centrality distribution truly follows a power law do not confirm this to be significant against null models (see \ref{app:subsubsec:centrality}).     
	
	\item  \textbf{Power law coefficients in alignment with the literature:} For the baseline case, the fitted power law coefficients range between 1.3-1.6 (\ref{app:subsubsec:centrality}), which is consistent with stylized facts reported by the literature and indicates greater consistency with other empirically described production networks compared to the values observed in the previous data.
	
	\item \textbf{Ranking of sectors by influence:} Investigating the top ranks of industries by centrality points to potential issues with the payment data. Public administration, telecommunications, financial intermediation, and business service support activities dominate the top ranks by centrality (Table \ref{tab:top10_influence_vector}). The top rank by public administration arises from the data construction, connecting this sector to almost any business in the UK that pays fees to public authorities or receives public benefits. Such transactions are differently treated in national accounts (see Sec. \ref{subsec:industry_similarity} and \citet{hotte2025national}).
	In addition, various top ranks are taken by sectors attached with an `not elsewhere classified' (n.e.c.) residual classification code, which might be used by a very heterogeneous range of businesses connected to diverse customers and input suppliers, depending on the good being produced and traded. However, this does not imply that an individual good being traded represents an essential input to the full range of customers. At higher levels of aggregation, this issue disappears when `n.e.c' subclasses are merged with aggregates.\footnote{Other research on industry classification has shown that n.e.c.s are often associated with emerging technologies. See \url{https://escoe-website.s3.amazonaws.com/wp-content/uploads/2022/06/13151521/SS-E-Russell.pdf}}
	
    When removing services from the data, we observed that energy, construction, and machinery manufacturing score high. This can be relevant for applied research on supply chain disruptions as these sectors are often considered origins of macro-level fluctuations. It seems that the payment network captures their influential role, despite their relative underreporting by scale in relation to national accounts (Sec. \ref{subsec:industry_similarity}). 
\end{itemize}

\FloatBarrier
\section{Concluding remarks and remaining challenges}
\label{sec:conclusion}
\subsection{Wrapping up}
This study provides a systematic empirical update on the 2025 public release of the UK inter-industry payment dataset, building on and extending earlier work and methodology \citep{hotte2025national}. For conceptual discussions and methodological detail, the reader is referred to the original article. 

The key advances of the updated data include (1) the broader inclusion of organizations ($>$3.1 million, exceeding 50\% of all registered UK businesses), (2) the integration of transactions made through the FPS, and (3) improvements in industry classification. Further, the increased scale of the data enabled the public release of a highly disaggregated 5-digit version of the data, without breaching SDC. These updates have significantly increased the dataset's representativeness, analytical utility, and consistency with official national accounts. 

The empirical validation exercises in this paper demonstrate that the data exhibit strong correlations with key macroeconomic aggregates, including GDP, monetary aggregates, and inflation. Compared to the previous version, these raw correlations are stronger, indicating an increased usefulness of the data in macro- and industry-level economic forecasting and modeling. 
Observations such as a lower average transaction value suggest an increased coverage of SMEs, which represent $>$99\% of businesses registered in the UK. Benchmarking against industry-level national accounts data reveals an improved coverage and consistency of the data in relation to official IOTs, while issues of underreporting and conceptual challenges remain. 
At the most granular 5-digit level, the data align well with documented stylized facts from the production network literature, suggesting the potential future use of the data in a real-time tracing of the granular origins of aggregate fluctuations \citep{acemoglu2012network}. Overall, the novel data mark a major step forward in constructing real-time economic indicators from bottom-up collected naturally occurring data \citep{buda2023national}.  
The novel data release offers a promising base for highly granular, real-time applied economic research, developing novel empirical methodologies and advancing economic theory.

\subsection{Remaining Challenges}
Despite these advances, various challenges and conceptual considerations remain. 
First, a considerable share of transactions (approximately 60\%) are linked to bank accounts which either remain unclassified or where associated SIC codes cannot be mapped to CPA codes used in official national accounts IOTs. This reduces the effective coverage of the cleaned data for national accounts benchmarking and introduces potential biases in empirical applications and theoretical models that rely on the assumption of a closed economic cycle and full classification. Additional biases arise from potential industry- and firm-specific payment behaviors as discussed in \citet{hotte2025national}. 

Second and relatedly, the presence of high-value transactions among non-identifiable entities is a puzzle. These payments may stem from foreign organizations, dormant and non-trading companies, or other institutions not registered in CH or registered with placeholder SIC codes (e.g., `74990' `99000', `99999'). Some of these entities might be involved in international trade, yet their structural roles in the economy remain unclear due to classification gaps. Future work might investigate their role systematically by analyzing the distribution of payment flows for which only one side (payer or payee) remains unclassified. Since input and output patterns are idiosyncratic across industries, insights could be gained into the nature of these semi-unclassified payments. As an interim solution, applied econometric modeling at the sector- and macro-level might consider including such semi-unclassified payment flows as control variables in the analyses. 

Third, another open challenge relates to the `time of recording' \citep[see][]{hotte2025national}. The real-time nature of the data opens a window into real-time economic measurement, yet little is known about the exact timing of firms' payment behavior and the role of industrial heterogeneity and financial intermediation in this process. This issue might be hard to address systematically during data processing. However, it is plausible to expect businesses that run into economic difficulties to be no longer able to pay their bills, which would be statistically reflected in the data, given the data's scale and coverage. It will be up to applied economic research to demonstrate the capacity of the data to provide real-time economic insights. 

Fourth, a related issue arises from adjustments in prices and rigidities in this process. The relationship of payments to inflationary patterns is to be investigated in future work, whereby the high level of granularity and the availability of payment counts as a novel indicator may open entirely new opportunities for economic research. 

Various other conceptual considerations discussed in \citet{hotte2023demand} remain valid for the novel data. Researchers using payment data need to keep these in mind when developing theoretical models and drawing empirical inference based on the data. 

\subsection{Outlook}
Some of the issues listed above are on the agenda of ongoing research and ONS continues its efforts of further developing, improving and extending the data and its portfolio of related data publications, including a regional breakdown. 
This paper gives an update on recent achievements and should serve as a primer on what can be expected soon. 


\newpage
\renewcommand \bibname{References}
\printbibliography

\newpage
\appendix
\renewcommand{\appendixname}{Appendix}
\renewcommand{\thesection}{\Alph{section}} \setcounter{section}{0}
\renewcommand{\thefigure}{\Alph{section}.\arabic{figure}} \setcounter{figure}{0}
\renewcommand{\thetable}{\Alph{section}.\arabic{table}} \setcounter{table}{0}
\renewcommand{\theequation}{\Alph{section}.\arabic{table}} \setcounter{equation}{0}

\FloatBarrier

\section{Supplementary statistics and previous results}
\label{app:sec:old_results}
This appendix provides additional descriptive statistics and shows figures and tables from \citet{hotte2025national} to supplement the comparison. 
\subsection{Benchmarking UK payments}
\label{app:subsec:payment_benchmarking}
\begin{figure}
	
	\caption{OLD DATA: Monthly time series of payment data and major UK schemes}
	\label{fig:ts_UKpayments_benchmarking_old} 
	\centering
	\begin{subfigure}[b]{\textwidth}
		\includegraphics[width=\textwidth]{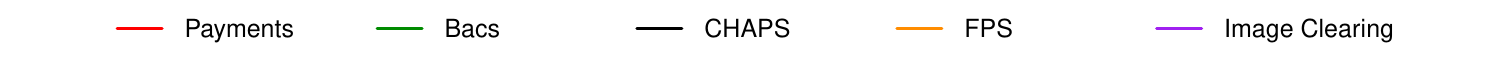}
	\end{subfigure}
	
	\begin{subfigure}[t]{0.49\textwidth}
		\centering     
		\includegraphics[width=0.8\textwidth]{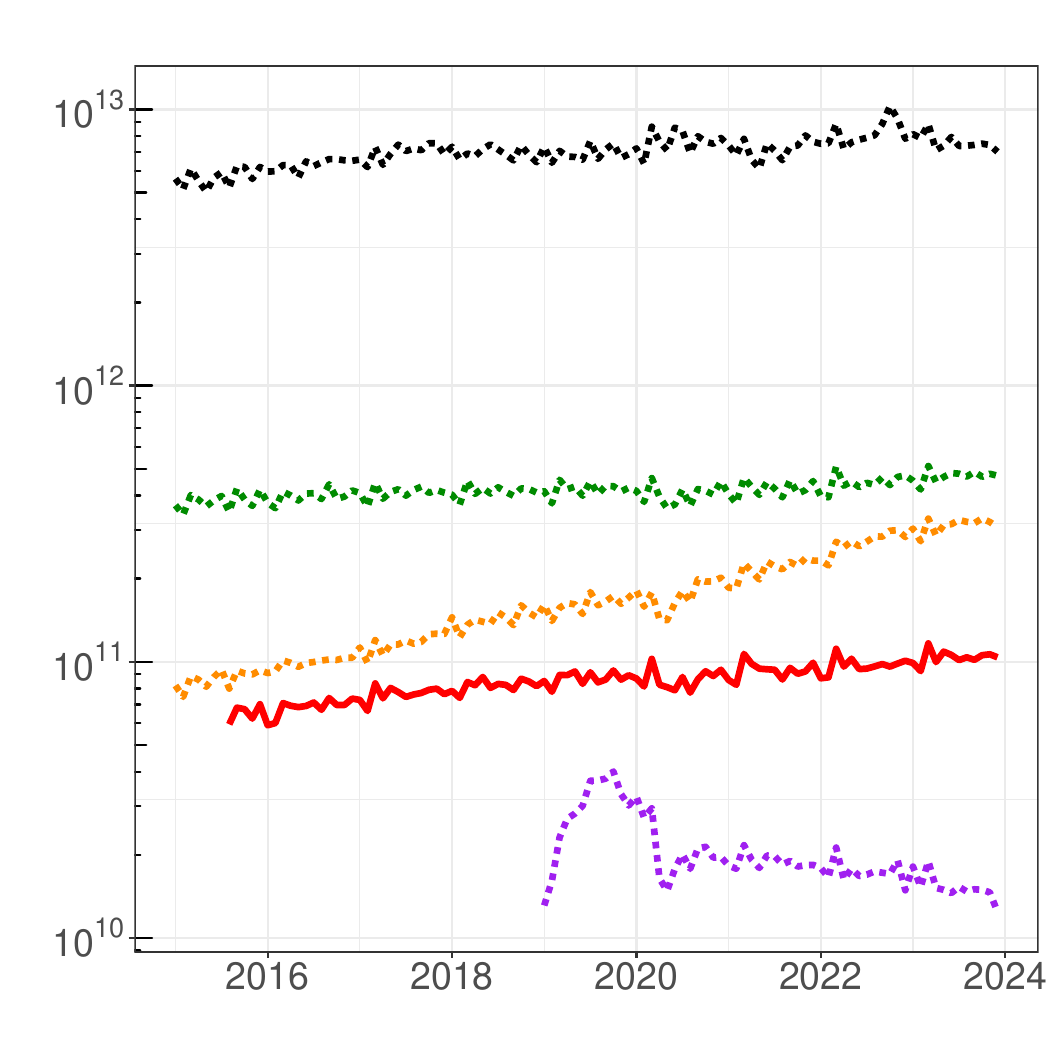}
		\caption{Values}
		\label{fig:ts_UKschemes_values_old}
	\end{subfigure}  
	\begin{subfigure}[t]{0.49\textwidth}
		\centering        
		\includegraphics[width=0.8\textwidth]{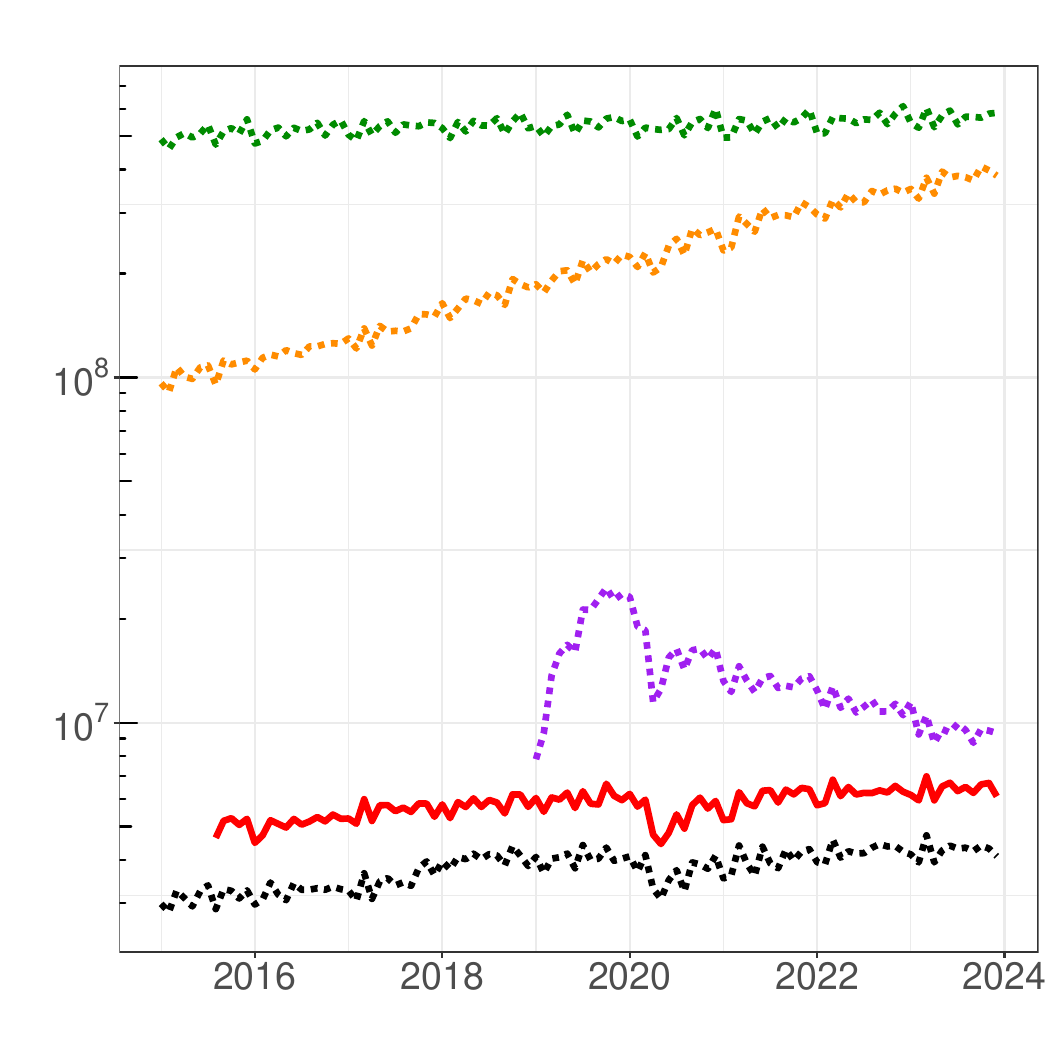}
		\caption{Counts}
		\label{fig:ts_UKschemes_counts_old}
	\end{subfigure}
	
	\justifying \scriptsize \noindent
	Notes: The vertical axis is scaled at a log-10 scale. Payments (red) are monthly aggregates of our data. The Bacs, CHAPS, FPS, and Image Clearing System data are downloaded from \citet{payuk2023historicaldata}. 
\end{figure}

\begin{figure}[!h]
	\centering   
	
	\caption{OLD DATA: Monthly payment data, direct debits and direct credits}
	\label{old:app:fig:ts_Bacs_instruments_benchmarking}
	
	\begin{subfigure}[b]{\textwidth}
		\centering
		\includegraphics[width=\textwidth]{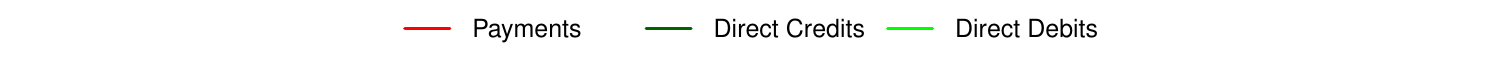}        
	\end{subfigure}
	
	\vspace{-0.225cm}

	\begin{subfigure}[b]{0.49\textwidth}
		\centering 
		\includegraphics[width=0.8\textwidth]{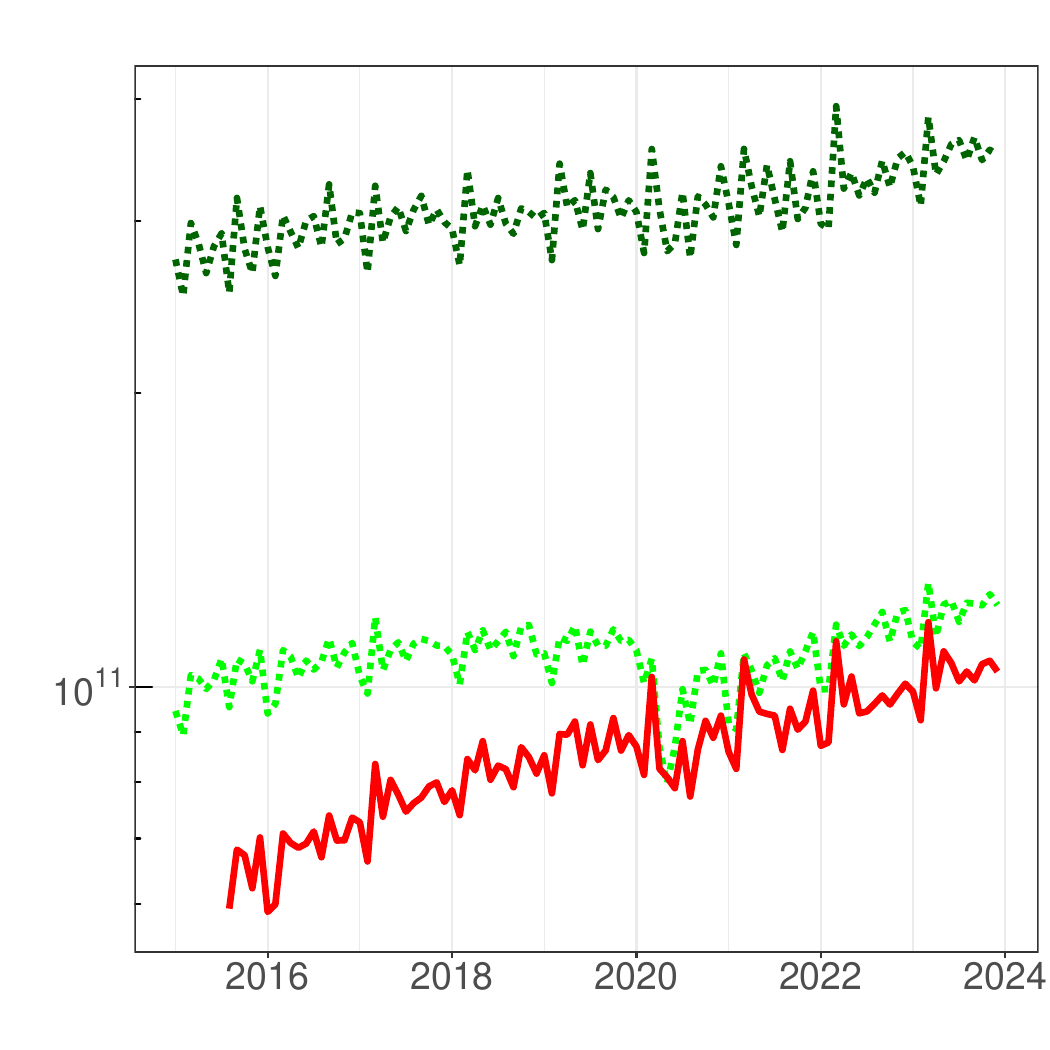}
		\caption{Values}
		\label{app:fig:ts_bacs_values_old}
	\end{subfigure}
	\begin{subfigure}[b]{0.49\textwidth}
		\centering
		\includegraphics[width=0.8\textwidth]{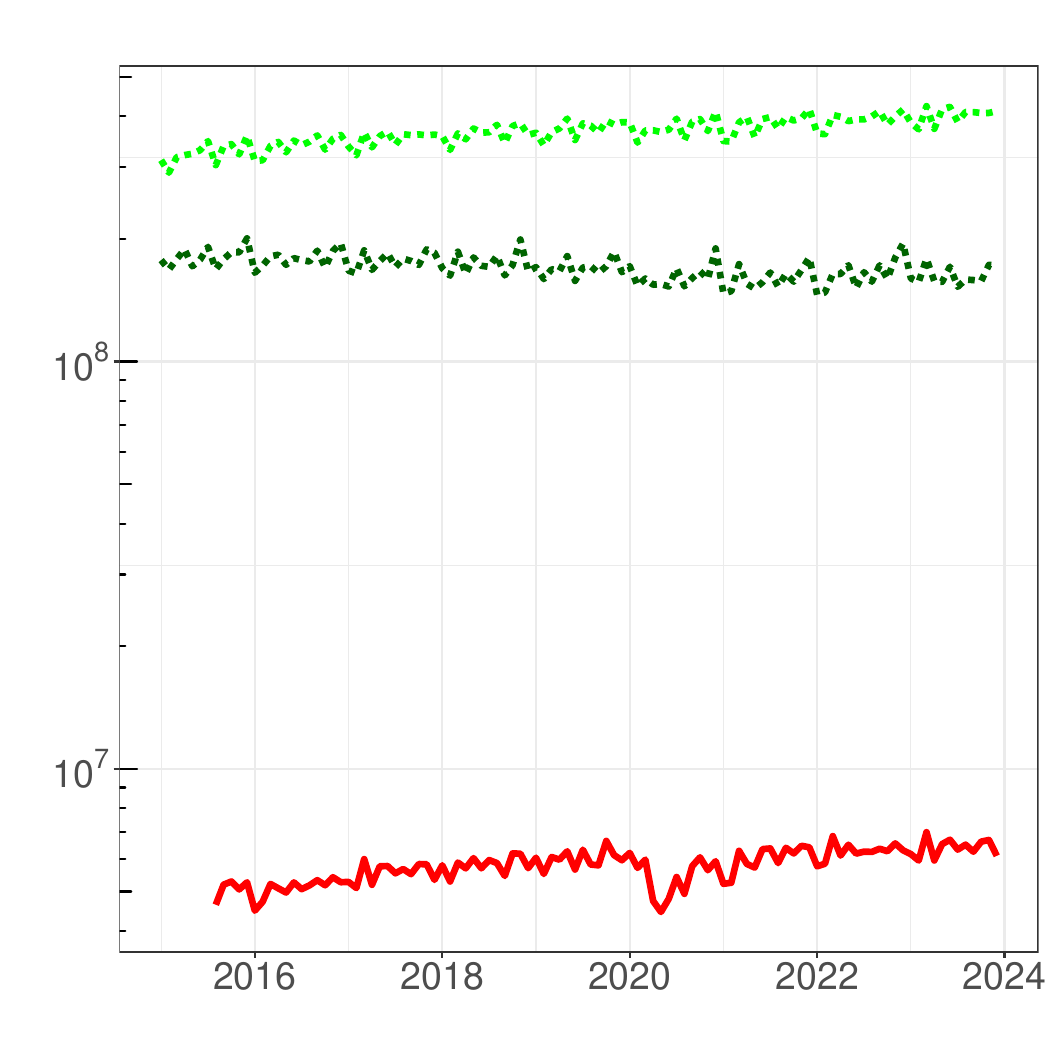}
		\caption{Counts}
		\label{app:fig:ts_bacs_counts_old}
	\end{subfigure}    
	
	\justifying \scriptsize \noindent
	Notes: The vertical axis is scaled at a log-10 scale. Payments (red) are monthly aggregates of our data. Bacs Direct Debit and Direct Credit data is downloaded from \citet{payuk2023historicaldata}. 
	
\end{figure}
\begin{figure}[!h]
	\centering   
	
	\caption{Monthly payment data, direct debits and direct credits}
	\label{app:fig:ts_Bacs_instruments_benchmarking}
	
	\begin{subfigure}[b]{\textwidth}
		\centering
		\includegraphics[width=\textwidth]{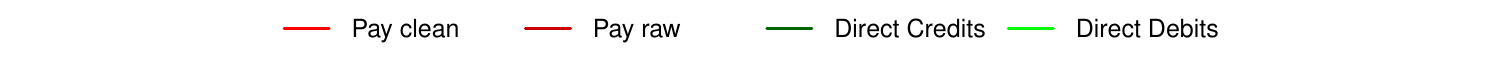}        
	\end{subfigure}
	
	\vspace{-0.225cm}

	\begin{subfigure}[b]{0.33\textwidth}
		\centering 
		\includegraphics[width=0.8\textwidth]{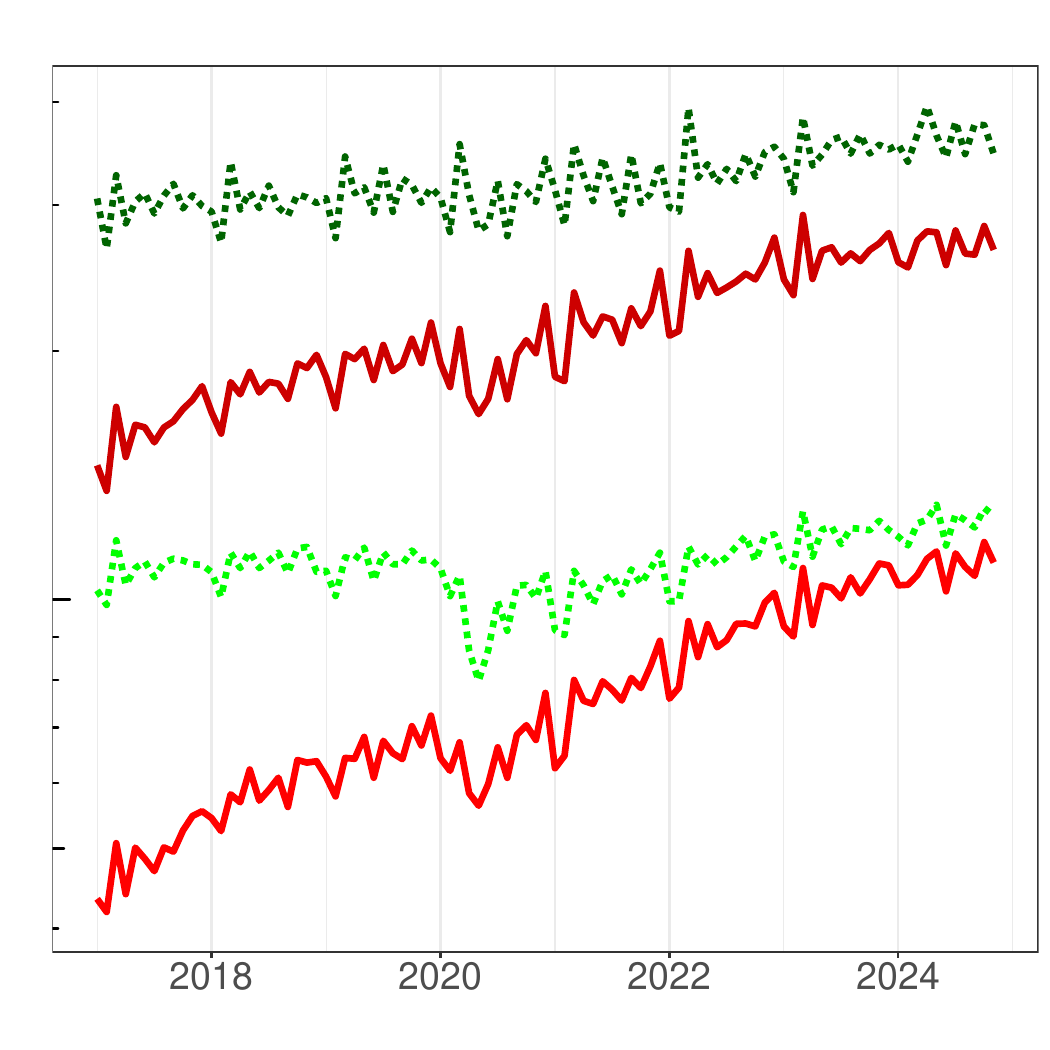}
		\caption{Values}
		\label{app:fig:ts_bacs_values}
	\end{subfigure}
	\begin{subfigure}[b]{0.33\textwidth}
		\centering 
		\includegraphics[width=0.8\textwidth]{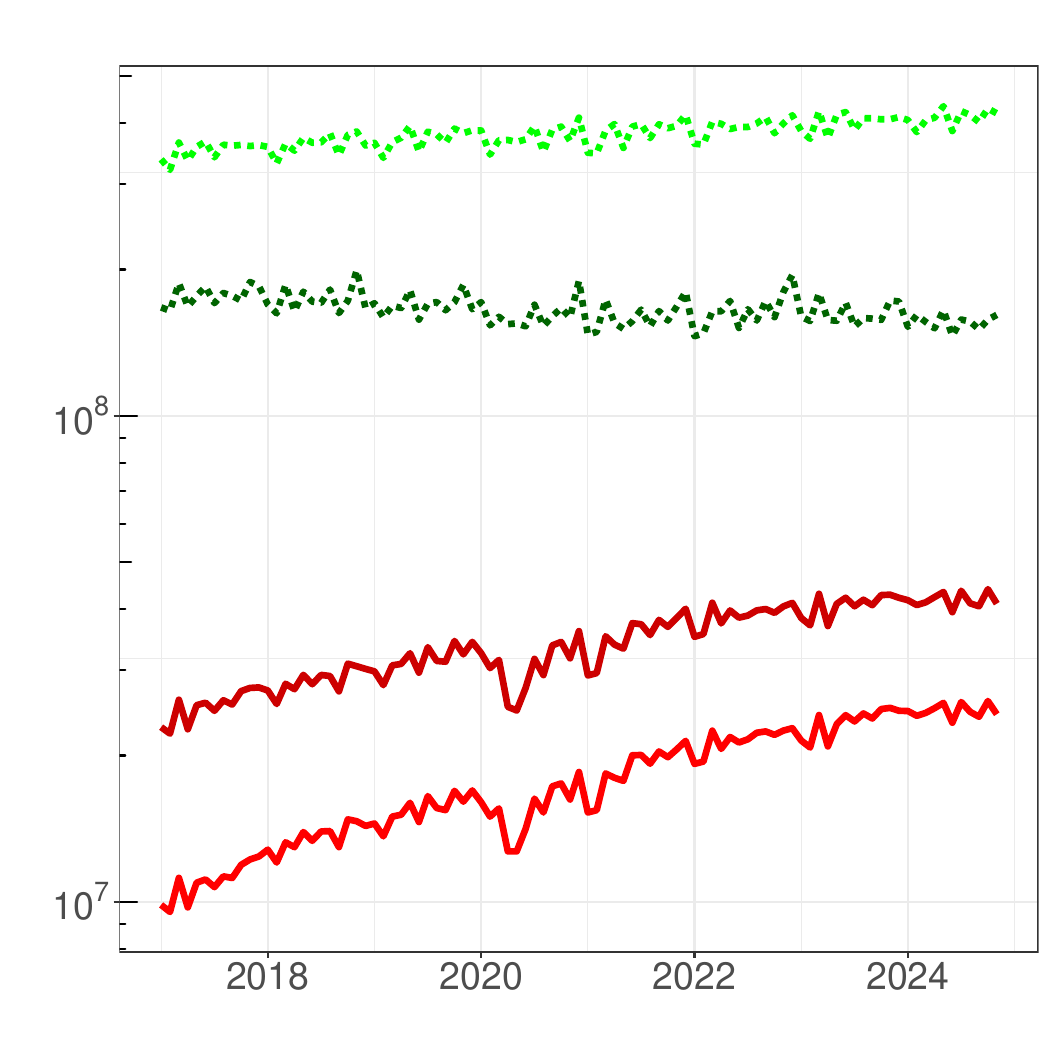}
		\caption{Counts}
		\label{app:fig:ts_bacs_counts}
	\end{subfigure}    
	\begin{subfigure}[b]{0.33\textwidth}
		\centering 
		\includegraphics[width=0.8\textwidth]{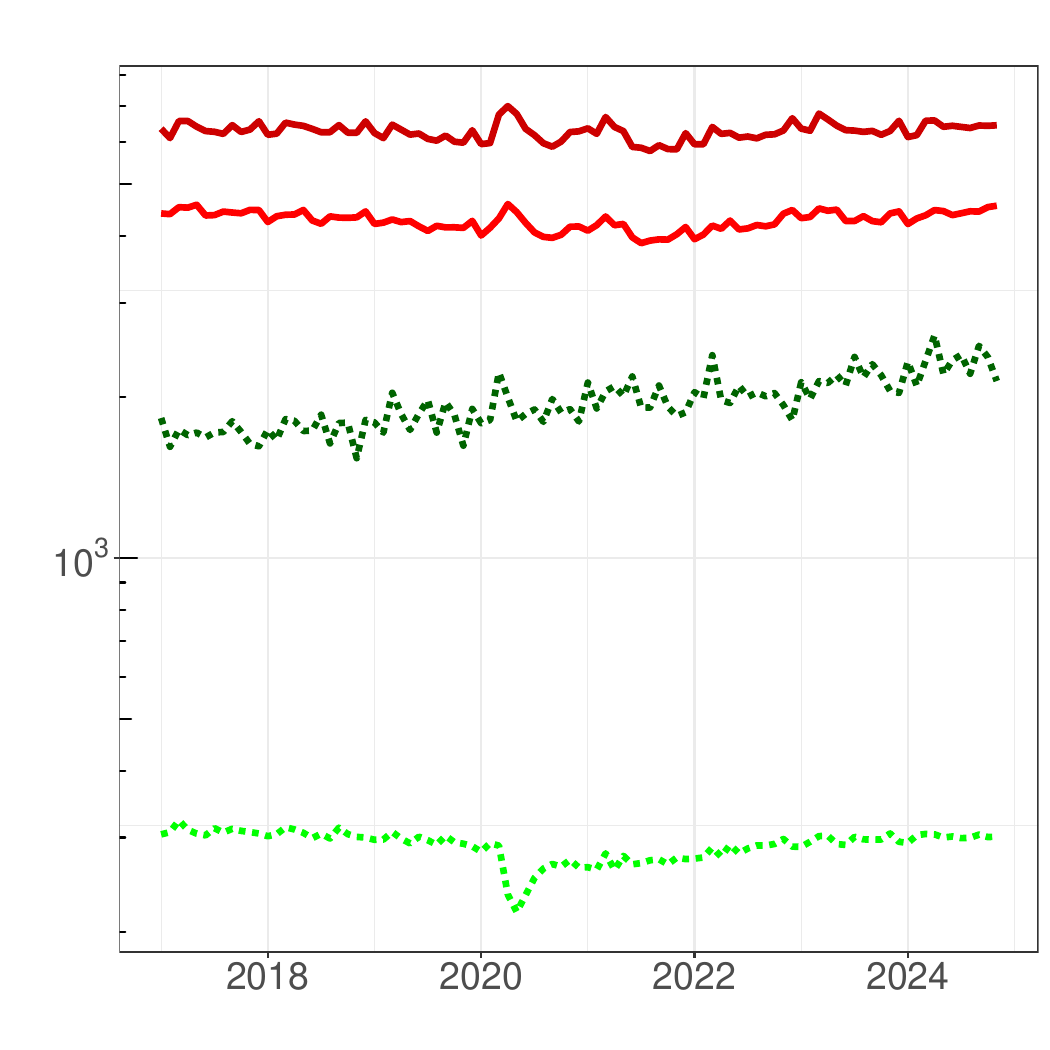}
		\caption{Average}
		\label{app:fig:ts_avg_counts}
	\end{subfigure} 
	
	\justifying \scriptsize \noindent
	Notes: The vertical axis is scaled at a log-10 scale. Payments (red) are monthly aggregates of our data. Bacs Direct Debit and Direct Credit data is downloaded from \citet{payuk2023historicaldata}. 
	
\end{figure}

\FloatBarrier
\subsection{Macroeconomic benchmarking}
\label{app:subsec:macro_benchmarking}
\begin{table}[!h] \centering 
	\caption{OLD DATA: Correlations with other payments and macro aggregates (excluding Covid-19 period)} 
	\label{OLD:tab:macro_benchmarking_3_noCovid} 
	\scriptsize 
	\begin{tabular}{@{\extracolsep{5pt}} ccccccccc} 
		\\[-1.8ex]\hline 
		\hline \\[-1.8ex] 
		& Bacs & FPS & CHAPS & GVA nsa & GVA sa & M1 nsa & M3 nsa & Prices \\ 
		\hline \\[-1.8ex] 
		
		\hline \\[-1.8ex] 
		\multicolumn{8}{l}{\emph{Raw data in levels}}\\
		\hline \\[-1.8ex] 
		Yearly (value) & $0.967$ & $0.962$ & $0.926$ & $0.998$ & $0.998$ & $0.996$ & $0.948$ & $0.999$ \\ 
		Monthly (value) & $0.874$ & $0.926$ & $0.794$ & $0.865$ & $0.915$ & $0.911$ & $0.921$ & $0.898$ \\ 
		Yearly (count) & $0.972$ & $0.884$ & $0.949$ & $0.988$ & $0.996$ & $0.993$ & $0.969$ & $0.990$ \\ 
		Monthly (count) & $0.817$ & $0.800$ & $0.934$ & $0.867$ & $0.854$ & $0.783$ & $0.806$ & $0.825$ \\ 
		Yearly (avg) & $0.948$ & $-0.190$ & $-0.487$ & $0.992$ & $0.978$ & $0.979$ & $0.894$ & $0.991$ \\ 
		Monthly (avg) & $0.696$ & $-0.095$ & $-0.409$ & $0.632$ & $0.858$ & $0.923$ & $0.919$ & $0.786$ \\ 
		
		\hline \\[-1.8ex] 
		\multicolumn{8}{l}{\emph{Growth rates}}\\
		\hline \\[-1.8ex] 
		Yearly (value) & $0.050$ & $-0.223$ & $0.462$ & $0.682$ & $0.066$ & $0.955$ & $0.787$ & $-0.837$ \\ 
		Monthly (value) & $0.882$ & $0.695$ & $0.565$ & $0.720$ & $0.365$ & $0.579$ & $0.630$ & $0.189$ \\ 
		Yearly (count) & $0.047$ & $-0.321$ & $-0.251$ & $0.686$ & $0.071$ & $0.956$ & $0.789$ & $-0.835$ \\ 
		Monthly (count) & $0.619$ & $0.137$ & $0.382$ & $0.785$ & $0.328$ & $0.067$ & $0.138$ & $0.240$ \\ 
		Yearly (avg) & $0.957$ & $0.732$ & $0.340$ & $0.415$ & $-0.261$ & $0.856$ & $0.616$ & $-0.905$ \\ 
		Monthly (avg) & $0.667$ & $0.548$ & $0.138$ & $-0.162$ & $0.132$ & $0.738$ & $0.735$ & $-0.108$ \\ 
		\hline \\[-1.8ex] 
		
	\end{tabular} 
	
	\justifying \noindent \scriptsize
	Notes: 
	This table shows Pearson correlations between annual (monthly) payments and other UK payment schemes and macroeconomic aggregates (GDP, M1, M3, Prices) during 2016 and 2023 (08/2015 and 12/2023), excluding the Covid-19 period, proxied by 2020 to 2022 (03/2020 to 12/2022). `sa' (`nsa') is short for (non-)seasonally adjusted. Our payment data and other payment aggregates are compared by aggregate values, counts, and average values (short `avg') given by value divided by count. 
	Growth rates are calculated as percentage growth compared to the (same month of the) previous year (for monthly data).\footnote{Bacs, FPS, and CHAPS data are obtained from \citet{payuk2023historicaldata}. 
		Monthly GDP is proxied by indicative (non-)seasonally adjusted monthly `Total Gross Value Added' index data published by the ONS \citep{ons2023indicativeGDPdata, ons2023indicativeGDPadjusted}. 
		`Prices' is short for Consumer prices index data obtained from the OECD Key Economic Indicators (KEI) dataset \citep{oecd2023KEIdata}. M1 (M3) are narrow (broad) monetary aggregates, and thus nominal indicators, obtained from the OECD Main Economic Indicators (MEI) dataset \citep{oecd2023MEIdata}.}
	
\end{table} 

\begin{table}[!h] \centering 
	\caption{OLD DATA: Correlations with other payments and macro aggregates (including Covid-19 period)} 
	\label{OLD:app:tab:macro_benchmarking_inclCovid} 
	\scriptsize 
	\begin{tabular}{@{\extracolsep{5pt}} ccccccccc} 
		\\[-1.8ex]\hline 
		\hline \\[-1.8ex] 
		& Bacs & FPS & CHAPS & GDP nsa & GDP sa & M1 nsa & M3 nsa & Prices \\ 
		\hline \\[-1.8ex] 
		
		Share in 2019 & $0.207$ & $0.540$ & $0.013$ & $0.469$ & $0.469$ & $0.578$ & $0.363$ &\\ 
		Share in 2021 & $0.221$ & $0.431$ & $0.013$ & $0.490$ & $0.514$ & $0.471$ & $0.321$ &\\ 
		
		\hline \\[-1.8ex] 
		\multicolumn{8}{l}{\emph{Raw data in levels}}\\
		\hline \\[-1.8ex] 
		Yearly (value) & $0.885$ & $0.964$ & $0.797$ & $0.159$ & $-0.380$ & $0.887$ & $0.916$ & $0.988$ \\ 
		Monthly (value) & $0.824$ & $0.907$ & $0.696$ & $0.394$ & $0.484$ & $0.832$ & $0.868$ & $0.816$ \\ 
		Yearly (count) & $0.941$ & $0.842$ & $0.948$ & $0.733$ & $-0.168$ & $0.629$ & $0.697$ & $0.756$ \\ 
		Monthly (count) & $0.768$ & $0.739$ & $0.928$ & $0.799$ & $0.747$ & $0.636$ & $0.674$ & $0.645$ \\ 
		Yearly (avg) & $0.476$ & $-0.525$ & $-0.024$ & $-0.472$ & $-0.507$ & $0.926$ & $0.904$ & $0.900$ \\ 
		Monthly (avg) & $0.371$ & $-0.439$ & $0.098$ & $-0.326$ & $-0.072$ & $0.752$ & $0.773$ & $0.625$ \\ 
		\hline \\[-1.8ex] 
		\multicolumn{8}{l}{\emph{Growth rates}}\\
		\hline \\[-1.8ex] 
		Yearly (value) & $0.226$ & $-0.197$ & $0.288$ & $0.352$ & $0.160$ & $0.196$ & $0.354$ & $-0.526$ \\ 
		Monthly (value) & $0.773$ & $0.613$ & $0.008$ & $0.709$ & $0.589$ & $-0.212$ & $-0.111$ & $0.010$ \\ 
		Yearly (count) & $0.429$ & $-0.361$ & $0.066$ & $0.391$ & $0.207$ & $0.154$ & $0.324$ & $-0.480$ \\ 
		Monthly (count) & $0.567$ & $0.526$ & $0.751$ & $0.893$ & $0.863$ & $-0.165$ & $-0.124$ & $0.199$ \\ 
		Yearly (avg) & $-0.626$ & $-0.737$ & $0.582$ & $-0.947$ & $-0.656$ & $0.693$ & $0.533$ & $-0.771$ \\ 
		Monthly (avg) & $-0.131$ & $-0.445$ & $0.590$ & $-0.813$ & $-0.823$ & $0.121$ & $0.141$ & $-0.343$ \\ 
		\hline \\[-1.8ex] 
		
	\end{tabular} 
	
	\justifying \noindent \scriptsize
	Notes: 
	This table shows Pearson correlations between annual (monthly) payments and other UK payment schemes and macroeconomic aggregates (GDP, M1, M3, Prices) during 2016 and 2023 (08/2015 and 12/2023), including the Covid-19 period. `sa' (`nsa') is short for (non-)seasonally adjusted. Our payment data and other payment aggregates are compared by aggregate values, counts, and average values (short `avg') given by value divided by count. 
	Growth rates are calculated as percentage growth compared to the (same month of the) previous year (for monthly data). Bacs, FPS, and CHAPS data are obtained from \citet{payuk2023historicaldata}. 
	Monthly GDP is proxied by indicative (non-)seasonally adjusted monthly `Total Gross Value Added' index data published by the ONS \citep{ons2023indicativeGDPdata, ons2023indicativeGDPadjusted}. 
	`Prices' is short for Consumer prices index data obtained from the OECD Key Economic Indicators (KEI) dataset \citep{oecd2023KEIdata}. M1 (M3) are narrow (broad) monetary aggregates, and thus nominal indicators, obtained from the OECD Main Economic Indicators (MEI) dataset \citep{oecd2023MEIdata}.
\end{table}

\begin{figure}[!h]
	\centering
	\begin{subfigure}[b]{0.9\textwidth}
		\includegraphics[width=\textwidth]{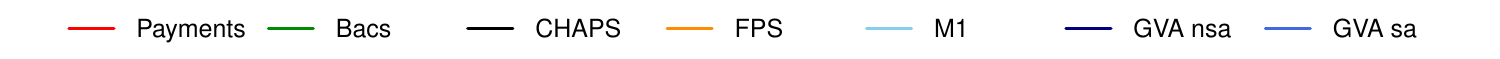}
	\end{subfigure}
	
	\begin{subfigure}[b]{0.32\textwidth}
		\centering
		\includegraphics[width=\textwidth]{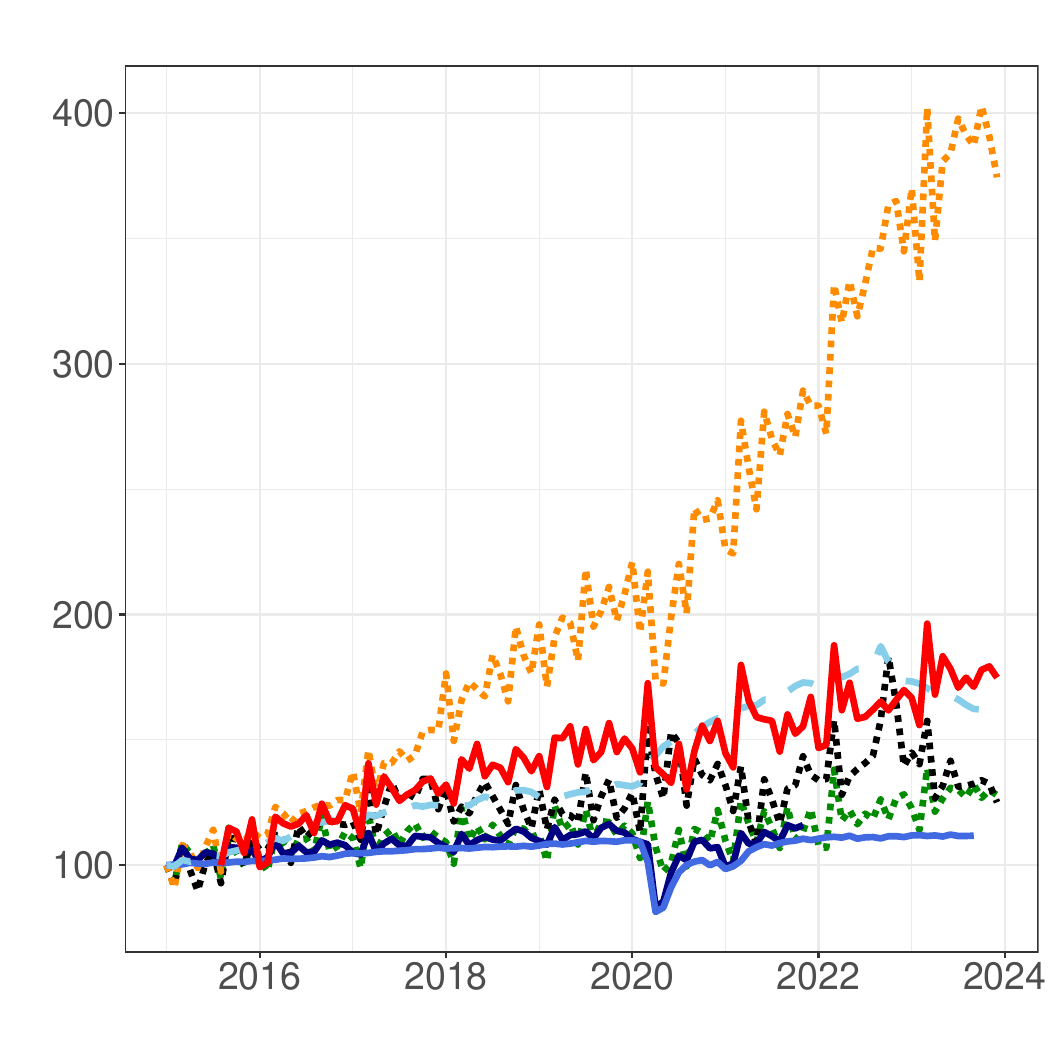}
		\caption{Value}
	\end{subfigure}
	\begin{subfigure}[b]{0.32\textwidth}
		\centering
		\includegraphics[width=\textwidth]{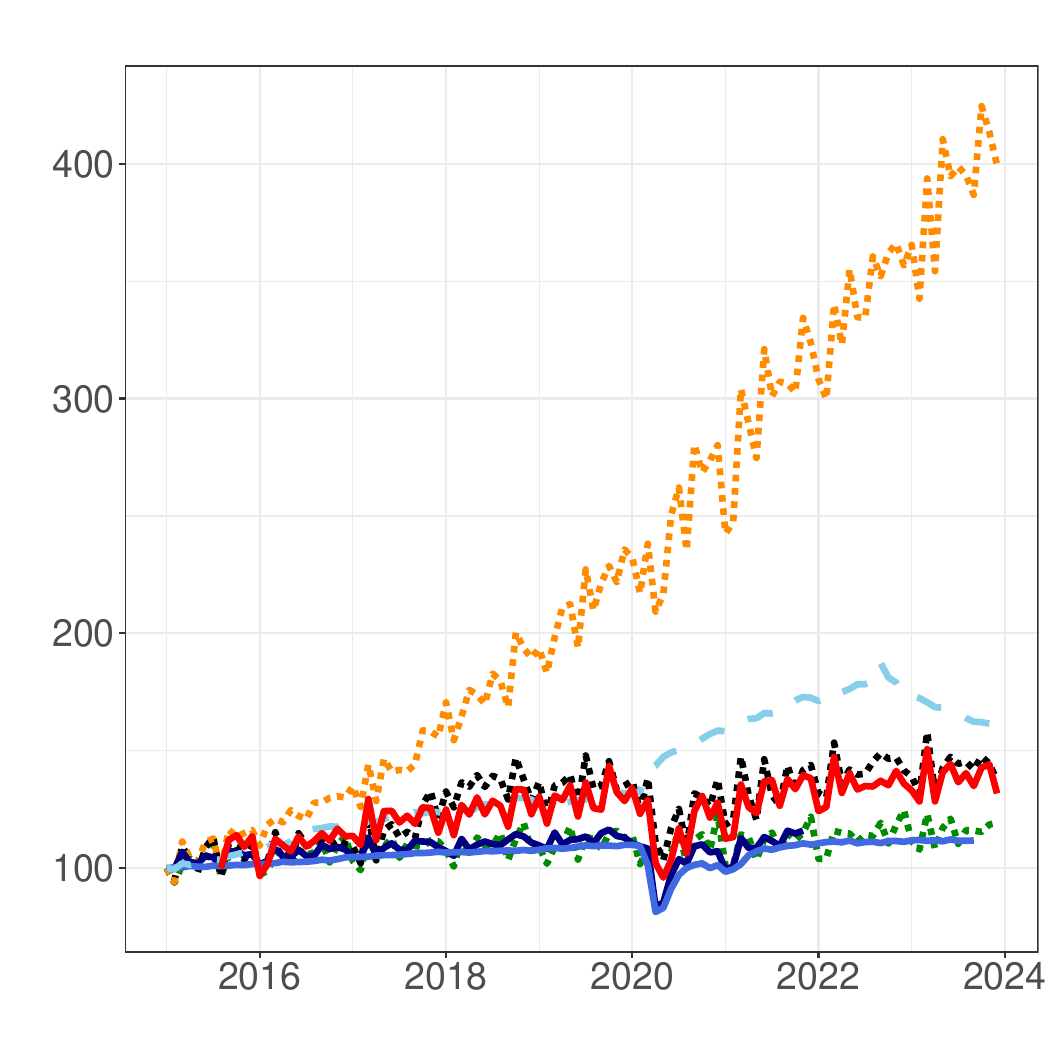}
		\caption{Counts}
	\end{subfigure}
	\begin{subfigure}[b]{0.32\textwidth}
		\centering
		\includegraphics[width=\textwidth]{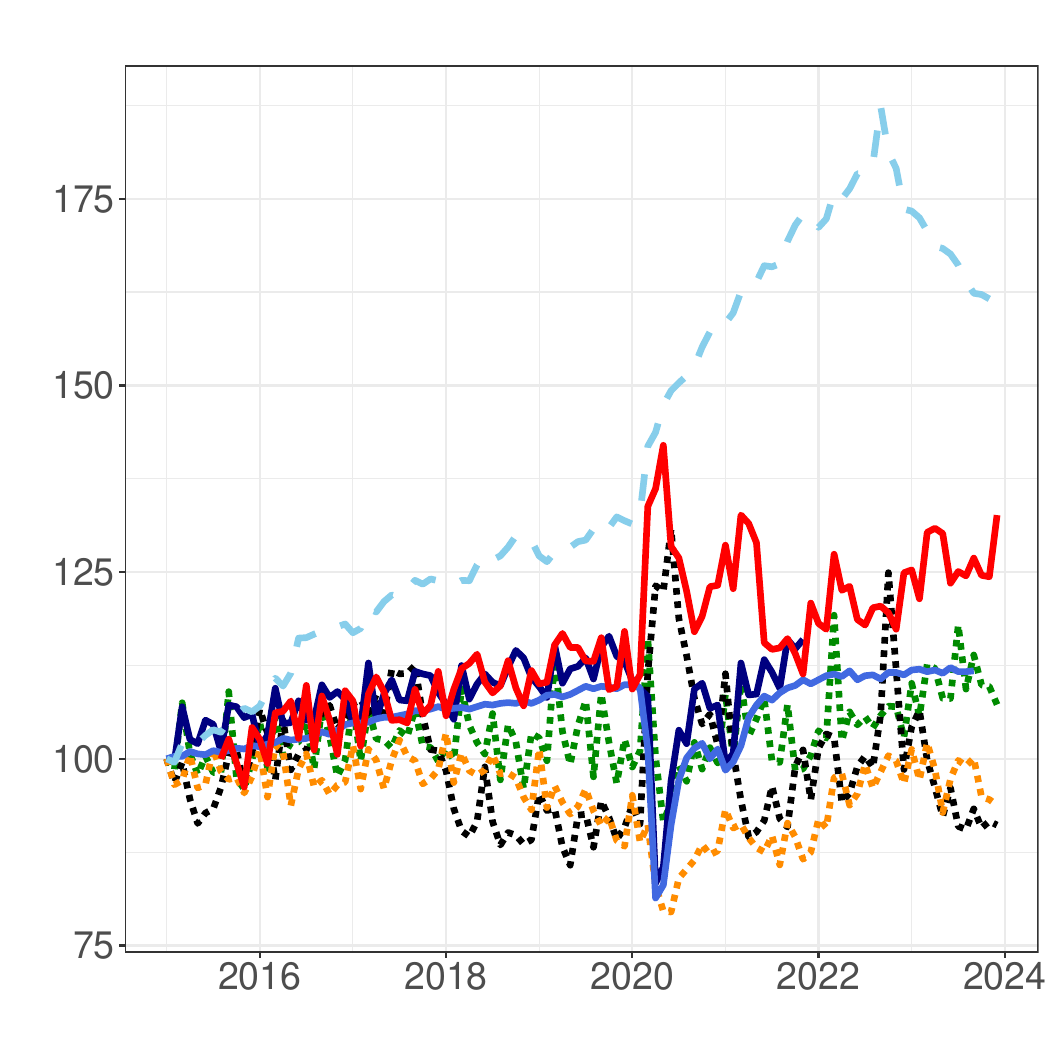}
		\caption{Average value}
	\end{subfigure}
	\caption{OLD DATA: Monthly UK payments, GDP and M1}
	\label{OLD:fig:timeseries_aggr_index_payments_vs_gdp}

	\justifying \scriptsize \noindent
	Notes: These figures show monthly time series (indexed to 08/2015 = 100) for payments, the major UK payment schemes, and indicative (non-)seasonally adjusted monthly 'Total Gross Value Added' (GVA) data published by the ONS \citep{ons2023indicativeGDPadjusted, ons2023indicativeGDPdata}. Average values are obtained by dividing total values by counts.
	
\end{figure}

\begin{figure}[h!]
	\centering
	\begin{subfigure}[b]{0.9\textwidth}
		\includegraphics[width=\textwidth]{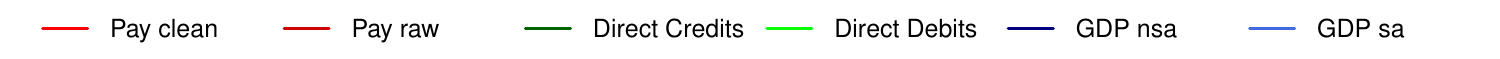}
	\end{subfigure}
	\begin{subfigure}[b]{0.32\textwidth}
		\centering
		\includegraphics[width=\textwidth]{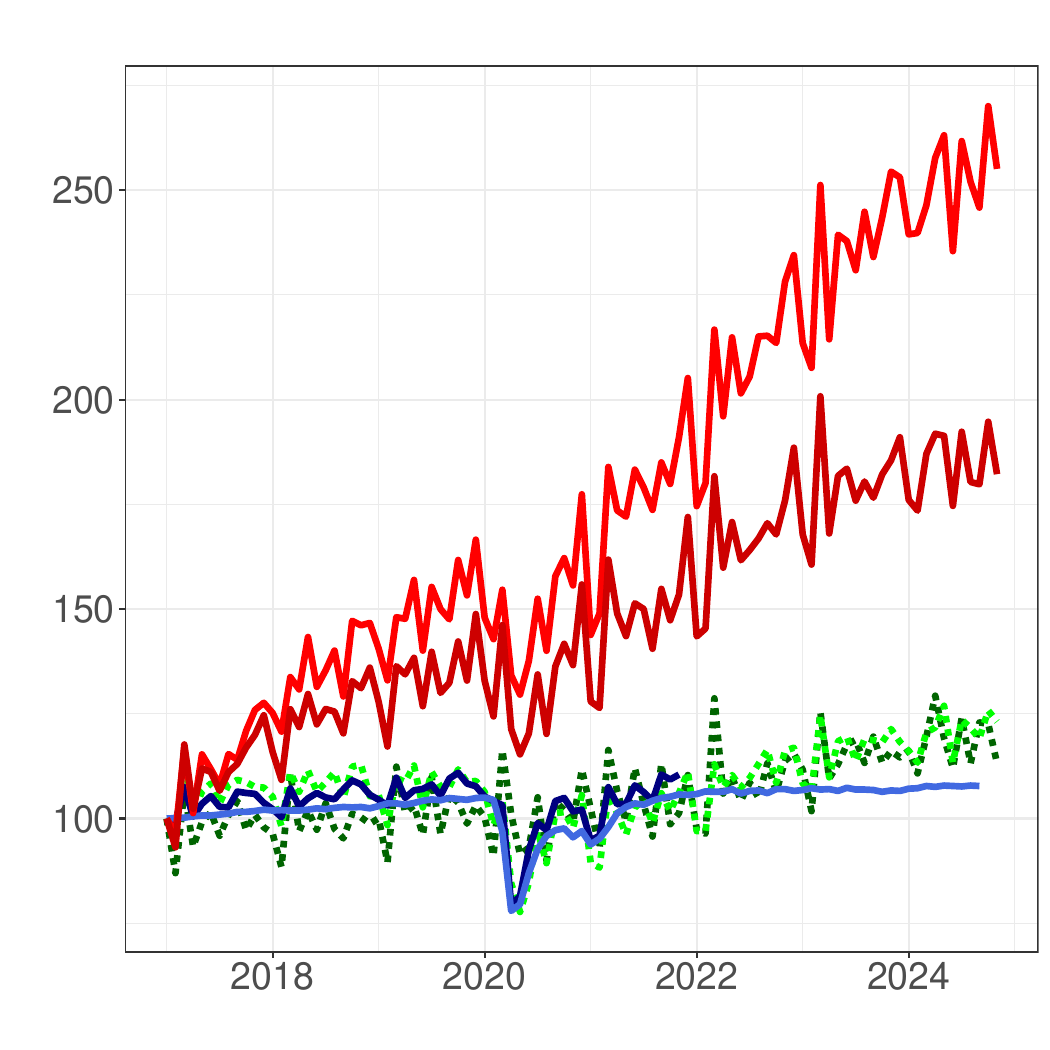}
		\caption{Value}
	\end{subfigure}
	\begin{subfigure}[b]{0.32\textwidth}
		\centering
		\includegraphics[width=\textwidth]{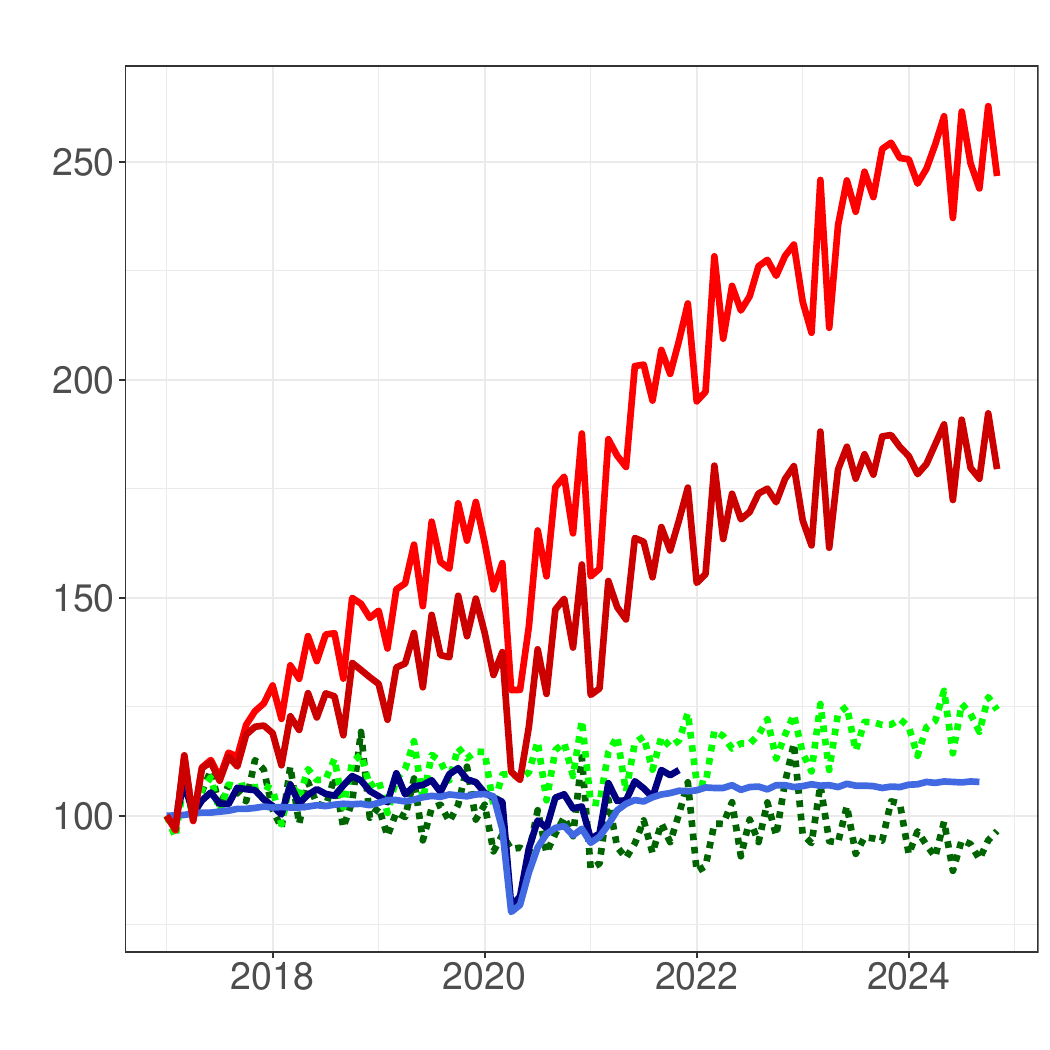}
		\caption{Counts}
	\end{subfigure}
	\begin{subfigure}[b]{0.32\textwidth}
		\centering
		\includegraphics[width=\textwidth]{inputs_Jan2025_data/aggr/Pay_value_vs_Bacs_DD_DC_month_idx2019=100.pdf}
		\caption{Average value}
	\end{subfigure}
	
	\caption{Monthly payments, Direct Debits and Credits}
	\label{app:fig:timeseries_aggr_index_payments_vs_bacs_instrumetns}
	
	\justifying \scriptsize
	\noindent
	Notes: These figures show monthly time series of the payment data, and Bacs transactions disaggregated by Direct Debits and Credits, indexed by 2019 = 100. The average value is calculated by dividing values by counts. The dark blue dashed line shows monthly deseasonalized GVA as a benchmark.

\end{figure}

\subsection{Comparison to national accounts}
\label{app:subsec:comparison_NA}
\subsubsection{Aggregate network statistics}
\begin{table}[!htp] \centering 
	\caption{OLD DATA: Properties of the payment and ONS input-output networks in 2019} 
	\label{OLD:tab:network_stats_2019_no_truncation} 
	\footnotesize 
	\begin{tabular}{@{\extracolsep{8pt}} lccccc} 
		\\[-1.8ex]\hline 
		\hline \\[-1.8ex] 
		& Value & Count & PxP & SUT & IxI \\ 
		\hline \\[-1.8ex]
		\underline{\emph{Raw transactions} }\\ 
		Density & $0.286$ & $0.286$ & $0.723$ & $0.474$ & $0.980$ \\ 
		Average degree & $28.550$ & $28.550$ & $75.202$ & $49.260$ & $101.885$ \\ 
		Average strength & $2,783.139$ & $239,906.400$ & $10,563.500$ & $12,741.830$ & $10,593.480$ \\ 
		Average weight & $97.483$ & $8,403.027$ & $140.468$ & $258.667$ & $103.975$ \\ 
		Reciprocity & $0.554$ & $0.554$ & $0.793$ & $0.534$ & $0.989$ \\ 
		Transitivity & $0.648$ & $0.648$ & $0.921$ & $0.787$ & $1$ \\ 
		Assortativity by degree & $-0.358$ & $-0.358$ & $-0.176$ & $-0.190$ & $-0.005$ \\ 
		\hline \\[-1.8ex] \underline{\emph{Input shares}} \\  
		Average strength & $0.885$ & $0.940$ & $0.840$ & $0.741$ & $0.846$ \\ 
		Average weight & $0.031$ & $0.033$ & $0.011$ & $0.015$ & $0.008$ \\ 
		
		\hline \\[-1.8ex]  \underline{\emph{Output shares}} \\ 
		
		Average strength & $0.839$ & $0.864$ & $0.812$ & $0.731$ & $0.828$ \\ 
		Average weight & $0.029$ & $0.030$ & $0.011$ & $0.015$ & $0.008$ \\ 
		\hline \\[-1.8ex] 
	\end{tabular} 
	
	\vspace{0.25cm}
	
	\justifying \noindent \scriptsize
	Notes: The first (second) column uses payment values (counts) as weights. The other columns represent official IOTs published by the ONS, where PxP is short for Product-by-Product, IxI for Industry-by-Industry, and SUT for Supply-and-Use Table. The data are aggregated into 105 distinct CPA codes \citep[see][Sec. 4.1]{hotte2025national}.
	Raw transaction data are shown in £ million. 
	
\end{table}

\begin{table}[!htbp] \centering 
	\caption{Properties of the payment and ONS-based input-output networks in 2019} 
	\label{tab:network_stats_2019_no_truncation} 
	\footnotesize 
	\begin{tabular}{@{\extracolsep{8pt}} lccccc} 
		\\[-1.8ex]\hline 
		\hline \\[-1.8ex] 
		Variable & Value & Count & SUT & PxP & IxI \\ 
		\hline \\[-1.8ex]
		\underline{\emph{Raw transactions} }\\ 
		Density & $0.534$ & $0.383$ & $0.475$ & $0.727$ & $0.979$ \\ 
		Average degree & $55.538$ & $39.476$ & $48.471$ & $74.157$ & $99.882$ \\ 
		Average strength & $6,529.642$ & $1,687,796.000$ & $12,526.810$ & $10,495.810$ & $10,500.010$ \\ 
		Average weight & $117.570$ & $42,755.290$ & $258.442$ & $141.535$ & $105.124$ \\ 
		Reciprocity & $0.783$ & $0.749$ & $0.534$ & $0.799$ & $0.989$ \\ 
		Transitivity & $0.747$ & $0.671$ & $0.750$ & $0.861$ & $0.991$ \\ 
		Assortativity by degree & $-0.338$ & $-0.436$ & $-0.189$ & $-0.180$ & $-0.005$ \\ 
		\hline \\[-1.8ex] \underline{\emph{Input shares}} \\  
		Average strength & $0.890$ & $0.956$ & $0.736$ & $0.837$ & $0.843$ \\ 
		Average weight & $0.016$ & $0.024$ & $0.015$ & $0.011$ & $0.008$ \\ 
		\hline \\[-1.8ex]  \underline{\emph{Output shares}} \\
		Average strength & $0.893$ & $0.941$ & $0.727$ & $0.808$ & $0.824$ \\ 
		Average weight & $0.016$ & $0.024$ & $0.015$ & $0.011$ & $0.008$ \\ 
		\hline \\[-1.8ex] 
	\end{tabular} 
	\vspace{0.25cm}
	
	\justifying \noindent \scriptsize
	Notes: The first (second) column uses payment values (counts) as weights. The other columns represent official IOTs published by the ONS, where PxP is short for Product-by-Product, IxI for Industry-by-Industry, and SUT for Supply-and-Use Table. The data are aggregated into 104 distinct CPA codes \citep[see][Sec. 4.1]{hotte2025national}.
	Raw transaction data are shown in £ million. 
	
\end{table}

\begin{figure}[h]
	\centering
	
	\caption{OLD DATA: Network density at different truncation thresholds}
	\label{OLD:fig:network_density_plot_3digit}
	
	\begin{subfigure}[b]{\textwidth}
		\centering
		\includegraphics[width=\textwidth]{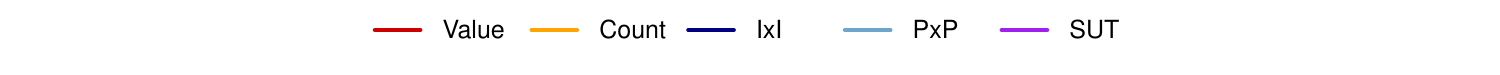}
	\end{subfigure}
	
	\begin{subfigure}[b]{0.44\textwidth}
		\centering
		\includegraphics[width=\textwidth]{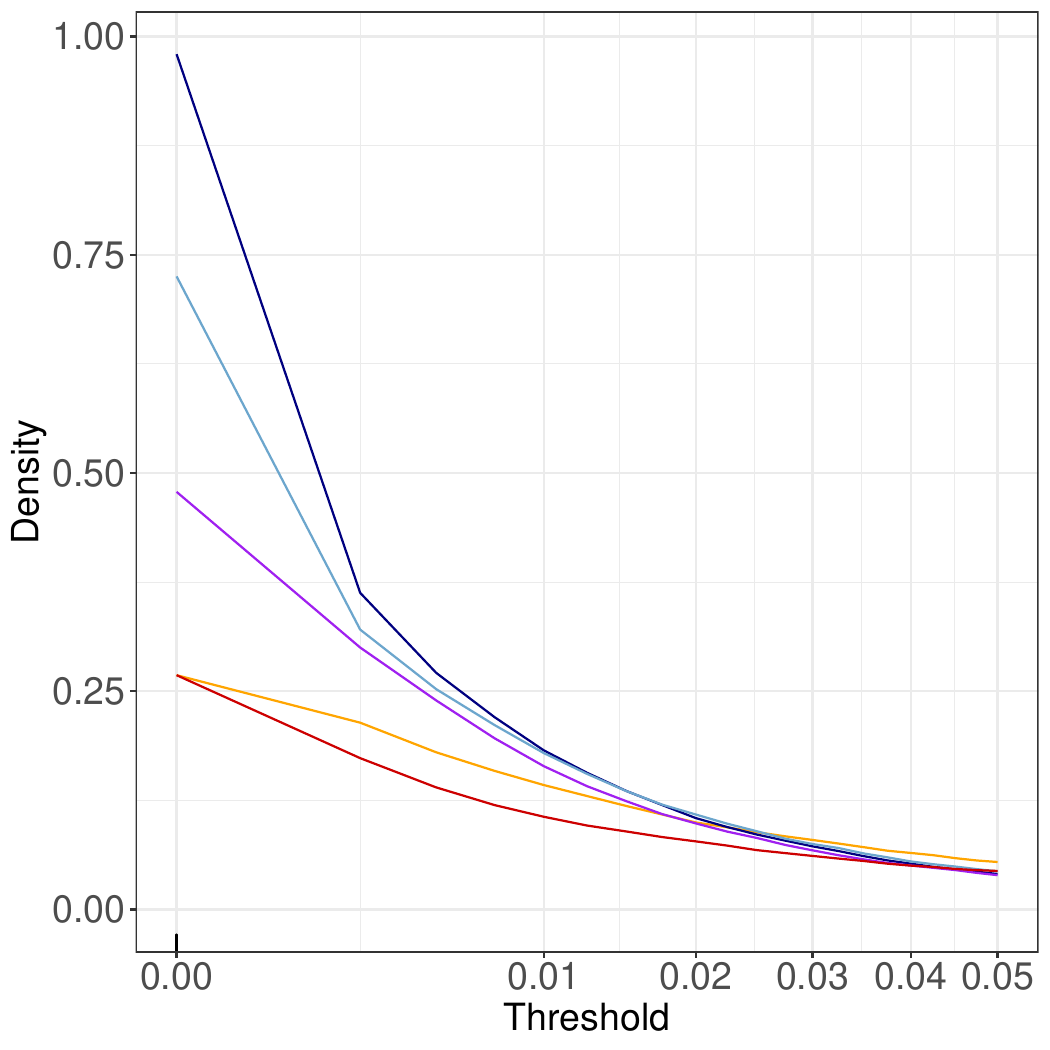}
		\caption{Input network}
	\end{subfigure}
	\begin{subfigure}[b]{0.44\textwidth}
		\centering
		\includegraphics[width=\textwidth]{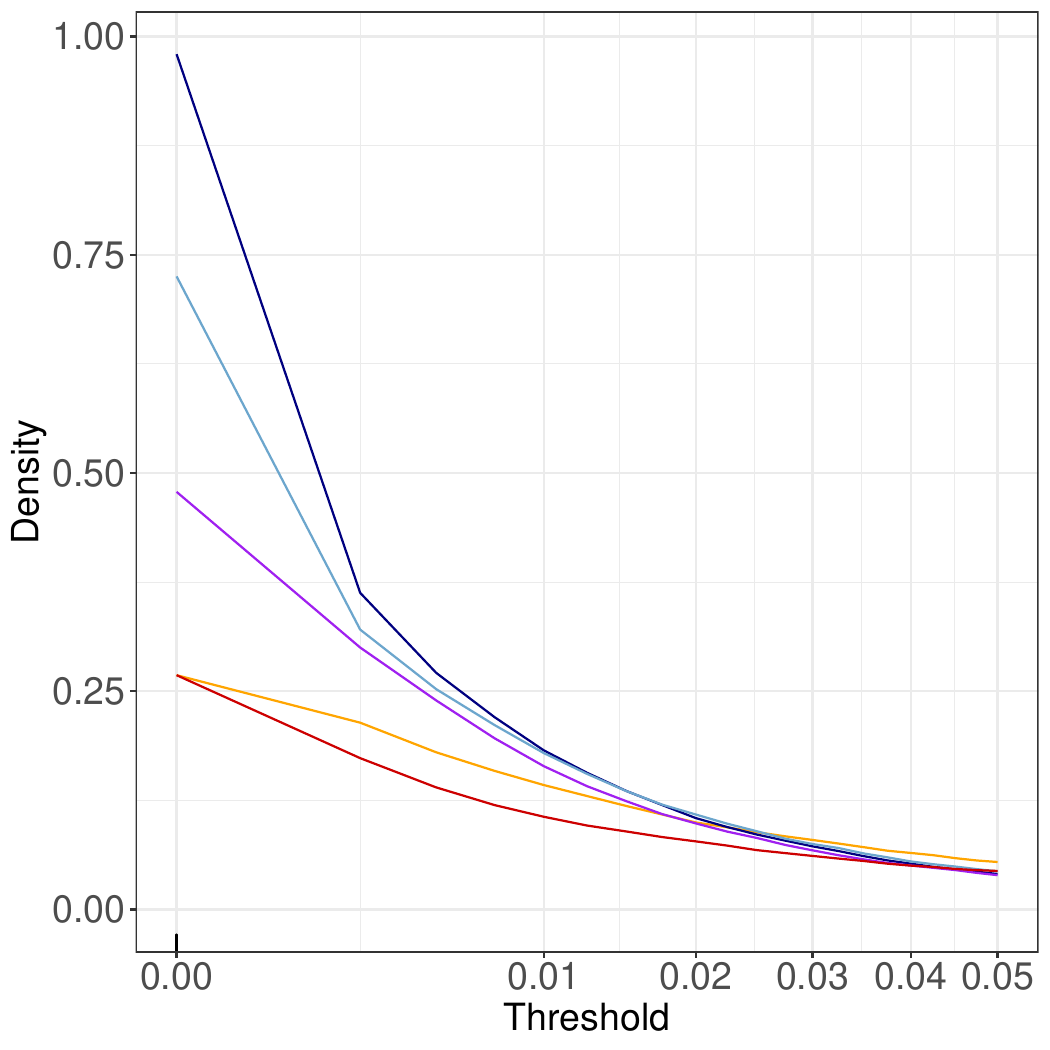}
		\caption{Output network}
	\end{subfigure}

	\justifying \scriptsize \noindent
	Notes: This figure shows the effect of network truncation thresholds (x-axis) on the network density (y-axix). In the left (right) figure, a link is removed if the input (output) share is smaller than the threshold value. 
	
\end{figure}

\FloatBarrier
\subsubsection{Edge-level correlations}
\label{app:subsec:auto_cross_correlations}
\begin{figure}[!h]
	
	\caption{OLD DATA: Auto- \& cross-correlations of input and output shares (2018-2019)}        
	\label{OLD:fig:correlations_edge_level_ONS_payments_IO_shares}
	
	\centering
	\begin{subfigure}[b]{0.49\textwidth}
		\includegraphics[width=\textwidth]{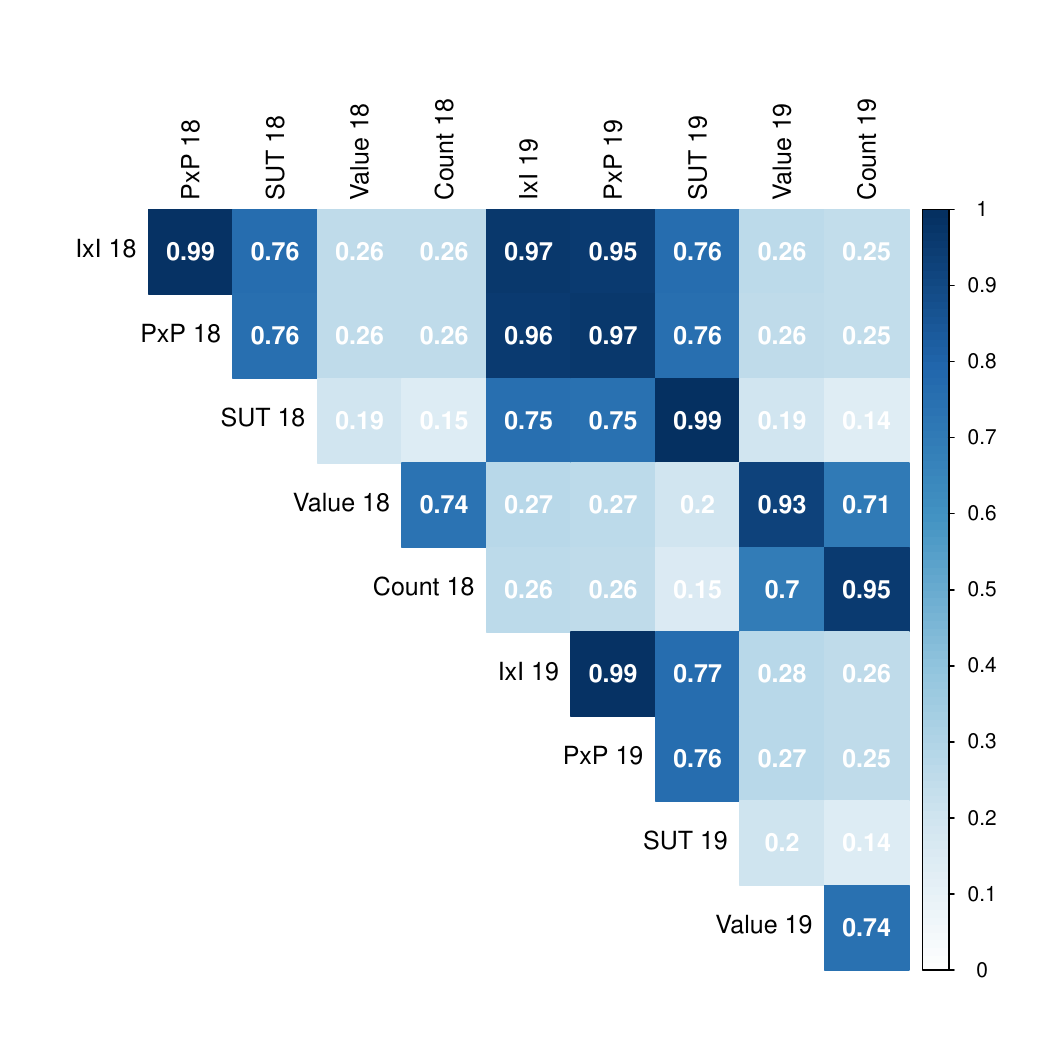}
		\caption{Input shares}
		\label{old:subfig:correlations_edge_level_ONS_payments_input_shares}
	\end{subfigure}
	\begin{subfigure}[b]{0.49\textwidth}
		\includegraphics[width=\textwidth]{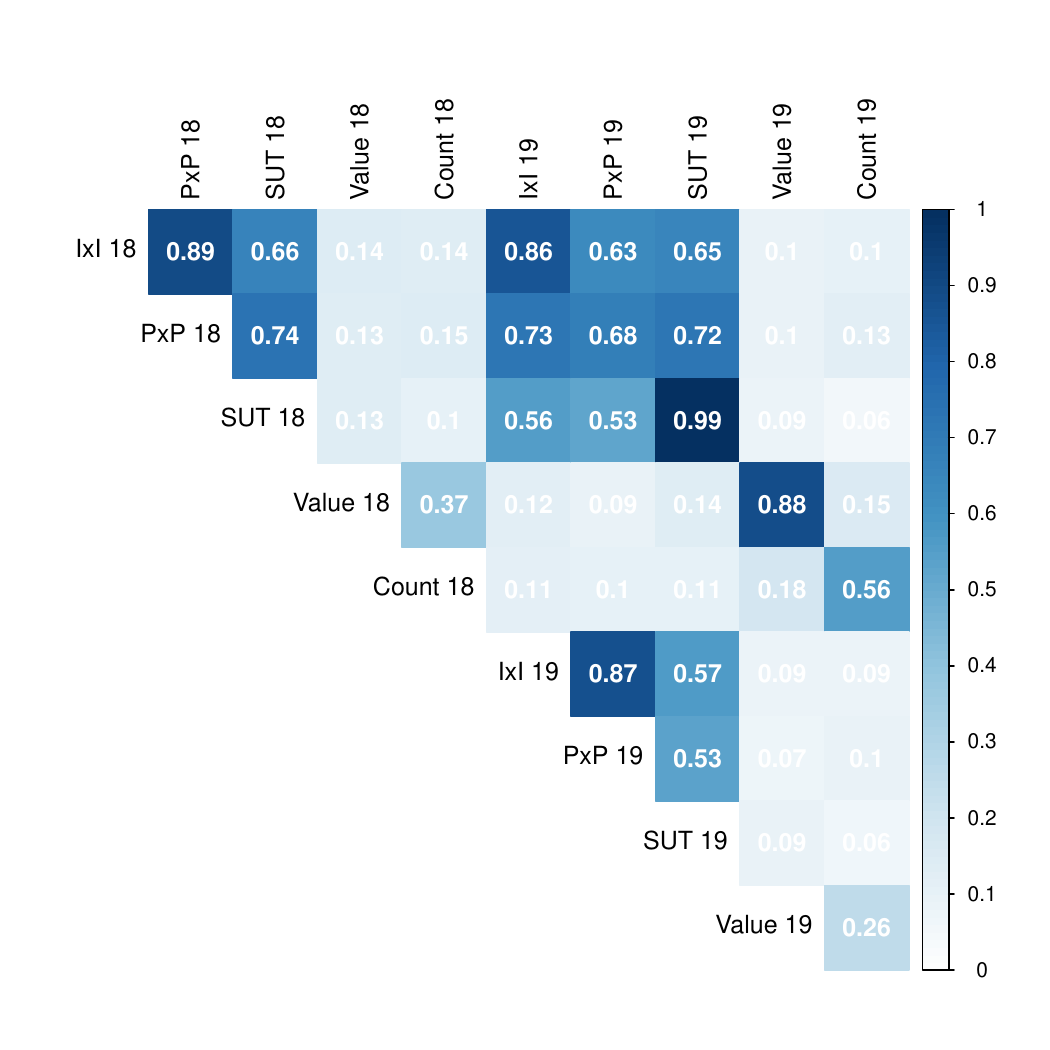}
		\caption{Output shares}
		\label{old:subfig:correlations_edge_level_ONS_payments_output_shares}
	\end{subfigure}
	
	\justifying \scriptsize \noindent
	Notes: The correlations are measured by the Pearson correlation coefficient between input and output shares in the payment-based IOTs (values and counts) and the IxI, PxP, and SUTs. 
	
\end{figure}

\begin{figure}[!h]
	
	\caption{Auto- \& cross-correlations at the edge level (2018-2019)}        
	\label{fig:correlations_edge_level_ONS_payments_IO_shares}
	
	\centering
	
	\begin{subfigure}[b]{0.49\textwidth}
		\includegraphics[width=\textwidth]{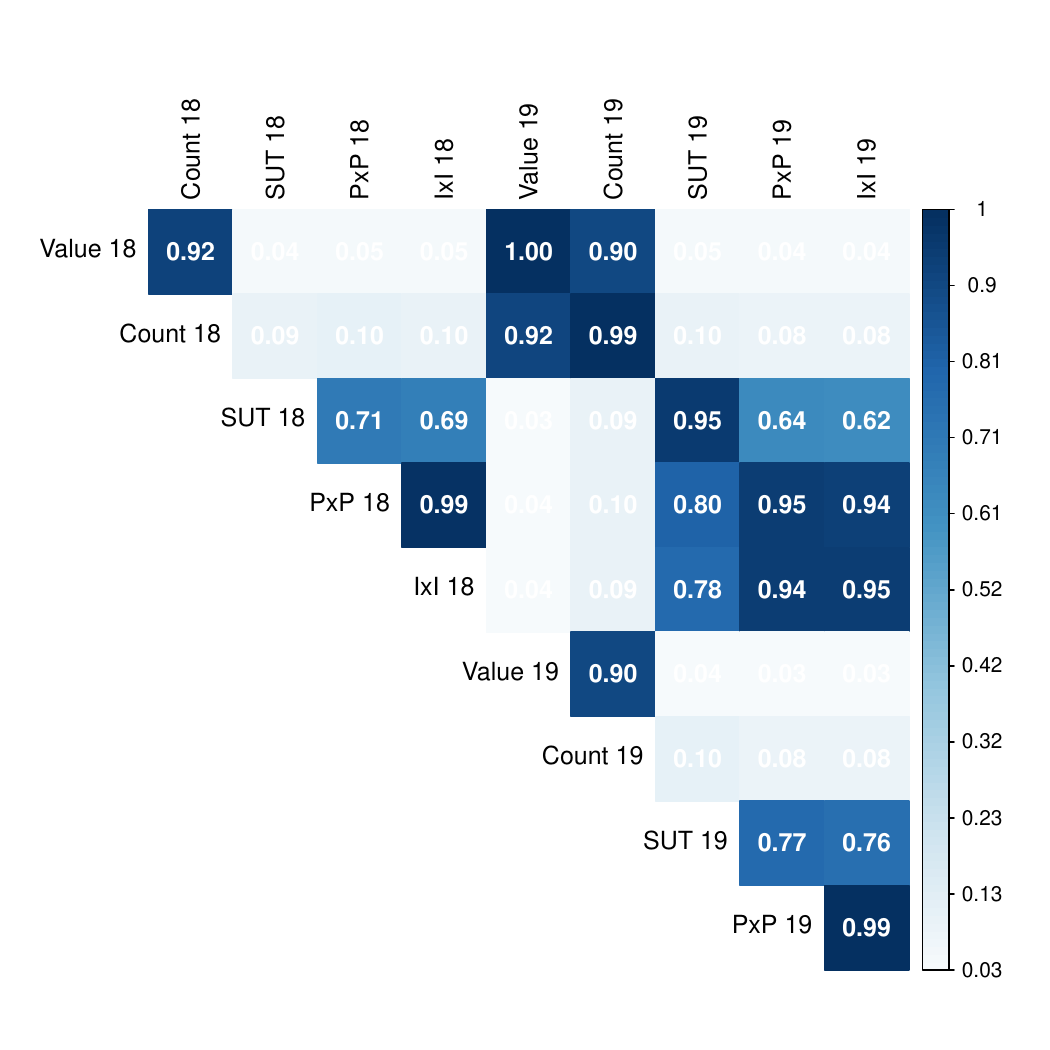}
		\caption{Input shares}
		\label{subfig:correlations_edge_level_ONS_payments_input_shares_1819}
	\end{subfigure}    
	\begin{subfigure}[b]{0.49\textwidth}
		\includegraphics[width=\textwidth]{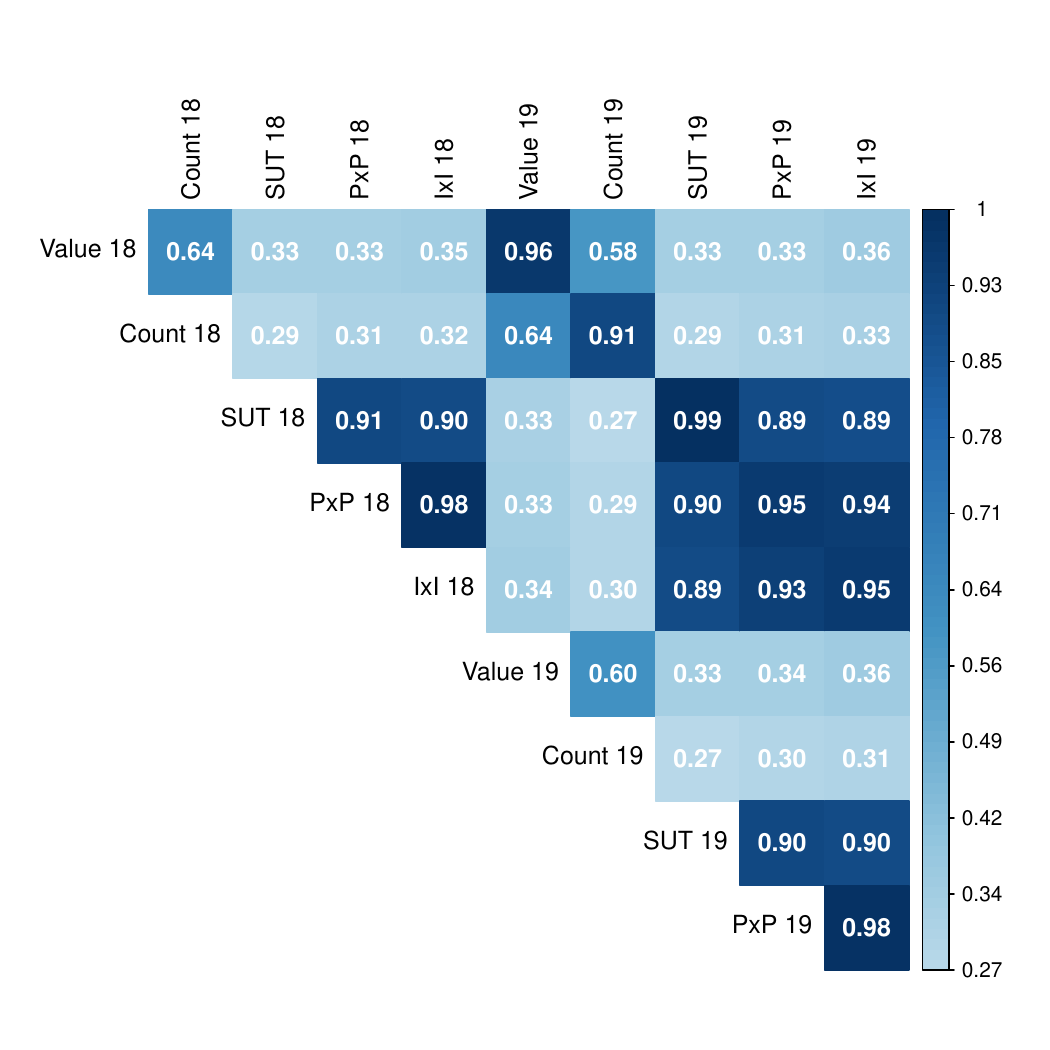}
		\caption{Output shares}
		\label{subfig:correlations_edge_level_ONS_payments_output_shares_1819}
	\end{subfigure}
	
	\justifying \scriptsize \noindent
	Notes: The correlations are measured by the Pearson correlation coefficient between raw transactions, input and output shares in the payment-based IOTs (values and counts) and the IxI, PxP, and SUTs. 
	
\end{figure}

\FloatBarrier
\subsubsection{Industry-level similarities}

\begin{figure}
	\centering
	
	\caption{OLD DATA: Auto- and cross-correlations of inputs \& outputs}
	\label{app:fig:correlations_industry_level_input_output_raw}
	
	\includegraphics[width=\textwidth]{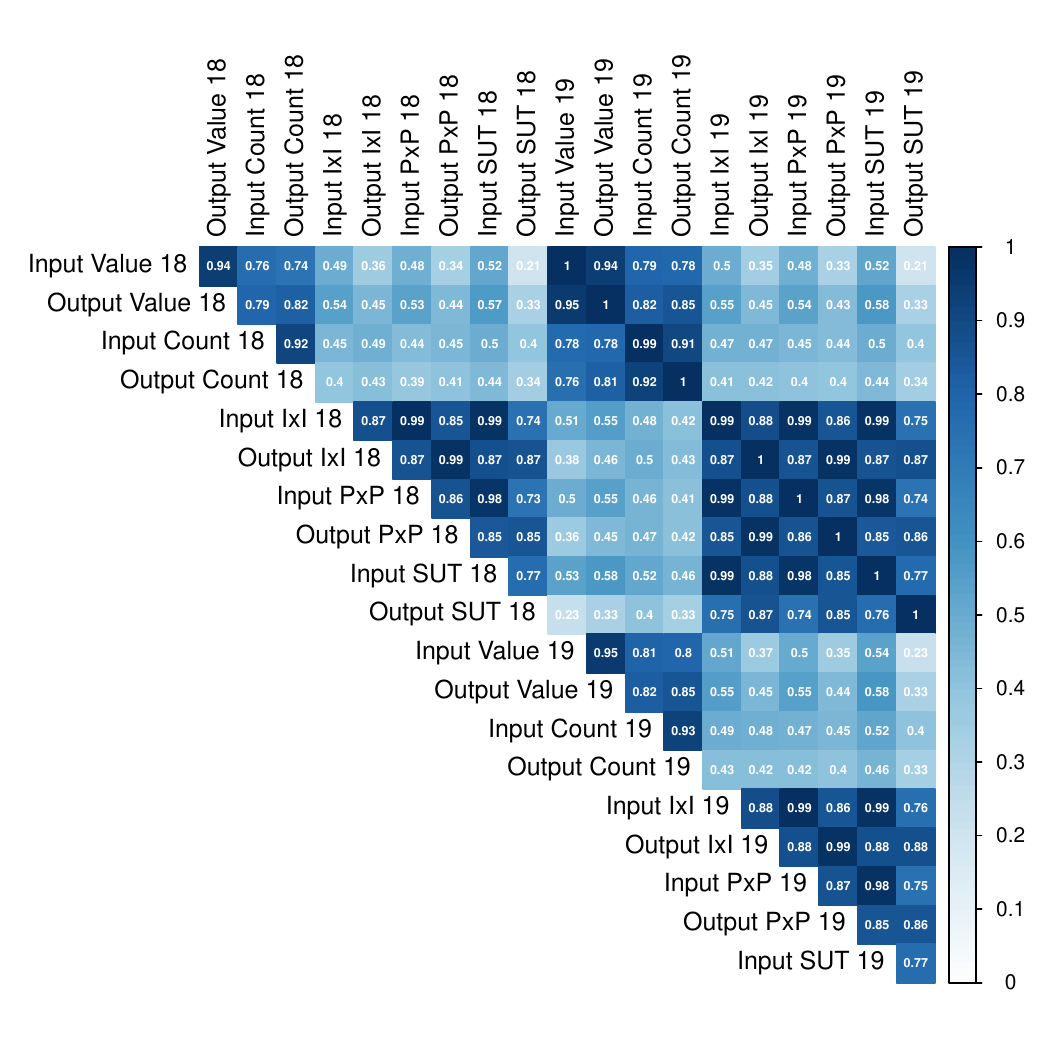}
	
	\justifying \scriptsize \noindent	
	Notes: The correlations are measured by the Pearson correlation coefficient between industry-level annual outputs and inputs in 2018-19 calculated by using raw transaction values and counts of the payment data and the row- and column sums of ONS IxI, PxP, and SUTs. 
\end{figure}

\begin{figure}[]
	\centering
	
	\caption{Auto- and cross-correlations of inputs \& outputs growth rates}
	\label{fig:correlations_industry_level_input_output_growth}
	
	\includegraphics[width=0.8\textwidth]{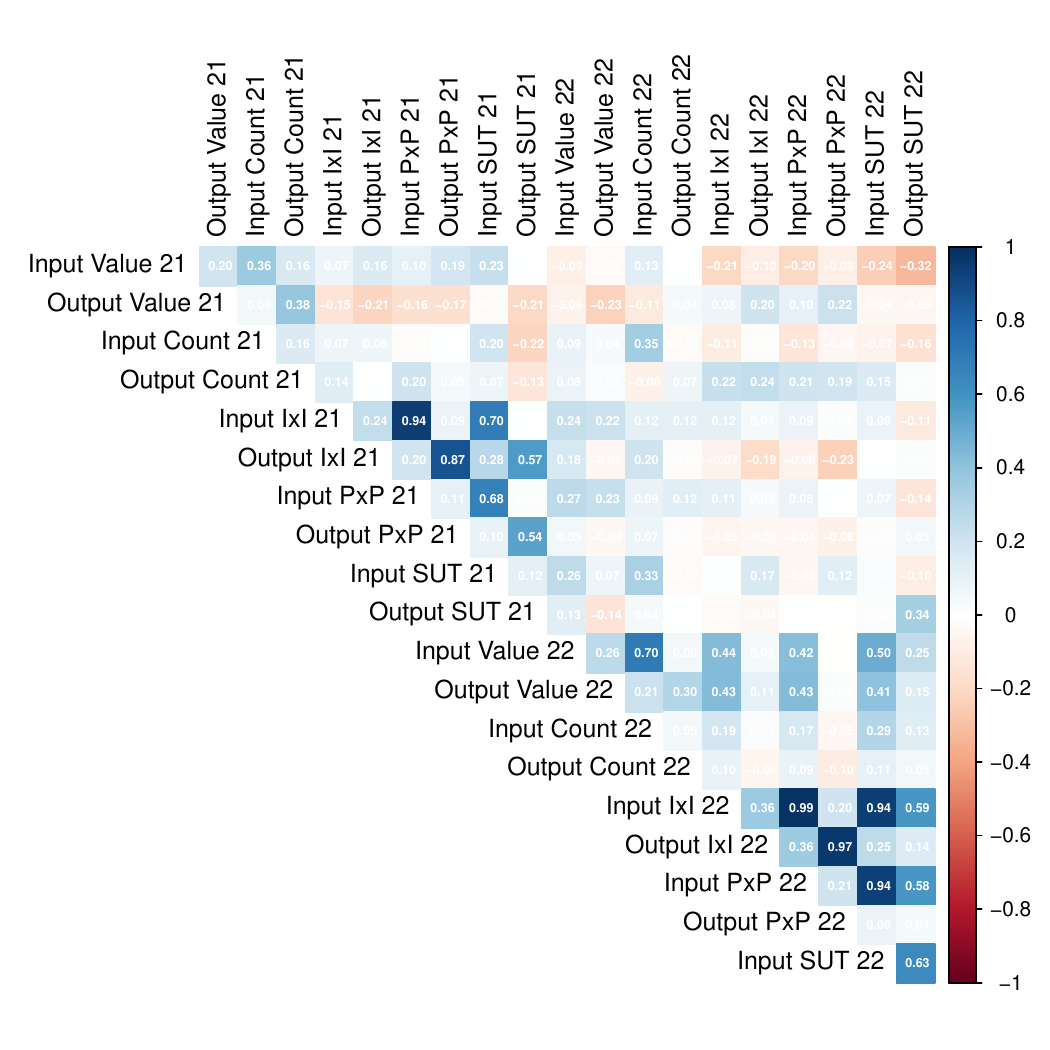}
	
	\justifying \scriptsize \noindent	
	Notes: The correlations are measured by the Pearson correlation coefficient between industry-level annual outputs and inputs in 2021-22 calculated by using raw transaction values and counts of the payment data and the row- and column sums of ONS IxI, PxP, and SUTs. 
\end{figure}

\begin{figure}
	
	\caption{OLD DATA: Auto- and cross-correlations of input \& output growth}
	\label{app:fig:correlations_industry_level_input_output_raw_growth}
	
	\centering
	\includegraphics[width=\textwidth]{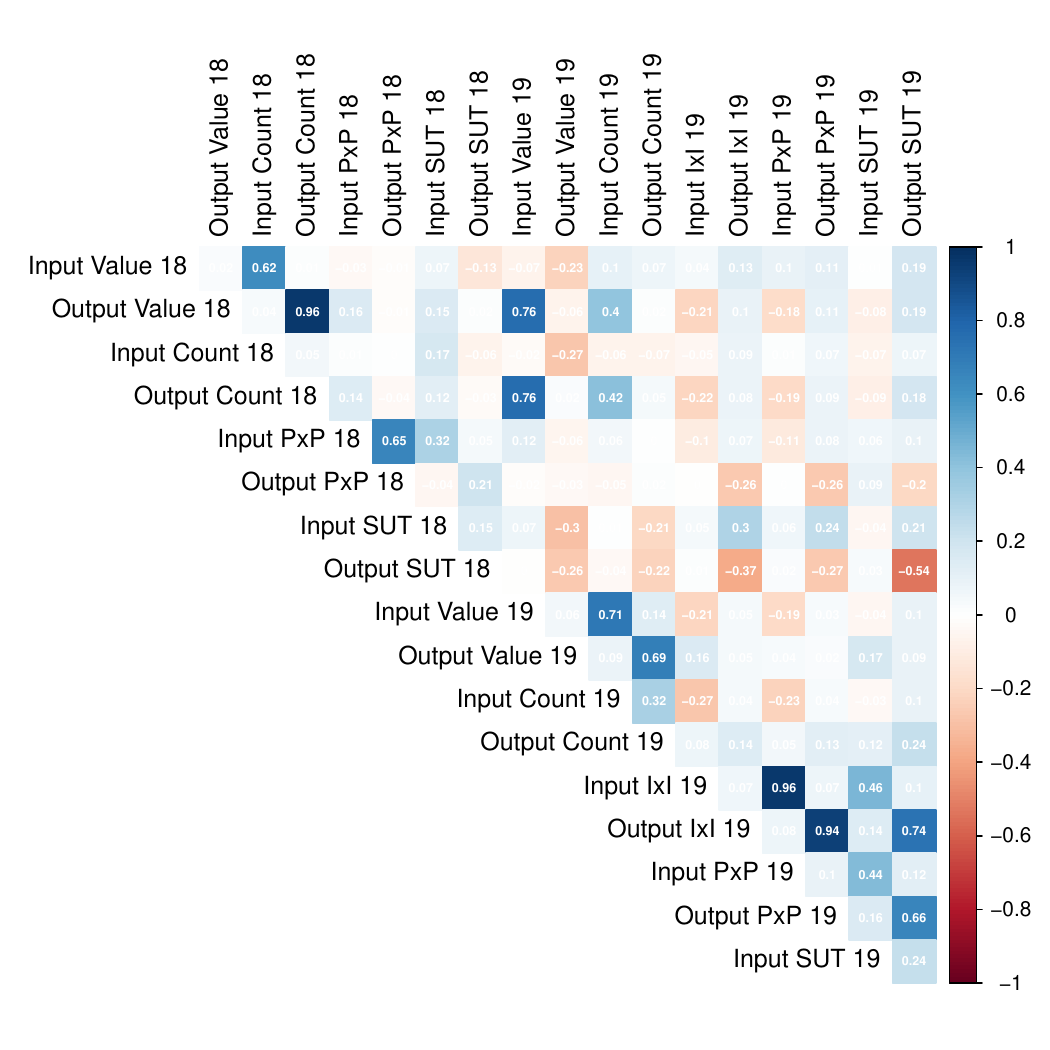}

	\justifying \scriptsize \noindent
	Notes: The correlations are measured by the Pearson correlation coefficient between industry-level annual growth rates of outputs and inputs in 2018 and 2019 calculated by using raw transaction values and counts of the payment data and the row- and column sums of ONS IxI, PxP, and SUTs. 
\end{figure}

\begin{figure}
	\centering
	
	\caption{Correlations with other national accounts variables}
	\label{fig:correlations_industry_level_SUT_raw}
	
	\includegraphics[width=\textwidth]{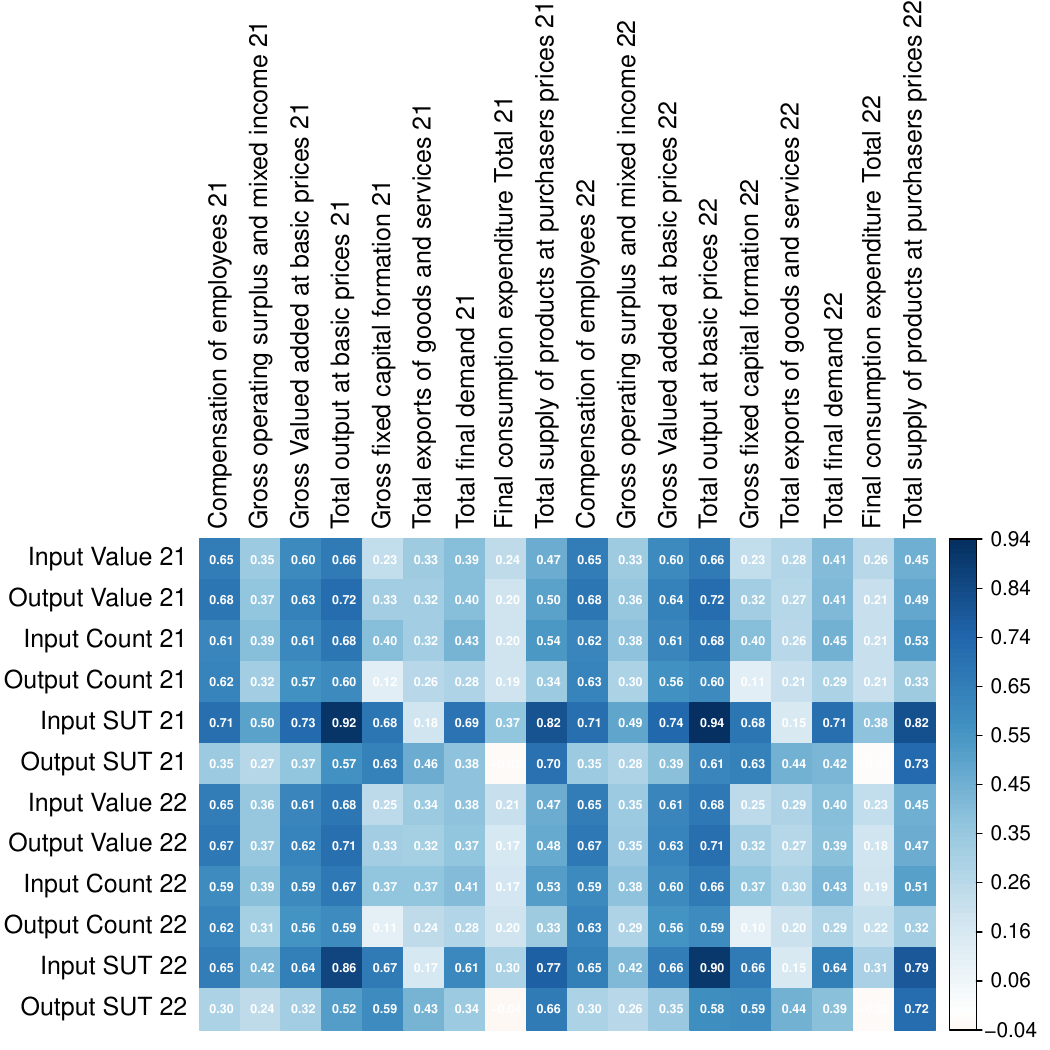}
	
	\justifying \scriptsize \noindent	
	Notes: The correlations are measured by the Pearson correlation coefficient between industry-level annual outputs and inputs in 2021-2022 calculated by using raw transaction values and counts of the payment data and various macro account data obtained from SUT. 
\end{figure}

\begin{figure}
	\centering
	
	\caption{Correlations with other national accounts variables (growth rates)}
	\label{fig:correlations_industry_level_SUT_growth}
	
	\includegraphics[width=\textwidth]{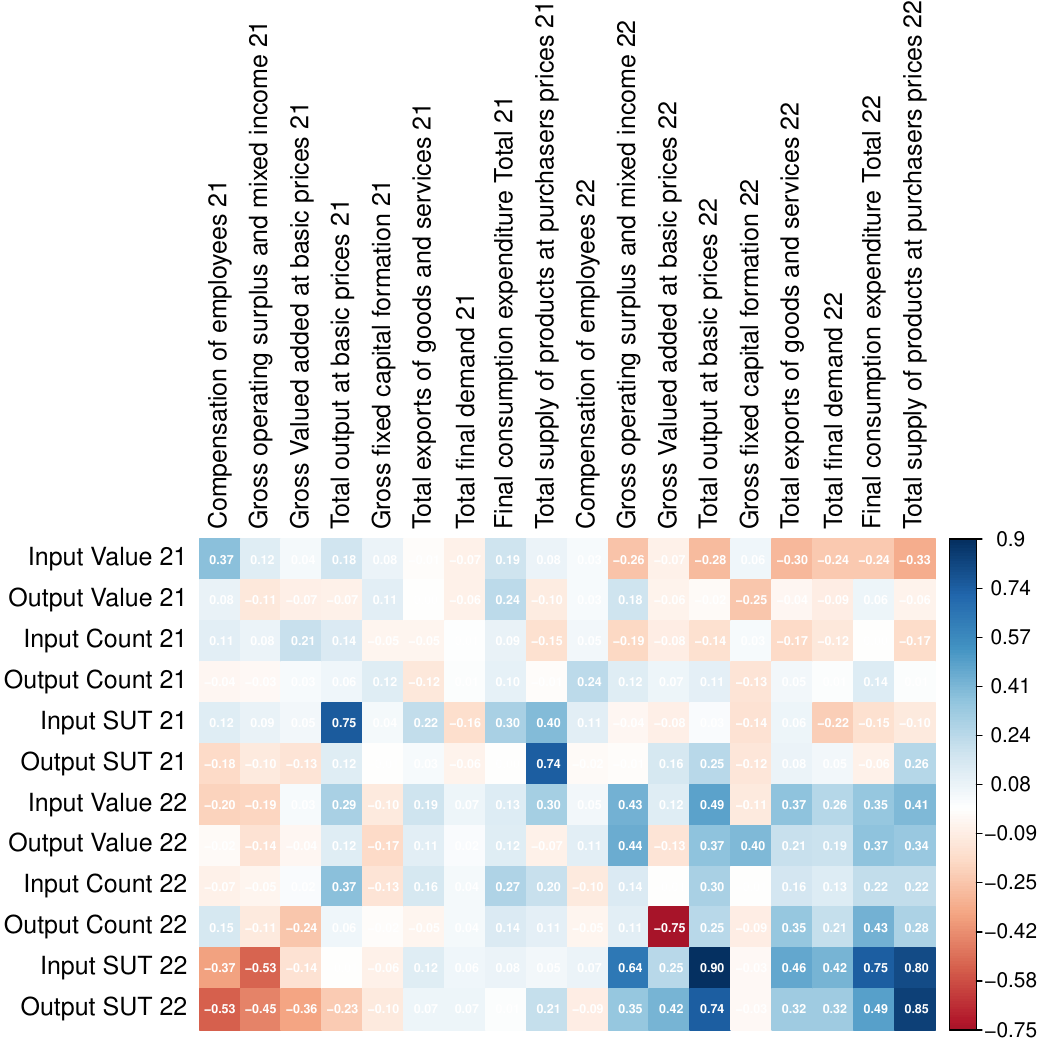}
	
	\justifying \scriptsize \noindent	
	Notes: The correlations are measured by the Pearson correlation coefficient between industry-level annual outputs and inputs in 2021-22 calculated by using raw transaction values and counts of the payment data and various macro account data obtained from SUT. 
\end{figure}

\begin{figure}[h]
	\centering
	\includegraphics[width=0.8\textwidth]{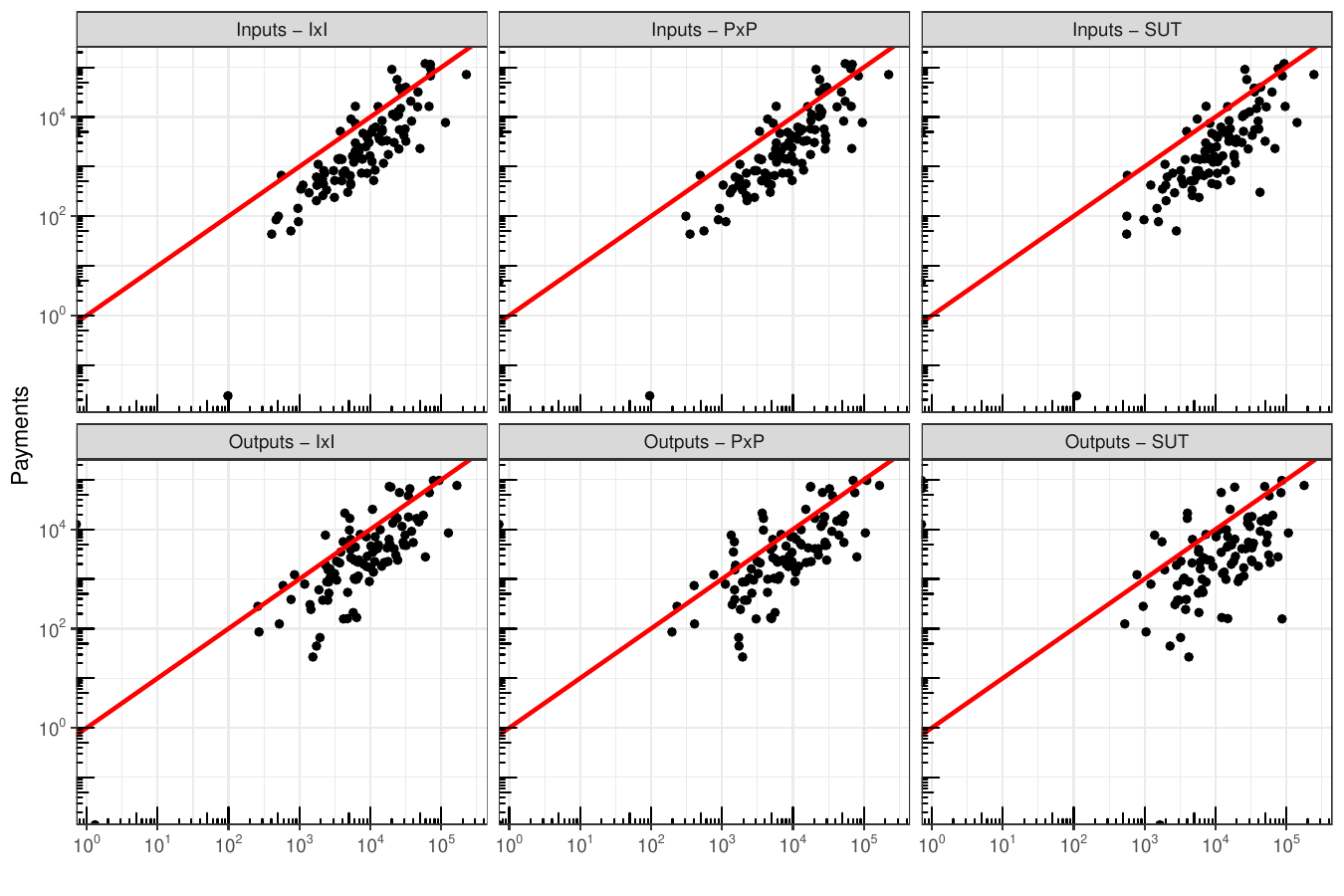}
	\caption{Comparison of industry sizes in 2022}
	\label{fig:input_output_scatterplot_2022}

	\justifying \scriptsize
	\noindent
	Notes: This figure shows the differences of industry-level aggregate inputs and outputs. Payment data values are shown at the vertical axis, and those for different ONS IOTs (IxI, PxP, SUT) at the horizontal. The red line shows at which the values in the payment data would be equal to those in the ONS table. 
\end{figure}

\begin{figure}[h]
	\centering
	\includegraphics[width=0.8\textwidth]{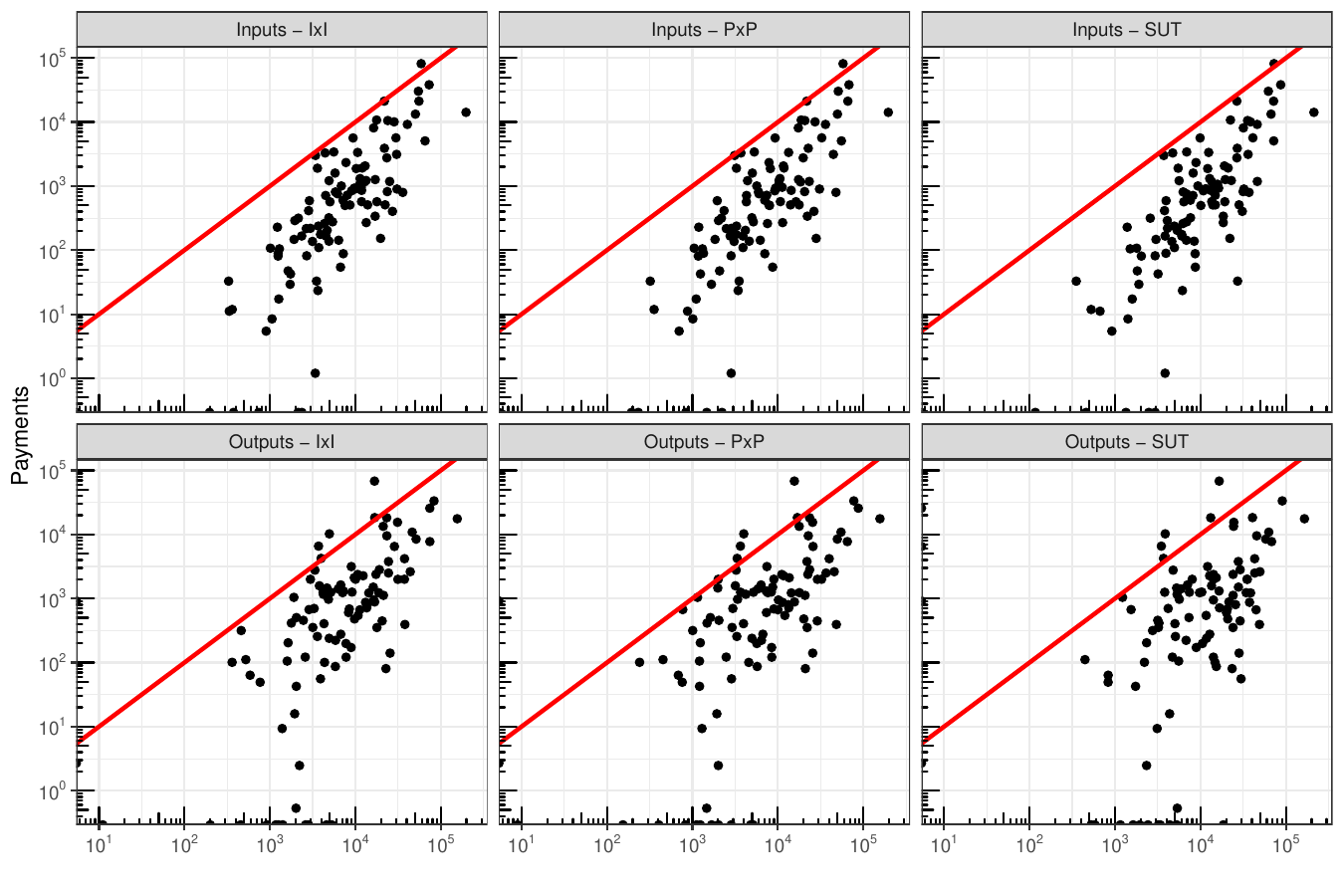}
	\caption{OLD DATA: Comparison of industry sizes in 2019}
	\label{OLD:app:fig:input_output_scatterplot_2019}

	\justifying \scriptsize
	\noindent
	Notes: This figure shows the differences of industry-level aggregate inputs and outputs. Payment data values are shown at the vertical axis, and those for different ONS IOTs (IxI, PxP, SUT) at the horizontal. The red line shows at which the values in the payment data would be equal to those in the ONS table. 
\end{figure}

\begin{figure}[h]
	\centering
	\includegraphics[width=\textwidth]{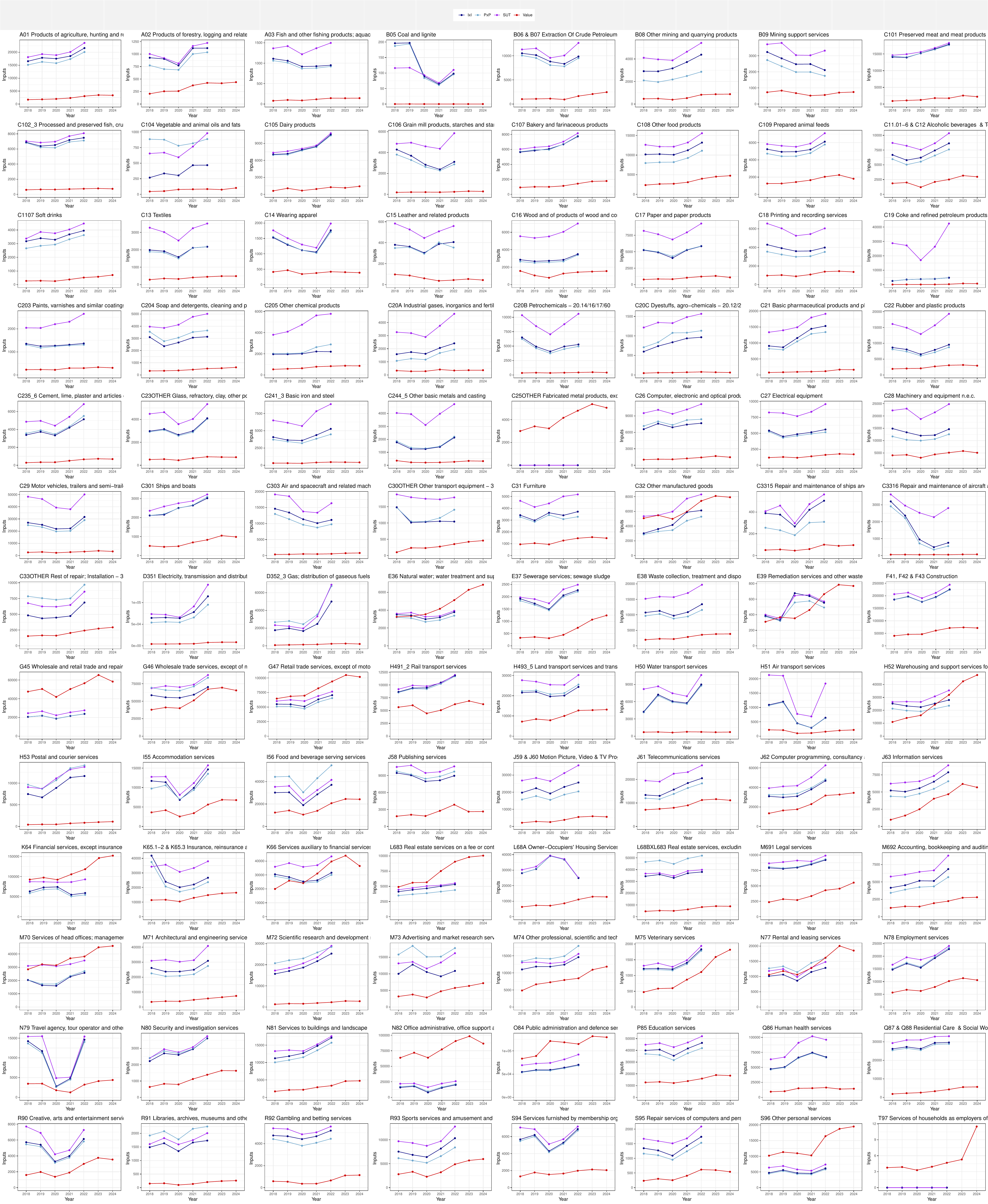}
	\caption{Comparison of aggregate industry sizes by sum of inputs over time}
	\label{fig:time_series_CPA_inputs}

	\justifying \scriptsize
	\noindent
	Notes: The figure shows how industry level aggregates evolved over time for different data sets. Industries are grouped into sectors. Aggregates are calculated as sum over all input links for an industry group. Scales of the y-axes differ across plots.
\end{figure}

\begin{figure}[h]
	\centering
	\includegraphics[width=\textwidth]{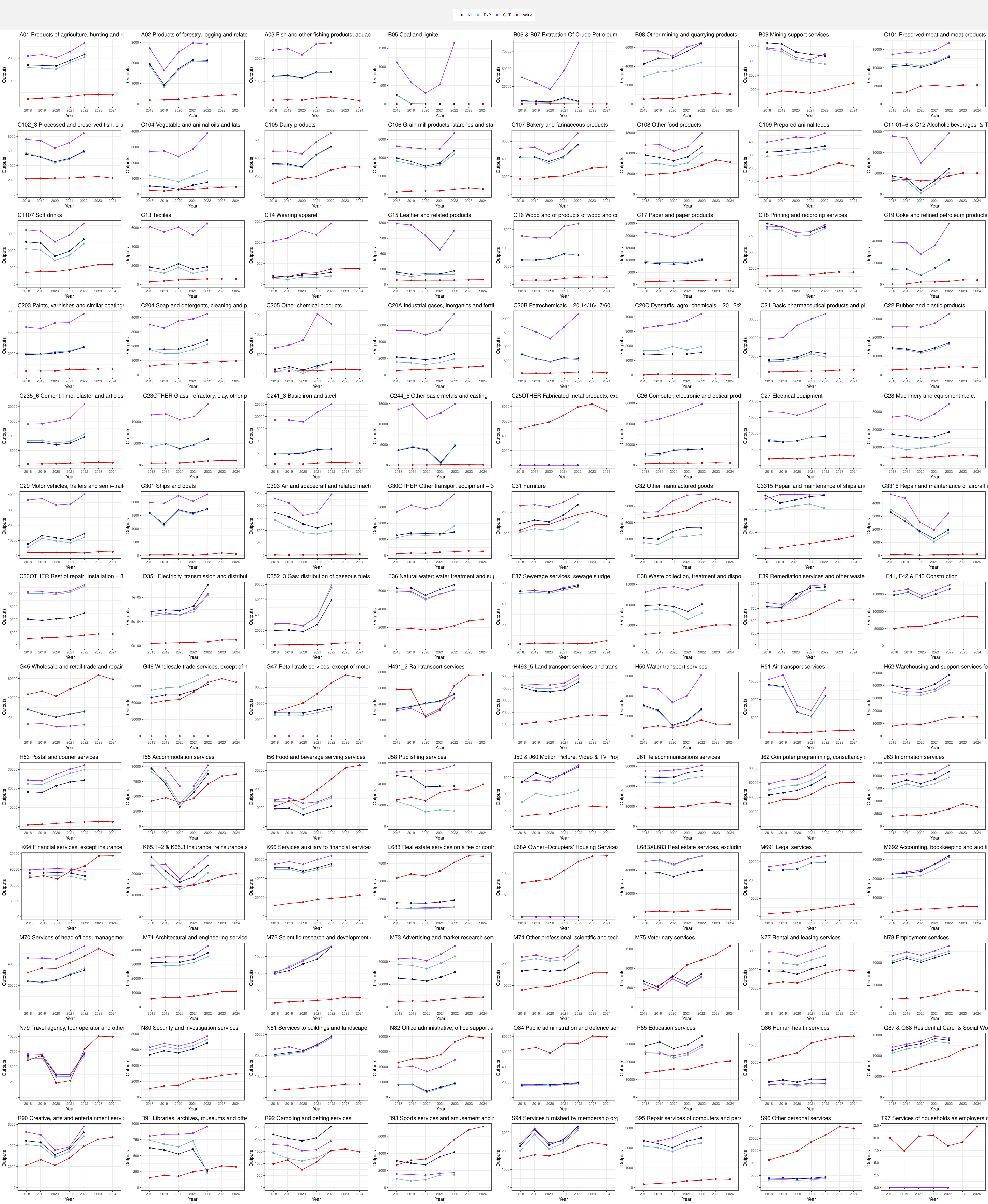}
	\caption{Comparison of aggregate industry sizes by sum of outputs over time}
	\label{fig:time_series_CPA_outputs}

	\justifying \scriptsize
	\noindent
	Notes: The figure shows how industry level aggregates evolved over time for different data sets. Industries are grouped into sectors. Aggregates are calculated as sum over all output links for an industry group. Scales of the y-axes differ across plots.
\end{figure}

\FloatBarrier
\subsubsection{Quantifying the edge-level difference}

\label{app:subsec:diff_quantification}
\label{subsec:difference_quantification}

\begin{figure}[h]
	\centering
	\caption{OLD DATA: Proportional differences between the ONS and payment-based IOTs}
	\label{OLD:fig:rel_diff_histogram_positives}

	\includegraphics[width=0.9\textwidth]{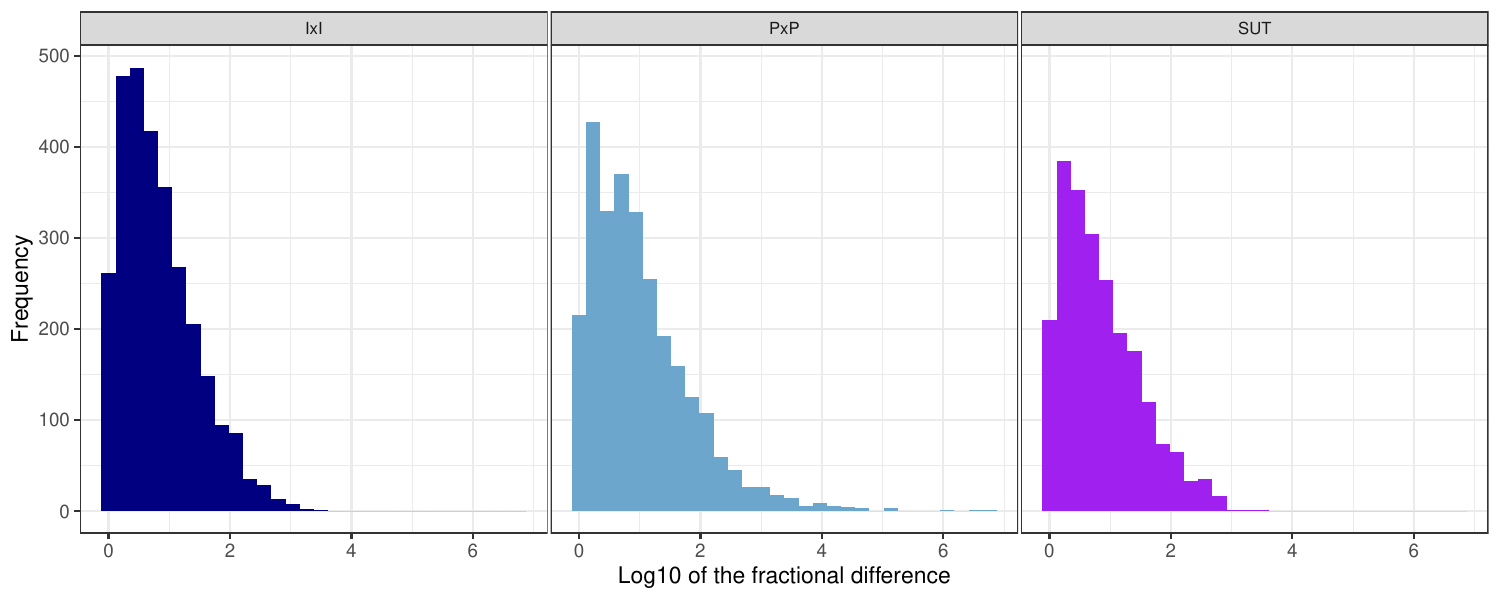}

	\justifying \scriptsize
	\noindent
	Notes: The figures show the distribution of the proportional edge-level differences (scaled at a log-10 basis) between the payment-based and ONS IOTs, using 2019 data. Industry pairs are removed if the transaction value is zero in one of the two datasets.\footnote{A comparison of differences with and without link removal are available in \ref{app:subsec:diff_quantification}, using a scaled percentage difference metric.} 
\end{figure}

\begin{figure}[h]
	\centering
	\caption{Proportional differences between the ONS and payment-based IOTs (2022 data)}
	\label{fig:rel_diff_histogram_positives}

	\includegraphics[width=0.9\textwidth]{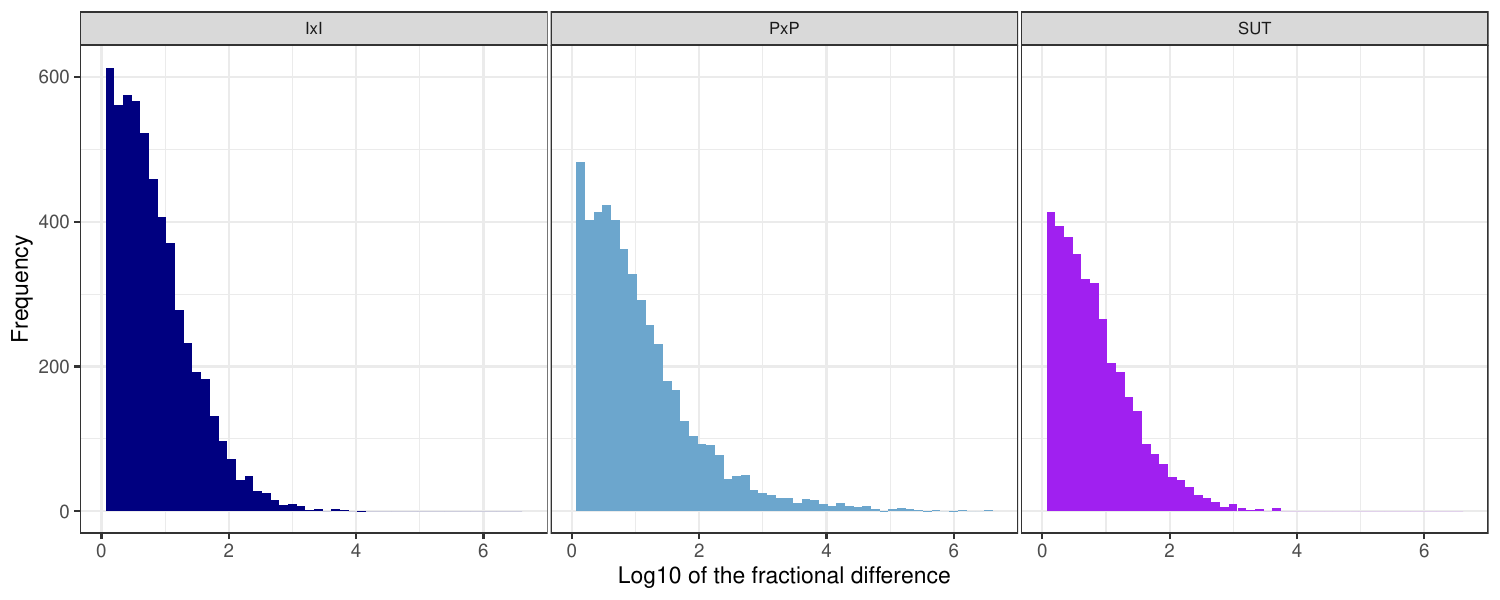}

	\justifying \scriptsize
	\noindent
	Notes: The figures show the distribution of the proportional edge-level differences (scaled at a log-10 basis) between the payment-based and ONS IOTs, using 2022 data. Industry pairs are removed if the transaction value is zero in one of the two datasets.\footnote{A comparison of differences with and without link removal are available in \ref{app:subsec:diff_quantification}, using a scaled percentage difference metric.} 
\end{figure}

\begin{table}[h]
	\centering
	\caption{OLD DATA: Quantiles of the proportional differences}
	\begin{tabular}{p{1.75cm}p{1.5cm}p{1.5cm}p{1.5cm}p{1.5cm}}
		\hline \hline \\[-1.8ex] 
		& 25\% & 50\% & 75\% & 100\% \\ 
		\hline \\[-1.8ex] 
		IxI  & 2.20 & 5.10 & 15.24 & 3851.66 \\
		PxP  & 2.40 & 6.76 & 27.74 & 582092.2  \\
		SUT  & 2.12 & 5.08 & 17.01 & 2911.96 \\
		
		\hline \hline \\[-1.8ex]
	\end{tabular}
	
	\justifying \scriptsize
	\noindent
	Notes: Quantiles of the proportional differences between the IxI, PxP, SUTs and the payment-based IOT in 2019. Unlike as in Fig. \ref{fig:rel_diff_histogram_positives}, the values are not log-scaled. 
	\label{tab:quantiles_diff_old}
\end{table}

\begin{table}[h]
	\centering
	\caption{Quantiles of the proportional differences}
	\begin{tabular}{p{3cm}p{1.5cm}p{1.5cm}p{1.5cm}p{1.5cm}}
		\hline \hline \\[-1.8ex] 
		& 25\% & 50\% & 75\% & 100\% \\ 
		\hline \\[-1.8ex] 
		IxI & $2.414$ & $6.420$ & $25.614$ & $3,365,800,927,989$ \\ 
		IxI (incl zero) & $1.000$ & $1.000$ & $1.000$ & $1.041$ \\ 
		PxP & $2.414$ & $6.420$ & $25.614$ & $3,365,800,927,989$ \\ 
		PxP (incl zero) & $1.000$ & $1.000$ & $1.000$ & $1.041$ \\ 
		SUT & $2.085$ & $4.773$ & $13.963$ & $4,749.393$ \\ 
		SUT (incl zero) & $1.000$ & $1.000$ & $1.000$ & $1.031$ \\
		\hline \hline \\[-1.8ex]
	\end{tabular}
	
	\justifying \scriptsize
	\noindent
	Notes: Quantiles of the proportional differences between the IxI, PxP, SUTs and the payment-based IOT in 2022. Unlike as in Fig. \ref{fig:rel_diff_histogram_positives}, the values are not log-scaled. 
	\label{tab:quantiles_diff}
\end{table}

\begin{figure}[!h]
	\centering
	
	\caption{OLD DATA: Scaled percentage difference}
	\label{OLD:app:fig:difference_hist_percentage}

	\begin{subfigure}[]{0.9\textwidth}
		\centering
		
		\subcaption{One-sided zero-links included}
		\includegraphics[width=\textwidth]{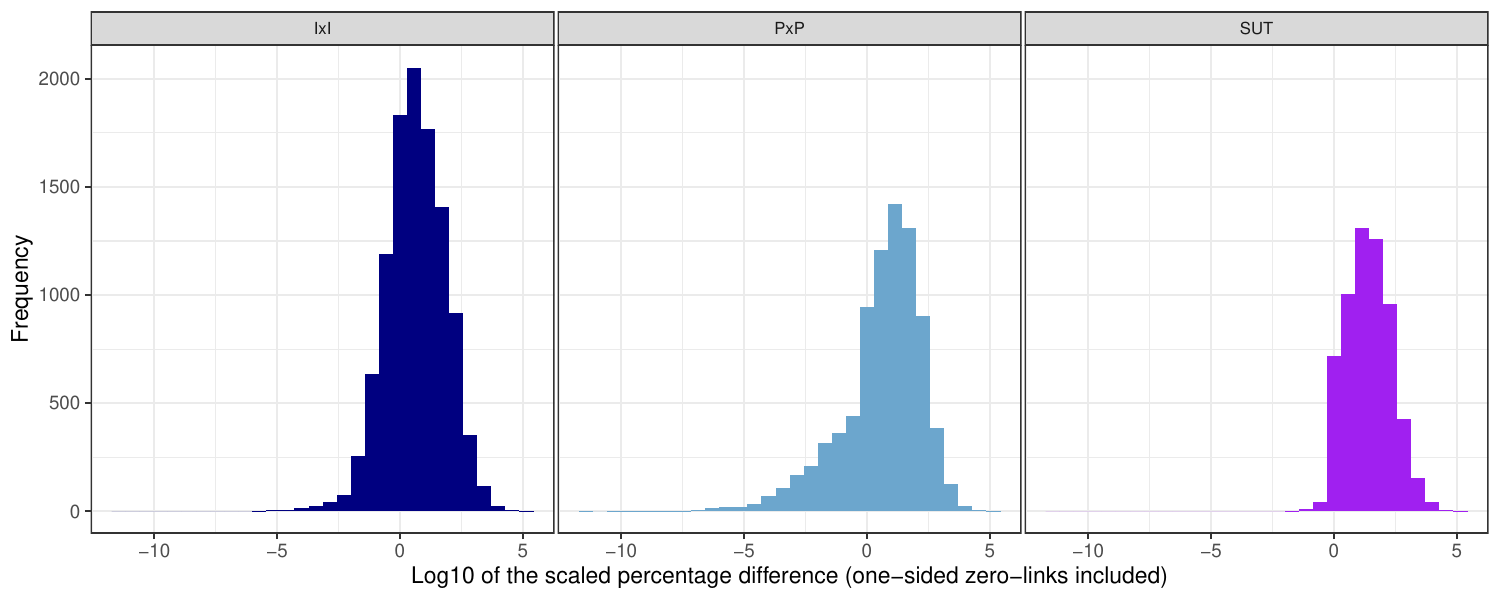}
	\end{subfigure}

	\begin{subfigure}[]{0.9\textwidth}
		\centering
		
		\subcaption{One-sided zero-links excluded}
		\includegraphics[width=\textwidth]{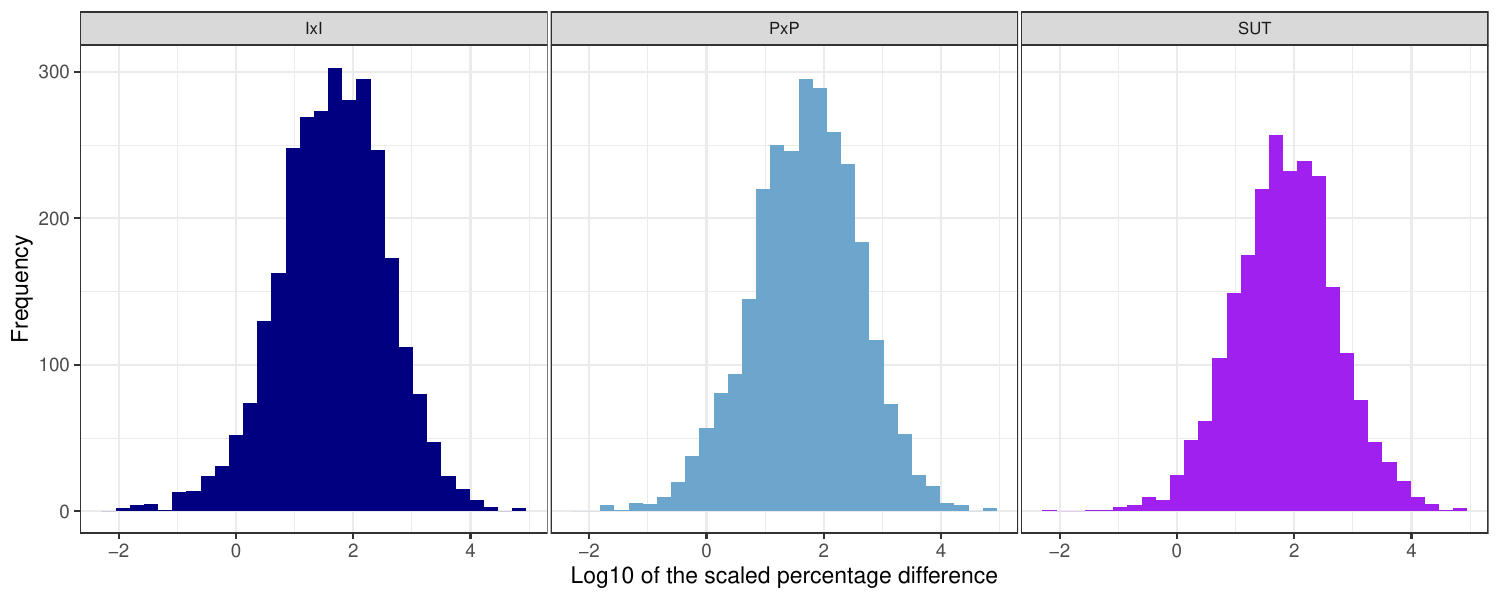}
		
	\end{subfigure}

	\justifying \scriptsize \noindent
	Notes: The figures show a comparison between the distribution of the scaled percentage edge-level differences (scaled at a log-10 basis) between the payment-based and ONS IOTs from 2019, when including or excluding links with a zero value in one of the two datasets. 
	
\end{figure}

\begin{figure}[!h]
	\centering
	
	\caption{Scaled percentage difference}
	\label{app:fig:difference_hist_percentage}

	\begin{subfigure}[]{0.9\textwidth}
		\centering
		
		\subcaption{One-sided zero-links included}
		\includegraphics[width=\textwidth]{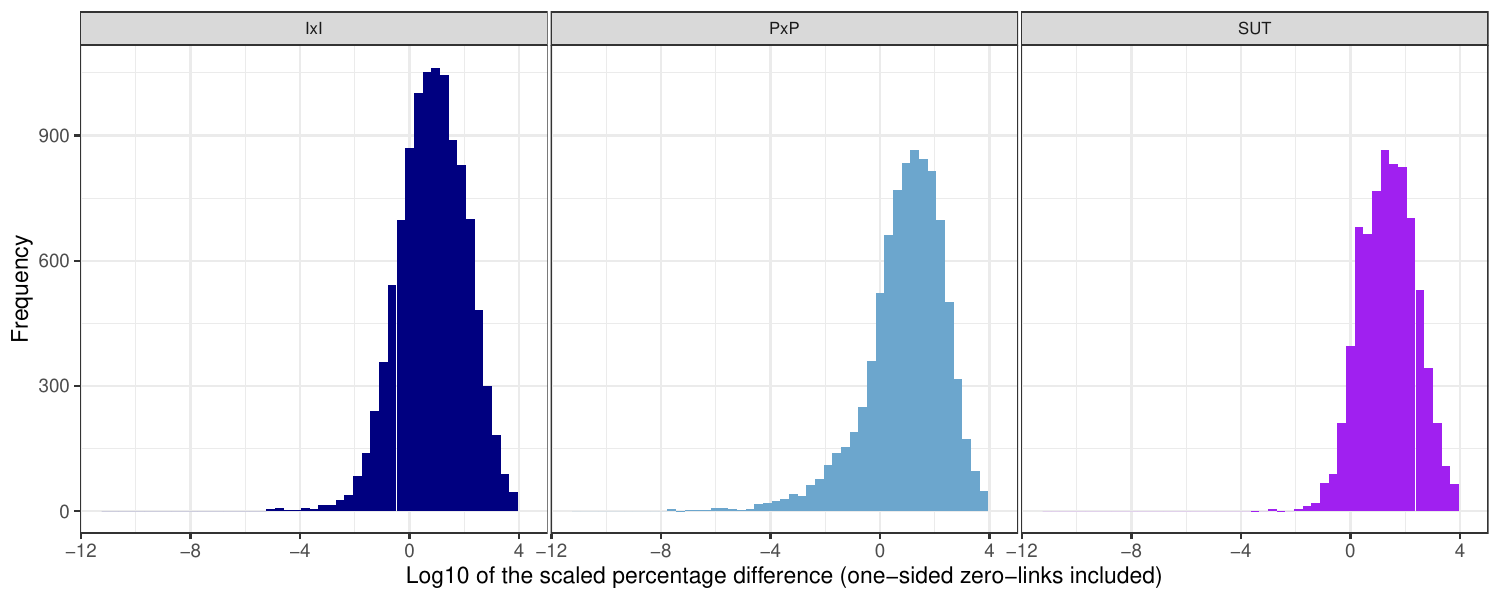}
	\end{subfigure}

	\begin{subfigure}[]{0.9\textwidth}
		\centering
		
		\subcaption{One-sided zero-links excluded}
		\includegraphics[width=\textwidth]{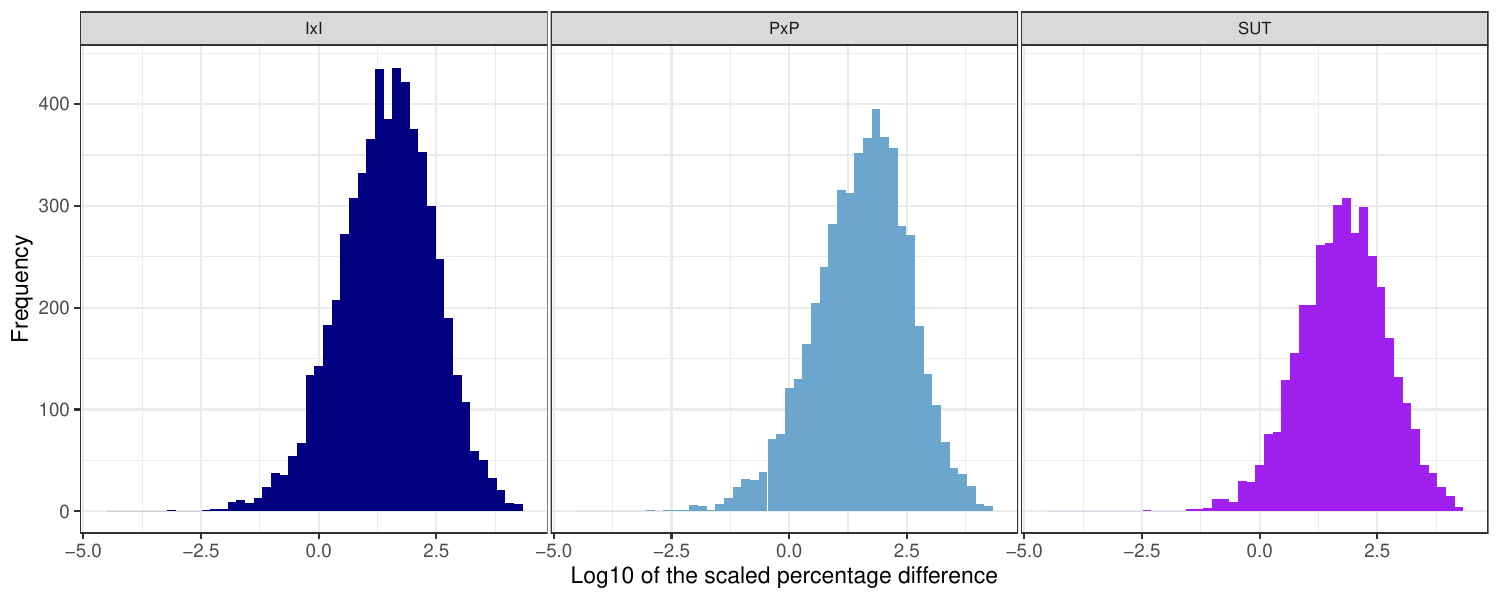}
		
	\end{subfigure}

	\justifying \scriptsize \noindent
	Notes: The figures show a comparison between the distribution of the scaled percentage edge-level differences (scaled at a log-10 basis) between the payment-based and ONS IOTs from 2022, when including or excluding links with a zero value in one of the two datasets. 
	
\end{figure}

\begin{table}[ht] \centering 
	\caption{OLD DATA: Quartiles of the scaled percentage differences} 
	\label{old:app:tab:differences_quartiles_zerolinks} 
	\begin{tabular}{ lp{1.75cm}p{1.75cm}p{1.75cm}p{1.75cm}} 
		\\[-1.8ex]\hline 
		\hline \\[-1.8ex] 
		& 25\% & 50\% & 75\% & 100\% \\ 
		\hline \\[-1.8ex] 
		IxI & $11.30$ & $47.81$ & $192.86$ & $62,674.01$ \\ 
		including zero & $0.72$ & $4.34$ & $30.52$ & $73,521.71$ \\ 
		\hline
		PxP & $12.31$ & $53.50$ & $213.27$ & $63,817.68$ \\ 
		including zero & $0.72$ & $8.05$ & $54.46$ & $73,543.50$ \\ 
		\hline
		SUT & $18.85$ & $72.75$ & $268.92$ & $81,455.65$ \\ 
		including zero & $5.36$ & $24.42$ & $113.21$ & $88,906.08$ \\ 
		\hline \hline \\[-1.8ex] 
	\end{tabular} 
	
	\justifying \scriptsize
	\noindent
	Notes: Quantiles of the scaled percentage differences between the IxI, PxP, SUTs and the payment-based IOT in 2019. Unlike as in Fig. \ref{app:fig:difference_hist_percentage}, the values are not log-scaled. The scale of the scaled percentage difference is not comparable to the proportional difference used in the main text. 
\end{table} 

\begin{table}[ht] \centering 
	\caption{Quartiles of the scaled percentage differences} 
	\label{app:tab:differences_quartiles_zerolinks} 
	\begin{tabular}{ lp{1.75cm}p{1.75cm}p{1.75cm}p{1.75cm}} 
		\\[-1.8ex]\hline 
		\hline \\[-1.8ex] 
		& 25\% & 50\% & 75\% & 100\% \\ 
		\hline \\[-1.8ex] 
		IxI & $7.867$ & $41.837$ & $171.152$ & $59,823.480$ \\ 
		IxI & $1.599$ & $12.901$ & $84.852$ & $59,595.450$ \\ 
		\hline
		PxP & $7.867$ & $41.837$ & $171.152$ & $59,823.480$ \\ 
		PxP & $1.599$ & $12.901$ & $84.852$ & $59,595.450$ \\
		\hline
		SUT & $14.375$ & $62.388$ & $254.859$ & $47,126.070$ \\ 
		SUT & $4.464$ & $25.256$ & $132.529$ & $53,446.630$ \\ 
		\hline \hline \\[-1.8ex] 
	\end{tabular} 
	
	\justifying \scriptsize
	\noindent
	Notes: Quantiles of the scaled percentage differences between the IxI, PxP, SUTs and the payment-based IOT in 2019. Unlike as in Fig. \ref{app:fig:difference_hist_percentage}, the values are not log-scaled. The scale of the scaled percentage difference is not comparable to the proportional difference used in the main text. 
\end{table}

This subsection illustrates a quantification of edge-level differences and their distribution following the approach in \citet{hotte2025national}. The key observations are: 

\begin{itemize}
	\item \textbf{Scale:} Edge-level differences can be still significant (Fig. \ref{fig:rel_diff_histogram_positives} and \ref{app:fig:difference_hist_percentage}) but they tend to be smaller by scale in relation to the previous version of the data (\ref{app:fig:difference_hist_percentage}). 
	\item \textbf{Distribution:} The distribution of edge-level differences indicates that there are more edges with small differences for all ONS IOTs. However, it also seems that there also more outliers, with transaction values between a pair of sectors being large in one data set and extremely tiny in the other. For plotting the data, a handful of very extreme outliers have been removed. Potentially, this arises from a lower number of one-sided zero links in the novel data, which has not been tested here but might be indicated by the overall higher density of the payment network compared to the previous data version (see Sec. \ref{subsec:aggregate_network}). 
\end{itemize}

\FloatBarrier

\FloatBarrier
\subsection{Stylized facts of granular networks}
\label{app:subsec:stylized}
\subsubsection{Correlations of growth rates}
\label{app:subsubsec:growth_corr}
\begin{figure}[!h]
	\centering
	
	\caption{OLD DATA: Correlations of growth rates}
	\label{OLD:fig:growth_correl_by_distance}
	
	\includegraphics[width=\textwidth]{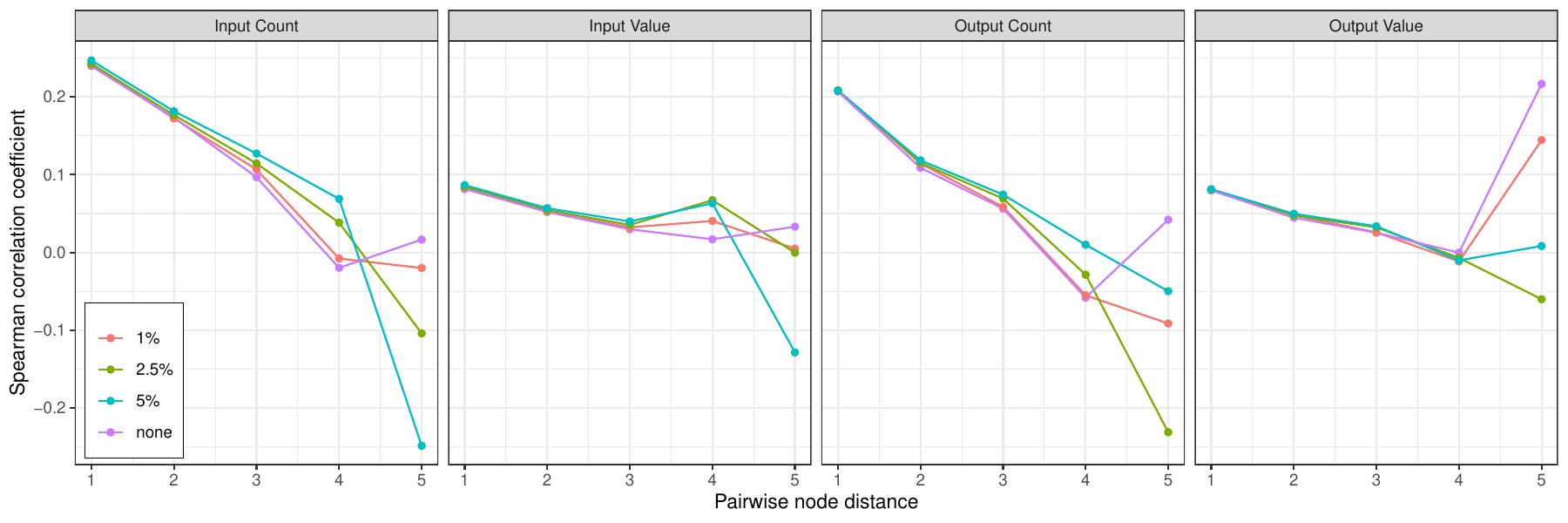}
	
	\justifying \scriptsize
	\noindent
	Notes: These figures illustrate the Spearman correlation coefficients between monthly (year on year) growth rates of directly and indirectly connected pairs of industries, using data from 2016 to 2019, excluding the Covid-19 period. 
	The x-axis shows the distance of the industry pairs in annual network aggregates.\footnote{The distances are the shortest path (lowest number of steps) that connects the pair of industries in the network. A value of one indicates a direct link (one step) between the pair.} 
	The colours indicate truncation thresholds imposed on the network before calculating the distances. Links with a weight (input share) below the threshold are removed (see also Section \ref{subsec:aggregate_network}).  
	
\end{figure}

\begin{figure}[!h]
	\centering
	
	\caption{Correlations of growth rates (intermediaries excluded)}
	\label{fig:growth_correl_by_distance_no_interm}
	
	\includegraphics[width=\textwidth]{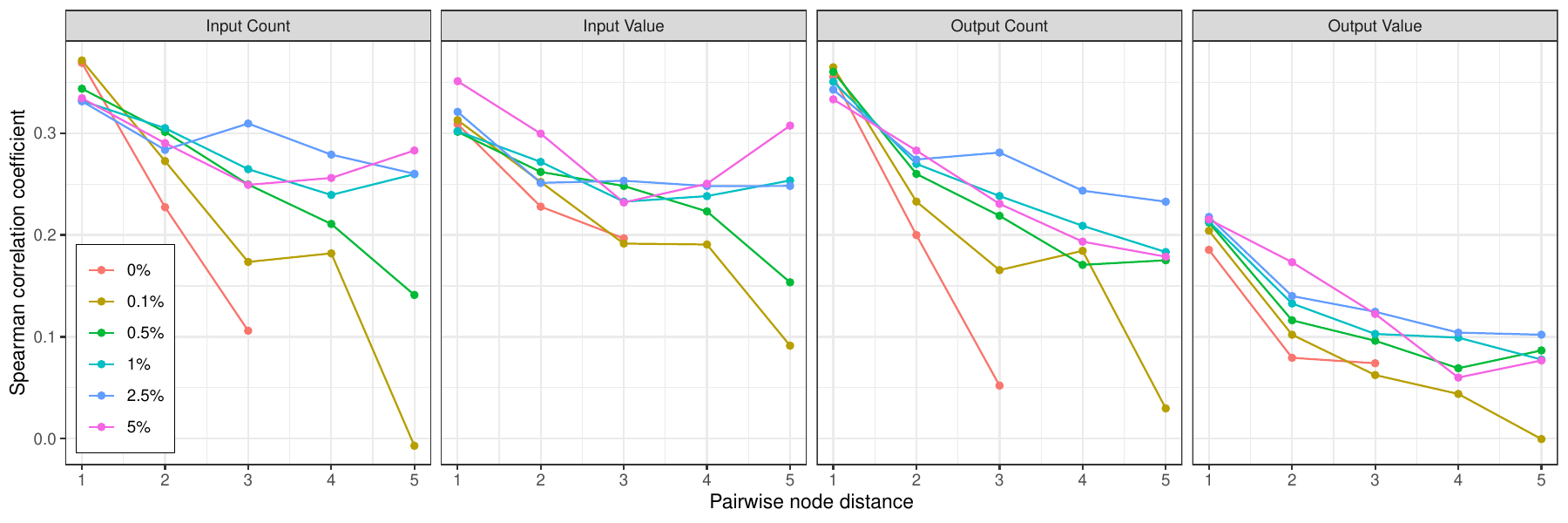}
	
	\justifying \scriptsize
	\noindent
	Notes: These figures illustrate the Spearman correlation coefficients between monthly (year on year) growth rates of directly and indirectly connected pairs of industries, using data from 2016 to 2024.  
	The x-axis shows the distance of the industry pairs in annual network aggregates.\footnote{The distances are the shortest path (lowest number of steps) that connects the pair of industries in the network. A value of one indicates a direct link (one step) between the pair.} 
	The colours indicate truncation thresholds imposed on the network before calculating the distances. Links with a weight (input share) below the threshold are removed (see also Section \ref{subsec:aggregate_network}). 
	
\end{figure}

\begin{figure}[!h]
	\centering
	
	\caption{Pearson correlations of growth rates}
	\label{fig:growth_correl_by_distance_pearson}
	
	\includegraphics[width=\textwidth]{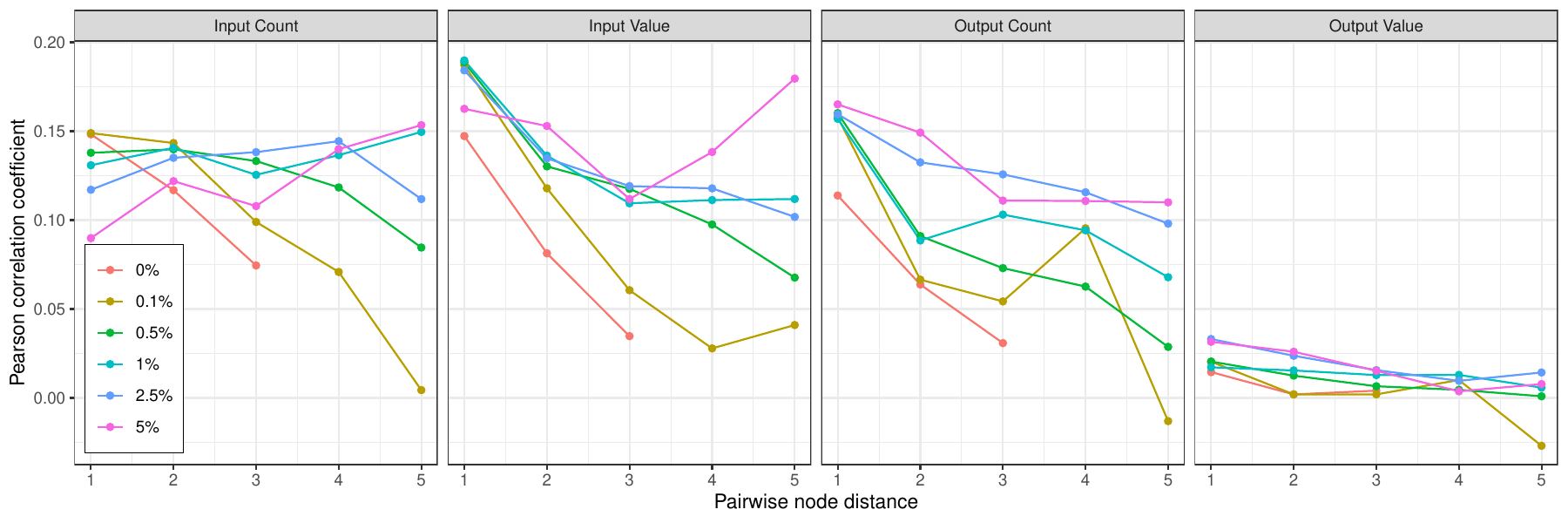}
	
	\justifying \scriptsize
	\noindent
	Notes: These figures illustrate the Pearson correlation coefficients between monthly (year on year) growth rates of directly and indirectly connected pairs of industries, using data from 2016 to 2018 and 2024. 
	The x-axis shows the distance of the industry pairs in annual network aggregates.\footnote{The distances are the shortest path (lowest number of steps) that connects the pair of industries in the network. A value of one indicates a direct link (one step) between the pair.} 
	The colors indicate truncation thresholds imposed on the network before calculating the distances. Links with a weight (input share) below the threshold are removed (see also Section \ref{subsec:aggregate_network}).
	
\end{figure}

\FloatBarrier

\subsubsection{Centrality and power law}
\label{app:subsubsec:centrality}
\begin{figure}[!h]
	\centering
	\caption{OLD DATA: CCDF of the Katz-Bonacich centrality}
	\label{OLD:fig:influence_vector}
	\includegraphics[width=0.8\textwidth]{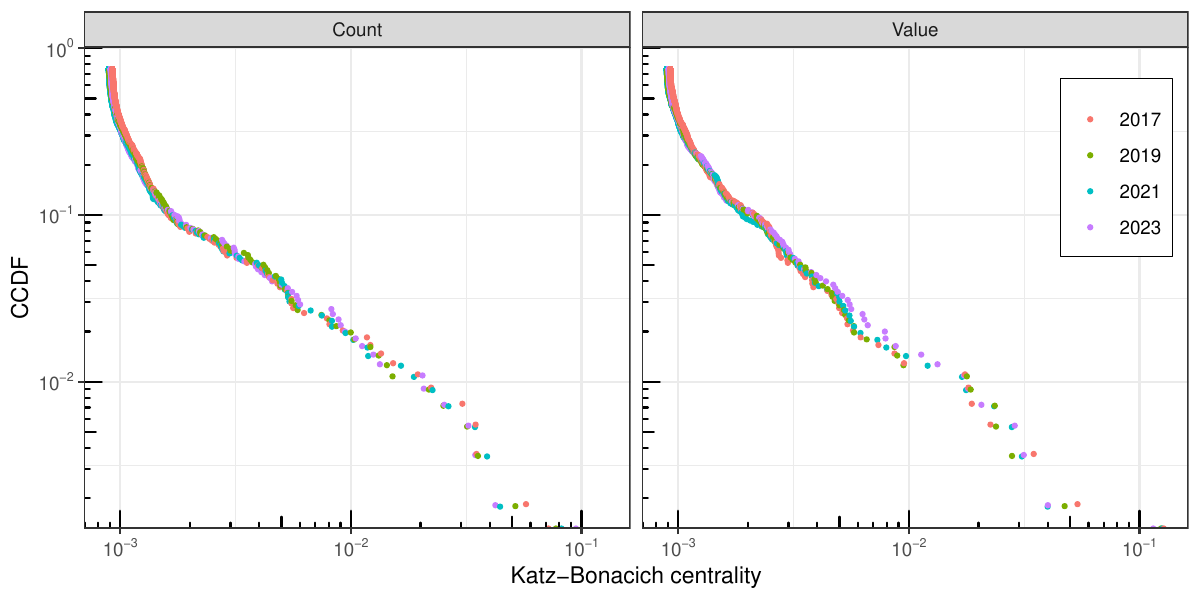}
	
	\justifying \scriptsize
	\noindent
	Notes: These figures illustrate the CCDF of the Katz-Bonacich centrality \citep[see][]{hotte2025national} for different years, using a labour share parameter of $\alpha_L = 0.5$ \citep{magerman2016heterogeneous} and payment-based input share matrices based on counts and values. 
	
\end{figure}

\begin{figure}[!h]
	\centering
	\caption{CCDF of the Katz-Bonacich centrality (truncated data)}
	\label{fig:influence_vector_truncated}
	\includegraphics[width=0.8\textwidth]{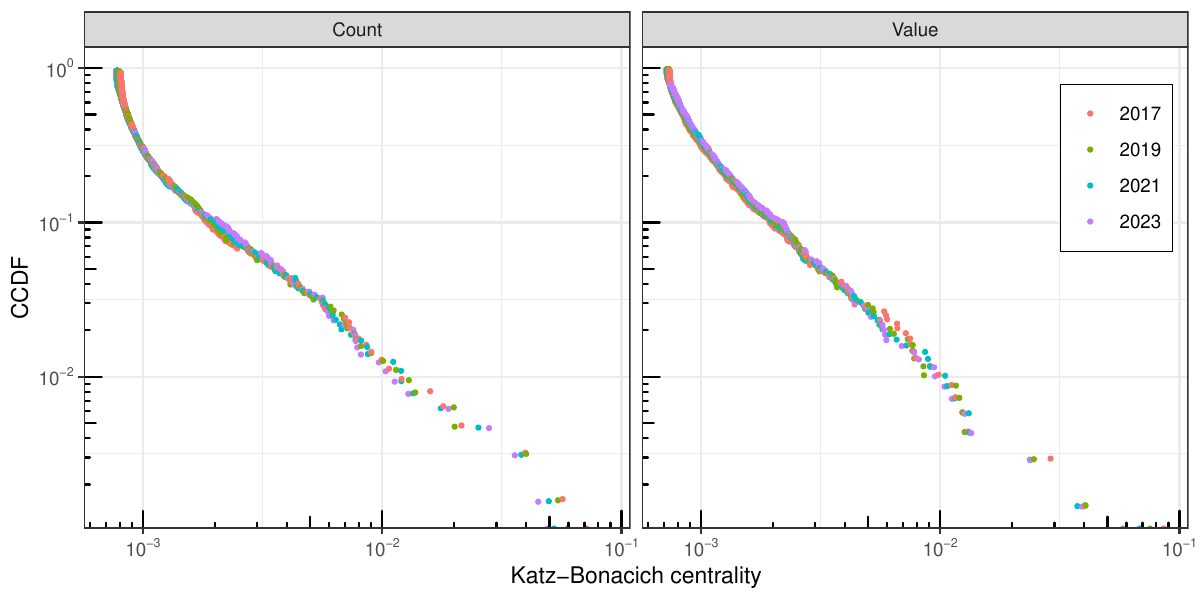}
	
	\justifying \scriptsize
	\noindent
	Notes: These figures illustrate the CCDF of the Katz-Bonacich centrality \citep[see][]{hotte2025national} for different years, using a labour share parameter of $\alpha_L = 0.5$ \citep{magerman2016heterogeneous} and payment-based input share matrices based on counts and values. 
	
\end{figure}

\begin{figure}[!h]
	\centering
	\caption{CCDF of the Katz-Bonacich centrality (excluding intermediary sectors)}
	\label{fig:influence_vector_no_intermediaries}
	\includegraphics[width=0.8\textwidth]{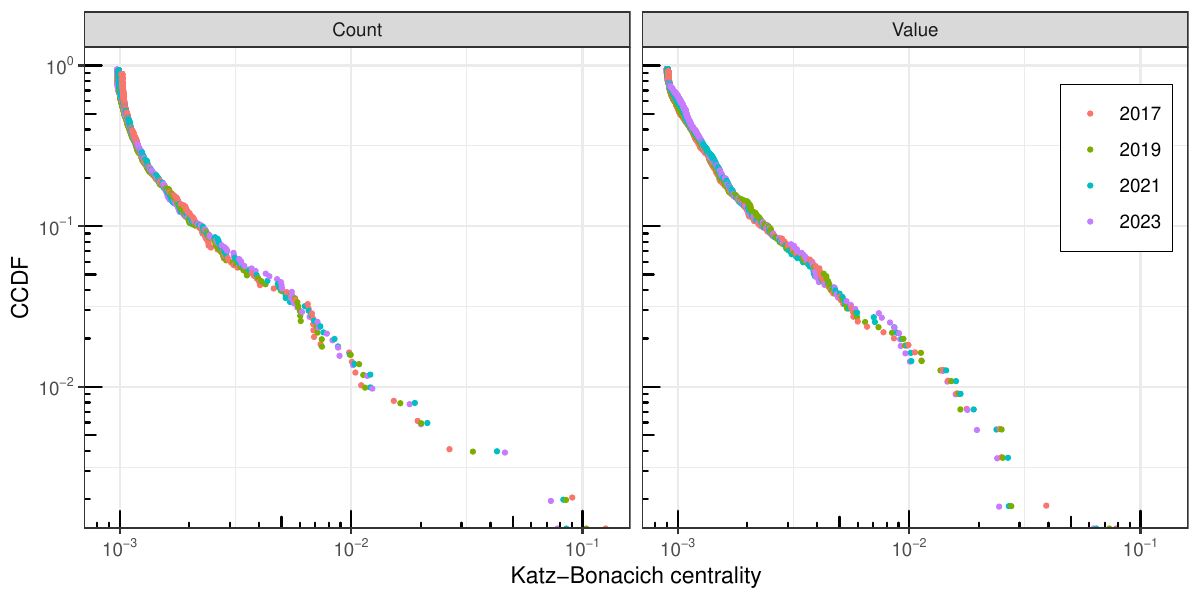}
	
	\justifying \scriptsize
	\noindent
	Notes: These figures illustrate the CCDF of the Katz-Bonacich centrality \citep[see][]{hotte2025national} for different years, using a labour share parameter of $\alpha_L = 0.5$ \citep{magerman2016heterogeneous} and payment-based input share matrices based on counts and values. Transaction links to intermediary sectors and public administration (G45, G46, K64, K65, K66, O84) have been removed from the data. 
	
\end{figure}

\begin{figure}[!h]
	\centering
	\caption{CCDF of the Katz-Bonacich centrality (excluding services)}
	\label{fig:influence_vector_no_services}
	\includegraphics[width=0.8\textwidth]{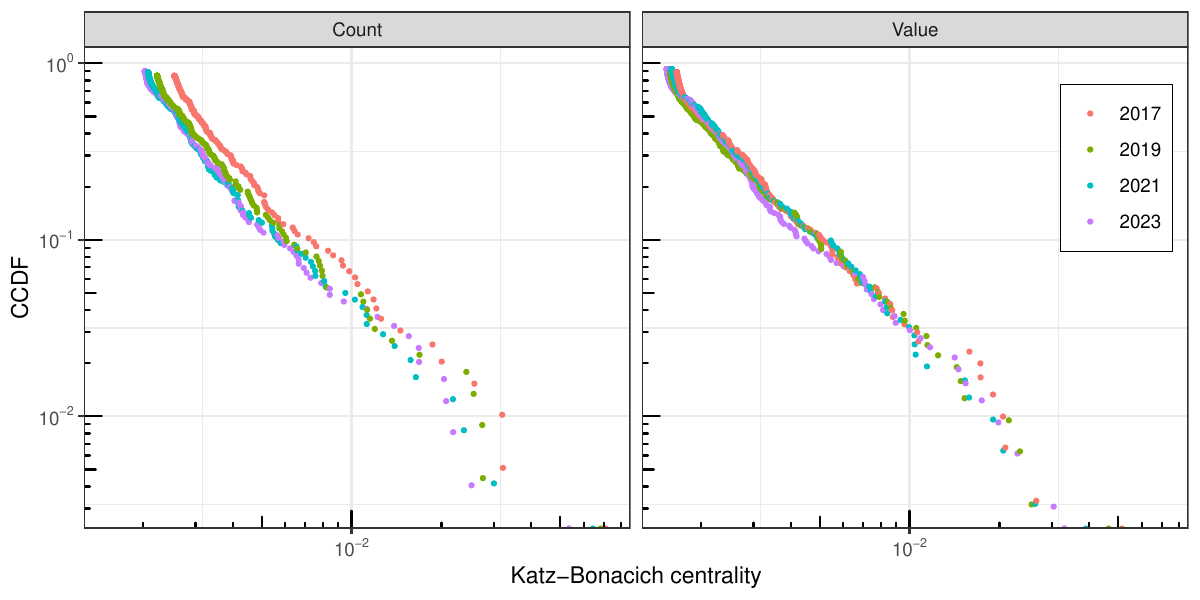}
	
	\justifying \scriptsize
	\noindent
	Notes: These figures illustrate the CCDF of the Katz-Bonacich centrality \citep[see][]{hotte2025national} for different years, using a labour share parameter of $\alpha_L = 0.5$ \citep{magerman2016heterogeneous} and payment-based input share matrices based on counts and values. Transaction links to service-related sectors (G45-Q88, S94-U99) have been removed from the data. 
	
\end{figure}

\FloatBarrier
\begin{table}[!htbp] \centering 
	\footnotesize
	\caption{OLD DATA: Power law fitting statistics} 
	\label{old:app:tab_plfit_stats} 
	\begin{tabular}{@{\extracolsep{5pt}} c|ccccc|ccccc} 
		\multicolumn{9}{l}{}\\[-1.8ex]\hline 
		\hline \multicolumn{9}{l}{}\\[-1.8ex] 
		\multicolumn{1}{c|}{}&\multicolumn{4}{c}{Value}&\multicolumn{4}{c}{Count}\\
		\hline \\[-1.8ex] 
		Year & $\gamma$ & xmin& logLik& KS.stat& p-value&$\gamma$& xmin& logLik& KS.stat& p-value\\ 
		\hline \\[-1.8ex] 
		2017 & 1.362 & 0.001 & 530.532 & 0.06 & 0.855 & 2.082 & 0.001 & 1570.413 & 0.167 & 0 \\ 
		2019 & 1.429 & 0.001 & 713.972 & 0.087 & 0.267 & 1.141 & 0.003 & 133.594 & 0.072 & 0.995 \\ 
		2021 & 1.615 & 0.001 & 1057.93 & 0.088 & 0.117 & 1.974 & 0.001 & 1539.515 & 0.17 & 0 \\ 
		2023 & 1.382 & 0.001 & 767.922 & 0.111 & 0.061 & 0.982 & 0.001 & 295.351 & 0.062 & 0.964 \\ 
		\hline \\[-1.8ex] 
		&\multicolumn{8}{c}{Data truncated at 10\% quantile of transaction value}  \\ 
		\hline \\[-1.8ex] 
		2017 & 1.343 & 0.001 & 474.976 & 0.054 & 0.952 & 2.207 & 0.001 & 1646.743 & 0.172 & 0 \\ 
		2019 & 1.455 & 0.001 & 677.768 & 0.086 & 0.305 & 1.882 & 0.001 & 1345.955 & 0.167 & 0 \\ 
		2021 & 1.689 & 0.001 & 1098.346 & 0.086 & 0.117 & 1.022 & 0.001 & 319.632 & 0.062 & 0.95 \\ 
		2023 & 1.49 & 0.001 & 857.5 & 0.117 & 0.03 & 1.022 & 0.001 & 309.591 & 0.058 & 0.977 \\ 
		\hline \multicolumn{9}{l}{}\\[-1.8ex] 
	\end{tabular}

	\justifying \noindent \scriptsize
	Notes: 
	This table shows the power law fitting statistics, where $\gamma$ is the fitted exponent, xmin is the minimum level of the influence vector beyond which a power law can be reasonably fitted (see \citet{Clauset_2009}), logLik shows the log-Likelihood, and KS is short for the Kolmogorov-Smirnov test statistic for significance. The p-value indicates the probability of rejecting the hypothesis that the distribution of the influence vector could have been drawn from a power law distribution. A p-value $<$0.05 supports the power law hypothesis.  
	
\end{table} 

\begin{table}[!htbp] \centering 
	\footnotesize
	\caption{Power law fitting statistics} 
	\label{app:tab_plfit_stats} 
	\begin{tabular}{@{\extracolsep{5pt}} c|ccccc|ccccc} 
		\multicolumn{9}{l}{}\\[-1.8ex]\hline 
		\hline \multicolumn{9}{l}{}\\[-1.8ex] 
		\multicolumn{1}{c|}{}&\multicolumn{4}{c}{Value}&\multicolumn{4}{c}{Count}\\
		\hline \\[-1.8ex] 
		Year & $\gamma$ & xmin& logLik& KS.stat& p-value&$\gamma$& xmin& logLik& KS.stat& p-value\\ 
		\hline \\[-1.8ex] 
		2017 & 1.356 & 0.002 & 486.138 & 0.033 & 0.94 & 1.394 & 0.001 & 890.361 & 0.06 & 0.105 \\ 
		2019 & 1.395 & 0.001 & 545.22 & 0.041 & 0.635 & 1.376 & 0.001 & 817.888 & 0.053 & 0.17 \\ 
		2021 & 1.573 & 0.001 & 1154.403 & 0.043 & 0.3 & 1.237 & 0.001 & 602.754 & 0.033 & 0.927 \\ 
		2023 & 1.483 & 0.001 & 890.667 & 0.035 & 0.733 & 1.243 & 0.001 & 606.849 & 0.034 & 0.93 \\ 
		\hline \\[-1.8ex] 
		&\multicolumn{8}{c}{Data truncated at 10\% quantile of transaction value}  \\ 
		\hline \\[-1.8ex] 
		2017 & 1.388 & 0.002 & 499.739 & 0.04 & 0.74 & 1.302 & 0.004 & 130.61 & 0.057 & 0.787 \\ 
		2019 & 1.43 & 0.001 & 560.129 & 0.039 & 0.713 & 1.368 & 0.001 & 766.544 & 0.054 & 0.162 \\ 
		2021 & 1.5 & 0.001 & 750.067 & 0.038 & 0.652 & 1.244 & 0.001 & 596.473 & 0.034 & 0.907 \\ 
		2023 & 1.543 & 0.001 & 952.907 & 0.036 & 0.662 & 1.269 & 0.001 & 668.004 & 0.032 & 0.95 \\
		\hline \multicolumn{9}{l}{}\\[-1.8ex] 
	\end{tabular}

	\justifying \noindent \scriptsize
	Notes: 
	This table shows the power law fitting statistics, where $\gamma$ is the fitted exponent, xmin is the minimum level of the influence vector beyond which a power law can be reasonably fitted (see \citet{Clauset_2009}), logLik shows the log-Likelihood, and KS is short for the Kolmogorov-Smirnov test statistic for significance. The p-value indicates the probability of rejecting the hypothesis that the distribution of the influence vector could have been drawn from a power law distribution. A p-value $<$0.05 supports the power law hypothesis.  
	
\end{table} 

\begin{table}[!htbp] \centering 
	\footnotesize
	\caption{Power law fitting statistics (intermediary sectors removed)} 
	\label{app:tab_plfit_stats_nointerm} 
	\begin{tabular}{@{\extracolsep{5pt}} c|ccccc|ccccc} 
		\multicolumn{9}{l}{}\\[-1.8ex]\hline 
		\hline \multicolumn{9}{l}{}\\[-1.8ex] 
		\multicolumn{1}{c|}{}&\multicolumn{4}{c}{Value}&\multicolumn{4}{c}{Count}\\
		\hline \\[-1.8ex] 
		Year & $\gamma$ & xmin& logLik& KS.stat& p-value&$\gamma$& xmin& logLik& KS.stat& p-value\\ 
		\hline \\[-1.8ex] 
		2017 & 1.334 & 0.003 & 207.251 & 0.054 & 0.755 & 1.551 & 0.001 & 752.531 & 0.074 & 0.018 \\ 
		2019 & 1.335 & 0.002 & 330.213 & 0.04 & 0.958 & 1.167 & 0.002 & 243.483 & 0.064 & 0.49 \\ 
		2021 & 1.326 & 0.002 & 419.443 & 0.044 & 0.85 & 1.173 & 0.002 & 290.15 & 0.049 & 0.833 \\ 
		2023 & 1.318 & 0.002 & 423.306 & 0.046 & 0.73 & 1.219 & 0.002 & 368.96 & 0.054 & 0.642 \\ 
		\hline \\[-1.8ex] 
		&\multicolumn{8}{c}{Data truncated at 10\% quantile of transaction value}  \\ 
		\hline \\[-1.8ex] 
		2017 & 1.316 & 0.002 & 346.369 & 0.047 & 0.723 & 1.567 & 0.001 & 740.562 & 0.074 & 0.015 \\ 
		2019 & 1.401 & 0.002 & 537.185 & 0.041 & 0.855 & 1.177 & 0.002 & 243.556 & 0.069 & 0.4 \\ 
		2021 & 1.322 & 0.002 & 374.95 & 0.047 & 0.8 & 1.185 & 0.002 & 290.887 & 0.048 & 0.833 \\ 
		2023 & 1.351 & 0.002 & 437.322 & 0.048 & 0.7 & 1.218 & 0.002 & 362.8 & 0.054 & 0.615 \\ 
		\hline \multicolumn{9}{l}{}\\[-1.8ex] 
	\end{tabular}

	\justifying \noindent \scriptsize
	Notes: 
	This table shows the power law fitting statistics, where $\gamma$ is the fitted exponent, xmin is the minimum level of the influence vector beyond which a power law can be reasonably fitted (see \citet{Clauset_2009}), logLik shows the log-Likelihood, and KS is short for the Kolmogorov-Smirnov test statistic for significance. The p-value indicates the probability of rejecting the hypothesis that the distribution of the influence vector could have been drawn from a power law distribution. A p-value $<$0.05 supports the power law hypothesis. Transaction links to intermediary sectors and public administration (G45, G45, G46, K64, K65, K66, O84) have been removed from the data. 
	
\end{table}

\begin{table}[!htbp] \centering 
	\footnotesize
	\caption{Power law fitting statistics (services removed)} 
	\label{app:tab_plfit_stats_noservices} 
	\begin{tabular}{@{\extracolsep{5pt}} c|ccccc|ccccc} 
		\multicolumn{9}{l}{}\\[-1.8ex]\hline 
		\hline \multicolumn{9}{l}{}\\[-1.8ex] 
		\multicolumn{1}{c|}{}&\multicolumn{4}{c}{Value}&\multicolumn{4}{c}{Count}\\
		\hline \\[-1.8ex] 
		Year & $\gamma$ & xmin& logLik& KS.stat& p-value&$\gamma$& xmin& logLik& KS.stat& p-value\\ 
		\hline \\[-1.8ex] 
		2017 & 1.777 & 0.002 & 1061.934 & 0.032 & 0.833 & 1.879 & 0.003 & 511.369 & 0.055 & 0.325 \\ 
		2019 & 1.592 & 0.002 & 492.461 & 0.04 & 0.708 & 1.892 & 0.002 & 668.283 & 0.048 & 0.392 \\ 
		2021 & 1.823 & 0.002 & 1254.179 & 0.039 & 0.388 & 2.115 & 0.002 & 1037.558 & 0.065 & 0.01 \\ 
		2023 & 1.887 & 0.002 & 1429.422 & 0.046 & 0.11 & 2.041 & 0.002 & 1059.163 & 0.043 & 0.322 \\ 
		\hline \\[-1.8ex] 
		&\multicolumn{8}{c}{Data truncated at 10\% quantile of transaction value}  \\ 
		\hline \\[-1.8ex] 
		2017 & 1.83 & 0.002 & 1088.138 & 0.028 & 0.902 & 1.94 & 0.003 & 482.776 & 0.058 & 0.275 \\ 
		2019 & 1.815 & 0.002 & 1070.375 & 0.043 & 0.31 & 1.848 & 0.003 & 451.254 & 0.049 & 0.51 \\ 
		2021 & 1.939 & 0.002 & 953.091 & 0.035 & 0.695 & 2.191 & 0.002 & 1055.467 & 0.07 & 0 \\ 
		2023 & 1.935 & 0.002 & 1429.714 & 0.044 & 0.168 & 2.186 & 0.002 & 1095.388 & 0.069 & 0.005 \\ 
		\hline \multicolumn{9}{l}{}\\[-1.8ex] 
	\end{tabular}

	\justifying \noindent \scriptsize
	Notes: 
	This table shows the power law fitting statistics, where $\gamma$ is the fitted exponent, xmin is the minimum level of the influence vector beyond which a power law can be reasonably fitted (see \citet{Clauset_2009}), logLik shows the log-Likelihood, and KS is short for the Kolmogorov-Smirnov test statistic for significance. The p-value indicates the probability of rejecting the hypothesis that the distribution of the influence vector could have been drawn from a power law distribution. A p-value $<$0.05 supports the power law hypothesis. Transaction links to service-related sectors (G45-Q88, S94-U99) have been removed from the data.
	
\end{table}

\FloatBarrier
\begin{table}[!htbp] \centering 
	\footnotesize
	\caption{OLD DATA: Top 10 industries by influence vector} 
	\label{old:app:tab:top10_influence_vector} 
	\begin{tabularx}{\textwidth}{@{\extracolsep{5pt}} ccX|ccX} 
		\\[-1.8ex]\hline 
		\hline \\[-1.8ex] 
		SIC & & Industry description & SIC & & Industry description\\ 
		\hline \\[-1.8ex] 
		\multicolumn{3}{l|}{2017}&&&\\ 
		\multicolumn{3}{c|}{Value}&\multicolumn{3}{c}{Count}\\ 
		\hline \\[-1.8ex] 
		84110 & 0.1273 & General public administration & 84110 & 0.0719 & General public administration \\ 
		82990 & 0.0538 & Other business support services n.e.c. & 82990 & 0.0574 & Other business support services n.e.c. \\ 
		64999 & 0.0347 & Financial intermediation n.e.c. & 64910 & 0.035 & Financial leasing \\ 
		64910 & 0.0226 & Financial leasing & 61900 & 0.0347 & Other telecommunications \\ 
		65110 & 0.0187 & Life insurance & 64999 & 0.0304 & Financial intermediation n.e.c. \\ 
		61900 & 0.0181 & Other telecommunications & 45111 & 0.0223 & Sale of new \& motor vehicles \\ 
		45111 & 0.0175 & Sale of new \& motor vehicles & 64191 & 0.0195 & Banks \\ 
		70100 & 0.0095 & of head offices & 65110 & 0.0152 & Life insurance \\ 
		62090 & 0.0087 & Other information technology services & 64921 & 0.0135 & Credit granting by non-deposit finance\\ 
		49410 & 0.0074 & Freight transport by road & 62090 & 0.0121 & Other information technology services \\ 
		\hline \\[-1.8ex] 
		\hline \\[-1.8ex] 
		\multicolumn{3}{l|}{2023}&&&\\ 
		\multicolumn{3}{c|}{Value}&\multicolumn{3}{c}{Count}\\ 
		\hline \\[-1.8ex] 
		84110 & 0.1146 & General public administration & 84110 & 0.0946 & General public administration \\ 
		82990 & 0.04 & Other business support services n.e.c. & 82990 & 0.0423 & Other business support services n.e.c. \\ 
		65110 & 0.0315 & Life insurance & 64910 & 0.0346 & Financial leasing \\ 
		64999 & 0.0287 & Financial intermediation n.e.c. & 61900 & 0.0323 & Other telecommunications \\ 
		64910 & 0.0206 & Financial leasing & 45111 & 0.0254 & Sale of new \& motor vehicles \\ 
		61900 & 0.018 & Other telecommunications & 64999 & 0.0207 & Financial intermediation n.e.c. \\ 
		45111 & 0.0173 & Sale of new \& motor vehicles & 62090 & 0.0204 & Other information technology services \\ 
		62090 & 0.0133 & Other information technology services & 65110 & 0.0133 & Life insurance \\ 
		35130 & 0.0113 & Distribution of electricity & 64921 & 0.0125 & Credit granting by non-deposit finance \\ 
		49410 & 0.0088 & Freight transport by road & 35130 & 0.0112 & Distribution of electricity \\ 
		\hline \\[-1.8ex] 
	\end{tabularx} 
\end{table}

\begin{table}[!htbp] \centering 
	\footnotesize
	\caption{Top 10 industries by influence vector} 
	\label{app:tab:top10_influence_vector} 
	\begin{tabularx}{\textwidth}{@{\extracolsep{5pt}} ccX|ccX} 
		\\[-1.8ex]\hline 
		\hline \\[-1.8ex] 
		SIC & & Industry description & SIC & & Industry description\\ 
		\hline \\[-1.8ex] 
		\multicolumn{3}{l|}{2017}&&&\\ 
		\multicolumn{3}{c|}{Value}&\multicolumn{3}{c}{Count}\\ 
		\hline \\[-1.8ex] 
		84110 & 0.0823 & General public administration & 61900 & 0.0722 & Other telecommunications activities \\ 
		82990 & 0.0394 & Other business support services n.e.c. & 82990 & 0.0557 & Other business support services n.e.c. \\ 
		64999 & 0.0285 & Financial intermediation n.e.c. & 84110 & 0.0388 & General public administration \\ 
		61900 & 0.0176 & Other telecommunications activities & 64999 & 0.0212 & Financial intermediation n.e.c. \\ 
		65110 & 0.0135 & Life insurance & 65110 & 0.0176 & Life insurance \\ 
		62090 & 0.0123 & Other information technology services & 62090 & 0.0157 & Other information technology services \\ 
		49410 & 0.0119 & Freight transport by road & 96090 & 0.0121 & Other service activities n.e.c. \\ 
		70100 & 0.0112 & Activities of head offices & 64910 & 0.0109 & Financial leasing \\ 
		62020 & 0.0092 & Information technology consultancy & 65120 & 0.0098 & Non-life insurance \\ 
		96090 & 0.0078 & Other service activities n.e.c. & 49410 & 0.0093 & Freight transport by road \\ 
		\hline \\[-1.8ex] 
		\hline \\[-1.8ex] 
		\multicolumn{3}{l|}{2023}&&&\\ 
		\multicolumn{3}{c|}{Value}&\multicolumn{3}{c}{Count}\\ 
		\hline \\[-1.8ex] 
		84110 & 0.0557 & General public administration & 82990 & 0.0486 & Other business support services n.e.c. \\ 
		82990 & 0.0386 & Other business support services n.e.c. & 61900 & 0.0456 & Other telecommunications activities \\ 
		64999 & 0.0234 & Financial intermediation n.e.c. & 84110 & 0.0356 & General public administration \\ 
		62090 & 0.0136 & Other information technology services & 62090 & 0.0279 & Other information technology services \\ 
		49410 & 0.0124 & Freight transport by road & 64999 & 0.0186 & Financial intermediation n.e.c. \\ 
		70100 & 0.0111 & Activities of head offices & 96090 & 0.0128 & Other service activities n.e.c. \\ 
		61900 & 0.0107 & Other telecommunications activities & 64910 & 0.0115 & Financial leasing \\ 
		96090 & 0.0103 & Other service activities n.e.c. & 49410 & 0.0106 & Freight transport by road \\ 
		62020 & 0.0095 & Information technology consultancy & 65110 & 0.0095 & Life insurance \\ 
		65110 & 0.0094 & Life insurance & 46900 & 0.008 & Non-specialised wholesale trade \\ 
		\hline \\[-1.8ex] 
	\end{tabularx} 
\end{table}

\begin{table}[!htbp] \centering 
	\footnotesize
	\caption{Top 10 industries by influence vector (intermediary sectors removed)} 
	\label{app:tab:top10_influence_vector_no_intermed} 
	\begin{tabularx}{\textwidth}{@{\extracolsep{5pt}} ccX|ccX} 
		\\[-1.8ex]\hline 
		\hline \\[-1.8ex] 
		SIC & & Industry description & SIC & & Industry description\\ 
		\hline \\[-1.8ex] 
		\multicolumn{3}{l|}{2017}&&&\\ 
		\multicolumn{3}{c|}{Value}&\multicolumn{3}{c}{Count}\\ 
		\hline \\[-1.8ex] 
		82990 & 0.0786 & Other business support services n.e.c. & 61900 & 0.126 & Other telecommunications activities \\ 
		61900 & 0.0392 & Other telecommunications activities & 82990 & 0.0903 & Other business support services n.e.c. \\ 
		49410 & 0.0251 & Freight transport by road & 62090 & 0.0266 & Other information technology services \\ 
		62090 & 0.0246 & Other information technology services & 96090 & 0.0194 & Other service activities n.e.c. \\ 
		62020 & 0.0177 & Information technology consultancy & 49410 & 0.0153 & Freight transport by road \\ 
		70100 & 0.0161 & Activities of head offices & 70229 & 0.0111 & Non-financial management consultancy \\ 
		96090 & 0.0148 & Other service activities n.e.c. & 62020 & 0.0104 & Information technology consultancy \\ 
		70229 & 0.014 & Non-financial management consultancy & 70100 & 0.0101 & Activities of head offices \\ 
		32990 & 0.0113 & Other manufacturing n.e.c. & 77110 & 0.0098 & Renting \& leasing of cars \\ 
		77110 & 0.0106 & Renting \& leasing of cars & 69201 & 0.0074 & Accounting and auditing activities \\ 
		\hline \\[-1.8ex] 
		\hline \\[-1.8ex] 
		\multicolumn{3}{l|}{2023}&&&\\ 
		\multicolumn{3}{c|}{Value}&\multicolumn{3}{c}{Count}\\ 
		\hline \\[-1.8ex] 
		82990 & 0.0628 & Other business support services n.e.c. & 61900 & 0.078 & Other telecommunications activities \\ 
		49410 & 0.0245 & Freight transport by road & 82990 & 0.073 & Other business support services n.e.c. \\ 
		62090 & 0.024 & Other information technology services & 62090 & 0.0462 & Other information technology services \\ 
		61900 & 0.0196 & Other telecommunications activities & 96090 & 0.0201 & Other service activities n.e.c. \\ 
		96090 & 0.0179 & Other service activities n.e.c. & 49410 & 0.0179 & Freight transport by road \\ 
		62020 & 0.0159 & Information technology consultancy & 62020 & 0.0124 & Information technology consultancy \\ 
		70100 & 0.0146 & Activities of head offices & 70229 & 0.0118 & Non-financial management consultancy \\ 
		70229 & 0.014 & Non-financial management consultancy & 77110 & 0.0102 & Renting \& leasing of cars \\ 
		32990 & 0.0101 & Other manufacturing n.e.c. & 69201 & 0.0089 & Accounting and auditing activities \\ 
		43999 & 0.0096 & Other specialised construction n.e.c. & 70100 & 0.0088 & Activities of head offices \\ 
		\hline \\[-1.8ex] 
	\end{tabularx}

	\justifying \noindent \scriptsize
	Notes: 
	Transaction links to intermediary sectors and public administration (G45, G45, G46, K64, K65, K66, O84) have been removed from the data. 
\end{table} 

\begin{table}[!htbp] \centering 
	\footnotesize
	\caption{Top 10 industries by influence vector (service sectors removed)} 
	\label{app:tab:top10_influence_vector_no_services} 
	\begin{tabularx}{\textwidth}{@{\extracolsep{5pt}} ccX|ccX} 
		\\[-1.8ex]\hline 
		\hline \\[-1.8ex] 
		SIC & & Industry description & SIC & & Industry description\\ 
		\hline \\[-1.8ex] 
		\multicolumn{3}{l|}{2017}&&&\\ 
		\multicolumn{3}{c|}{Value}&\multicolumn{3}{c}{Count}\\ 
		\hline \\[-1.8ex] 
		32990 & 0.0515 & Other manufacturing n.e.c. & 32990 & 0.071 & Other manufacturing n.e.c. \\ 
		43999 & 0.0266 & Other specialised construction n.e.c. & 35140 & 0.0321 & Trade of electricity \\ 
		35140 & 0.0209 & Trade of electricity & 33200 & 0.032 & Industrial machinery installation \\ 
		25990 & 0.0206 &Metal products manufacture n.e.c. & 43999 & 0.0258 & Other specialised construction n.e.c. \\ 
		35110 & 0.019 & Production of electricity & 36000 & 0.02 & Water collection, treatment and supply \\ 
		33200 & 0.0173 & Industrial machinery installation & 25990 & 0.0187 &Metal products manufacture n.e.c. \\ 
		22290 & 0.0173 & Manufacture of other plastic products & 43210 & 0.0146 & Electrical installation \\ 
		43210 & 0.0159 & Electrical installation & 35110 & 0.0126 & Production of electricity \\ 
		35130 & 0.0107 & Distribution of electricity & 29100 & 0.0121 & Manufacture of motor vehicles \\ 
		10910 & 0.0106 & Manufacture of feeds for farm animals & 22290 & 0.0118 & Manufacture of other plastic products \\ 
		\hline \\[-1.8ex] 
		\hline \\[-1.8ex] 
		\multicolumn{3}{l|}{2023}&&&\\ 
		\multicolumn{3}{c|}{Value}&\multicolumn{3}{c}{Count}\\ 
		\hline \\[-1.8ex] 
		32990 & 0.033 & Other manufacturing n.e.c. & 32990 & 0.0535 & Other manufacturing n.e.c. \\ 
		43999 & 0.0304 & Other specialised construction n.e.c. & 43999 & 0.0252 & Other specialised construction n.e.c. \\ 
		35140 & 0.023 & Trade of electricity & 36000 & 0.0219 & Water collection, treatment and supply \\ 
		35130 & 0.0198 & Distribution of electricity & 33200 & 0.0207 & Industrial machinery installation \\ 
		25990 & 0.0174 &Metal products manufacture n.e.c. & 35140 & 0.0204 & Trade of electricity \\ 
		35110 & 0.0154 & Production of electricity & 35220 & 0.0168 & Gas fuels distribution through mains \\ 
		35220 & 0.0146 & Gas fuels distribution through mains & 35130 & 0.0168 & Distribution of electricity \\ 
		43210 & 0.0142 & Electrical installation & 43210 & 0.0156 & Electrical installation \\ 
		22290 & 0.0117 & Manufacture of other plastic products & 25990 & 0.0139 &Metal products manufacture n.e.c. \\ 
		42990 & 0.0109 & Civil engineering construction n.e.c. & 43220 & 0.0122 & Plumbing/heat/air-condition installation \\  
		\hline \\[-1.8ex] 
	\end{tabularx}

	\justifying \noindent \scriptsize
	Notes: 
	Transaction links to service-related sectors (G45-Q88, S94-U99) have been removed from the data. 
\end{table}

\FloatBarrier
\section{Concordance table}
\label{app:concordance}
Table \ref{tab:concordance} shows how industries classified by 5-digit SIC codes are re-allocated to CPA codes used in the official ONS IOT and national accounts data \citep{ons2009sic, eurostat2015cpa}. The 5-digit SIC codes are aggregated into 104 CPA classes. The codes in the first column (SIC) are short for the first 2-4 digits of the 5-digit codes. All industries with these digits as leading digits are aggregated into the respective CPA category. The `$\cdot$'s in the columns of the table indicate which SIC codes belong to a more aggregate CPA category. 

\begingroup\tiny \centering
\begin{longtable}{|p{0.5cm}|p{6cm}|p{1cm}|p{6cm}|}
  \hline \\[-1.8ex] 
SIC & SIC names & CPA & CPA names \\[1ex]
  \hline  \\[-1.8ex] 
  
  \endhead
01 & Crop and animal production, hunting and related service activities & A01 & Products of agriculture, hunting and related services \\ 
  02 & Forestry and logging & A02 & Products of forestry, logging and related services \\ 
  03 & Fishing and aquaculture & A03 & Fish and other fishing products; aquaculture products; support services to fishing \\ 
  05 & Mining of coal and lignite & B05 & Coal and lignite \\ 
  06 & Extraction of crude petroleum and natural gas & B06-F7 & Extraction Of Crude Petroleum And Natural Gas \& Mining Of Metal Ores \\ 
  07 & Mining of metal ores &$\cdot$&$\cdot$\\ 
  08 & Other mining and quarrying & B08 & Other mining and quarrying products \\ 
  09 & Mining support service activities & B09 & Mining support services \\ 
  101 & Preserved meat and meat products & C101 & Preserved meat and meat products \\ 
  102 & Processing and preserving of fish, crustaceans and molluscs & C102-3 & Processed and preserved fish, crustaceans, molluscs, fruit and vegetables \\ 
  103 & Processing and preserving of fruit and vegetables &$\cdot$&$\cdot$\\ 
  104 & Vegetable and animal oils and fats & C104 & Vegetable and animal oils and fats \\ 
  105 & Dairy products & C105 & Dairy products \\ 
  106 & Grain mill products, starches and starch products & C106 & Grain mill products, starches and starch products \\ 
  107 & Bakery and farinaceous products & C107 & Bakery and farinaceous products \\ 
  108 & Other food products & C108 & Other food products \\ 
  109 & Prepared animal feeds & C109 & Prepared animal feeds \\ 
  1101 & Distilling, rectifying and blending of spirits & C11.01-6 \& C12 & Alcoholic beverages \& Tobacco products \\ 
  1102 & Manufacture of wine from grape &$\cdot$&$\cdot$\\ 
  1103 & Manufacture of cider and other fruit wines &$\cdot$&$\cdot$\\ 
  1104 & Manufacture of other non-distilled fermented beverages &$\cdot$&$\cdot$\\ 
  1105 & Manufacture of beer &$\cdot$&$\cdot$\\ 
  1106 & Manufacture of malt &$\cdot$&$\cdot$\\ 
  1107 & Manufacture of soft drinks & C1107 & Soft drinks \\ 
  12 & Manufacture of tobacco products &$\cdot$&$\cdot$\\ 
  13 & Manufacture of textiles & C13 & Textiles \\ 
  14 & Manufacture of wearing apparel & C14 & Wearing apparel \\ 
  15 & Manufacture of leather and related products & C15 & Leather and related products \\ 
  16 & Manufacture of wood and of products of wood and cork, except furniture; manufacture of articles of straw and plaiting materials & C16 & Wood and of products of wood and cork, except furniture; articles of straw and plaiting materials \\ 
  17 & Manufacture of paper and paper products & C17 & Paper and paper products \\ 
  18 & Printing and reproduction of recorded media & C18 & Printing and recording services \\ 
  19 & Manufacture of coke and refined petroleum products & C19 & Coke and refined petroleum products \\ 
  2011 & Manufacture of industrial gases & C20A & Industrial gases, inorganics and fertilisers (all inorganic chemicals) - 20.11/13/15 \\ 
  2012 & Manufacture of dyes and pigments & C20C & Dyestuffs, agro-chemicals - 20.12/20 \\ 
  2013 & Manufacture of other inorganic basic chemicals &$\cdot$&$\cdot$\\ 
  2014 & Manufacture of other organic basic chemicals & C20B & Petrochemicals - 20.14/16/17/60 \\ 
  2015 & Manufacture of fertilisers and nitrogen compounds &$\cdot$&$\cdot$\\ 
  2016 & Manufacture of plastics in primary forms &$\cdot$&$\cdot$\\ 
  2017 & Manufacture of synthetic rubber in primary forms &$\cdot$&$\cdot$\\ 
  2020 & Manufacture of pesticides and other agrochemical products &$\cdot$&$\cdot$\\ 
  203 & Paints, varnishes and similar coatings, printing ink and mastics & C203 & Paints, varnishes and similar coatings, printing ink and mastics \\ 
  204 & Soap and detergents, cleaning and polishing preparations, perfumes and toilet preparations & C204 & Soap and detergents, cleaning and polishing preparations, perfumes and toilet preparations \\ 
  205 & Other chemical products & C205 & Other chemical products \\ 
  2060 & Manufacture of man-made fibres &$\cdot$&$\cdot$\\ 
  21 & Manufacture of basic pharmaceutical products and pharmaceutical preparations & C21 & Basic pharmaceutical products and pharmaceutical preparations \\ 
  22 & Manufacture of rubber and plastic products & C22 & Rubber and plastic products \\ 
  231 & Manufacture of glass and glass products & C23 other & Glass, refractory, clay, other porcelain and ceramic, stone and abrasive products - 23.1-4/7-9 \\ 
  232 & Manufacture of refractory products &$\cdot$&$\cdot$\\ 
  233 & Manufacture of clay building materials &$\cdot$&$\cdot$\\ 
  234 & Manufacture of other porcelain and ceramic products &$\cdot$&$\cdot$\\ 
  235 & Manufacture of cement, lime and plaster & C235-6 & Cement, lime, plaster and articles of concrete, cement and plaster \\ 
  236 & Manufacture of articles of concrete, cement and plaster &$\cdot$&$\cdot$\\ 
  237 & Cutting, shaping and finishing of stone &$\cdot$&$\cdot$\\ 
  239 & Manufacture of abrasive products and non-metallic mineral products n.e.c. &$\cdot$&$\cdot$\\ 
  241 & Manufacture of basic iron and steel and of ferro-alloys & C241-3 & Basic iron and steel \\ 
  242 & Manufacture of tubes, pipes, hollow profiles and related fittings, of steel &$\cdot$&$\cdot$\\ 
  243 & Manufacture of other products of first processing of steel &$\cdot$&$\cdot$\\ 
  244 & Manufacture of basic precious and other non-ferrous metals & C244-5 & Other basic metals and casting \\ 
  245 & Casting of metals &$\cdot$&$\cdot$\\ 
  251 & Manufacture of structural metal products & C25 other & Fabricated metal products, incl. machinery and equipment and weapons \& ammunition - 25.1-9 \\ 
  252 & Manufacture of tanks, reservoirs and containers of metal &$\cdot$&$\cdot$\\ 
  253 & Manufacture of steam generators, except central heating hot water boilers &$\cdot$&$\cdot$\\ 
  254 & Weapons and ammunition &$\cdot$&$\cdot$\\  
  255 & Forging, pressing, stamping and roll-forming of metal; powder metallurgy &$\cdot$&$\cdot$\\ 
  256 & Treatment and coating of metals; machining &$\cdot$&$\cdot$\\ 
  257 & Manufacture of cutlery, tools and general hardware &$\cdot$&$\cdot$\\ 
  259 & Manufacture of other fabricated metal products &$\cdot$&$\cdot$\\ 
  26 & Manufacture of computer, electronic and optical products & C26 & Computer, electronic and optical products \\ 
  27 & Manufacture of electrical equipment & C27 & Electrical equipment \\ 
  28 & Manufacture of machinery and equipment n.e.c. & C28 & Machinery and equipment n.e.c. \\ 
  29 & Manufacture of motor vehicles, trailers and semi-trailers & C29 & Motor vehicles, trailers and semi-trailers \\ 
  301 & Ships and boats & C301 & Ships and boats \\ 
  302 & Manufacture of railway locomotives and rolling stock & C30 other & Other transport equipment - 30.2/4/9 \\ 
  303 & Air and spacecraft and related machinery & C303 & Air and spacecraft and related machinery \\ 
  304 & Manufacture of military fighting vehicles &$\cdot$&$\cdot$\\ 
  309 & Manufacture of transport equipment n.e.c. &$\cdot$&$\cdot$\\ 
  31 & Manufacture of furniture & C31 & Furniture \\ 
  32 & Other manufacturing & C32 & Other manufactured goods \\ 
  3311 & Repair of fabricated metal products & C33 other & Rest of repair; Installation - 33.11-14/17/19/20 \\ 
  3312 & Repair of machinery &$\cdot$&$\cdot$\\ 
  3313 & Repair of electronic and optical equipment &$\cdot$&$\cdot$\\ 
  3314 & Repair of electrical equipment &$\cdot$&$\cdot$\\ 
  3315 &$\cdot$& C3315 & Repair and maintenance of ships and boats \\ 
  3316 &$\cdot$& C3316 & Repair and maintenance of aircraft and spacecraft \\ 
  3317 & Repair and maintenance of other transport equipment &$\cdot$&$\cdot$\\ 
  3319 & Repair of other equipment &$\cdot$&$\cdot$\\ 
  332 & Installation of industrial machinery and equipment &$\cdot$&$\cdot$\\ 
  351 & Electricity, transmission and distribution & D351 & Electricity, transmission and distribution \\ 
  352 & Manufacture of gas; distribution of gaseous fuels through mains & D352-3 & Gas; distribution of gaseous fuels through mains; steam and air conditioning supply \\ 
  353 & Steam and air conditioning supply &$\cdot$&$\cdot$\\ 
  36 & Water collection, treatment and supply & E36 & Natural water; water treatment and supply services \\ 
  37 & Sewerage & E37 & Sewerage services; sewage sludge \\ 
  38 & Waste collection, treatment and disposal activities; materials recovery & E38 & Waste collection, treatment and disposal services; materials recovery services \\ 
  39 & Remediation activities and other waste management services. & E39 & Remediation services and other waste management services \\ 
  41 & Construction of buildings & F41-43 & Construction \\ 
  42 & Civil engineering &$\cdot$&$\cdot$\\ 
  43 & Specialised construction activities &$\cdot$&$\cdot$\\ 
  45 & Wholesale and retail trade and repair of motor vehicles and motorcycles & G45 & Wholesale and retail trade and repair services of motor vehicles and motorcycles \\ 
  46 & Wholesale trade, except of motor vehicles and motorcycles & G46 & Wholesale trade services, except of motor vehicles and motorcycles \\ 
  47 & Retail trade, except of motor vehicles and motorcycles & G47 & Retail trade services, except of motor vehicles and motorcycles \\ 
  491 & Passenger rail transport, interurban & H491-2 & Rail transport services \\ 
  492 & Freight rail transport &$\cdot$&$\cdot$\\ 
  493 & Other passenger land transport & H493-5 & Land transport services and transport services via pipelines, excluding rail transport \\ 
  494 & Freight transport by road and removal services &$\cdot$&$\cdot$\\ 
  495 & Transport via pipeline &$\cdot$&$\cdot$\\ 
  50 & Water transport & H50 & Water transport services \\ 
  51 & Air transport & H51 & Air transport services \\ 
  52 & Warehousing and support activities for transportation & H52 & Warehousing and support services for transportation \\ 
  53 & Postal and courier activities & H53 & Postal and courier services \\ 
  55 & Accommodation & I55 & Accommodation services \\ 
  56 & Food and beverage service activities & I56 & Food and beverage serving services \\ 
  58 & Publishing activities & J58 & Publishing services \\ 
  59 & Motion picture, video and television programme production, sound recording and music publishing activities & J59-60 & Motion Picture, Video \& TV Programme Production, Sound Recording \& Music Publishing Activities \& Programming And Broadcasting Activities \\ 
  60 & Programming and broadcasting activities &$\cdot$&$\cdot$\\ 
  61 & Telecommunications & J61 & Telecommunications services \\ 
  62 & Computer programming, consultancy and related activities & J62 & Computer programming, consultancy and related services \\ 
  63 & Information service activities & J63 & Information services \\ 
  64 & Financial service activities, except insurance and pension funding & K64 & Financial services, except insurance and pension funding \\ 
  651 & Insurance & K65.1-3 & Insurance, reinsurance and pension funding services, except compulsory social security \\ 
  652 & Reinsurance &$\cdot$& $\cdot$ \\ 
  653 & Pension funding &$\cdot$&$\cdot$\\ 
  66 & Activities auxiliary to financial services and insurance activities & K66 & Services auxiliary to financial services and insurance services \\ 
  681 & Buying and selling of own real estate & L68 BX L683 & Real estate services, excluding on a fee or contract basis and imputed rent \\ 
  682 & Owner-Occupiers' Housing Services & L68A & Owner-Occupiers' Housing Services \\ 
 $\cdot$& Renting and operating of own or leased real estate &$\cdot$&$\cdot$\\ 
  683 & Real estate services on a fee or contract basis & L683 & Real estate services on a fee or contract basis \\ 
  691 & Legal services & M691 & Legal services \\ 
  692 & Accounting, bookkeeping and auditing services; tax consulting services & M692 & Accounting, bookkeeping and auditing services; tax consulting services \\ 
  70 & Activities of head offices; management consultancy activities & M70 & Services of head offices; management consulting services \\ 
  71 & Architectural and engineering activities; technical testing and analysis & M71 & Architectural and engineering services; technical testing and analysis services \\ 
  72 & Scientific research and development & M72 & Scientific research and development services \\ 
  73 & Advertising and market research & M73 & Advertising and market research services \\ 
  74 & Other professional, scientific and technical activities & M74 & Other professional, scientific and technical services \\ 
  75 & Veterinary activities & M75 & Veterinary services \\ 
  77 & Rental and leasing activities & N77 & Rental and leasing services \\ 
  78 & Employment activities & N78 & Employment services \\ 
  79 & Travel agency, tour operator and other reservation service and related activities & N79 & Travel agency, tour operator and other reservation services and related services \\ 
  80 & Security and investigation activities & N80 & Security and investigation services \\ 
  81 & Services to buildings and landscape activities & N81 & Services to buildings and landscape \\ 
  82 & Office administrative, office support and other business support activities & N82 & Office administrative, office support and other business support services \\ 
  84 & Public administration and defence; compulsory social security & O84 & Public administration and defence services; compulsory social security services \\ 
  85 & Education & P85 & Education services \\ 
  86 & Human health activities & Q86 & Human health services \\ 
  87 & Residential care activities & Q87-88 & Residential Care \& Social Work Activities \\ 
  88 & Social work activities without accommodation &$\cdot$&$\cdot$\\ 
  90 & Creative, arts and entertainment activities & R90 & Creative, arts and entertainment services \\ 
  91 & Libraries, archives, museums and other cultural activities & R91 & Libraries, archives, museums and other cultural services \\ 
  92 & Gambling and betting activities & R92 & Gambling and betting services \\ 
  93 & Sports activities and amusement and recreation activities & R93 & Sports services and amusement and recreation services \\ 
  94 & Activities of membership organisations & S94 & Services furnished by membership organisations \\ 
  95 & Repair of computers and personal and household goods & S95 & Repair services of computers and personal and household goods \\ 
  96 & Other personal service activities & S96 & Other personal services \\ 
  97 & Activities of households as employers of domestic personnel & T97 & Services of households as employers of domestic personnel \\ 
   \hline
\hline
\caption{Concordance table from SIC to CPA codes.} 
\label{tab:concordance}
\end{longtable}
\endgroup

\FloatBarrier

\end{document}